\documentclass[traditabstract]{aa}
\usepackage[varg]{txfonts}
\usepackage{time}
\usepackage{color}
\usepackage{natbib}
\usepackage[percent]{overpic}
\usepackage{array}
\newcolumntype{C}{>{$}c<{$}} 
\usepackage{amstext,amsmath,mathtools}
\usepackage{graphicx}
\usepackage{textcomp}
\usepackage{txfonts}
\usepackage{url}
\usepackage{booktabs} 
\usepackage{lineno}
\newcommand{\gs}{}

\def\Fourier{\mathfrak{F}}
\def\abs#1{\left|#1\right|}
\def\expten#1#2{\ensuremath{#1\times 10^{#2}}}

\begin{document}
\titlerunning{Multi-aperture FPI systems for EST}
\title{Operating the Fabry-P\'erot systems of the European Solar Telescope in multi-aperture mode}

\author{G.B. Scharmer\inst{1,3}
\and
M.G. L{\"o}fdahl\inst{1} 
\and
J. de la Cruz Rodr\'iguez \inst{1}
\and
B. Lindberg\inst{2}
\and
H. Socas-Navarro\inst{3}
\and 
D. Kiselman \inst{1}
\and
M. Rempel\inst{4}
\and 
J. Leenaarts\inst{1}
}

\institute{Institute for Solar Physics, Dept. of Astronomy, Stockholm University,
AlbaNova University Center, SE 106\,91 Stockholm, Sweden \and
Lens Tech AB, Tallbackagatan 11, SE 931\,64 Skellefte\aa, Sweden \and
EST Foundation, C/ Via Láctea s/n 38200 La Laguna, Spain \and
High Altitude Observatory, NSF National Center for Atmospheric Research, P.O. Box 3000, Boulder, CO 80307, USA}
\date{Draft: \now\ \today}
\frenchspacing

\abstract{We discuss how to optimise the science output of the European Solar Telescope (EST), when used without the wide-field compensation for high-altitude seeing that the EST multi conjugate adaptive optics (MCAO) system will offer. This will be the mode of operating EST during its first year(s). Without MCAO, in particular with the Sun at low elevations or when observing at short wavelengths, high-altitude seeing will strongly attenuate the signal at high spatial frequencies over almost the entire 1 arc min science field-of-view (FOV) of EST. To mitigate the consequences of this, we propose to provide EST with an optional multi-aperture mode, by optically segmenting the 4.2 m aperture into six 1.4 m subapertures. Operating at smaller aperture diameter pushes down the root mean square wavefront errors from the high altitude seeing to levels that can more reliably be compensated for with conventional ground-layer adaptive optics and post processing of restored images using e.g., the MOMFBD methods developed for the Swedish 1-m Solar Telescope. This will significantly improve image quality over the entire FOV. In particular, the multi-aperture mode will provide the sustained stable high image quality needed for obtaining time sequences of spectropolarimetric data. The multi-aperture mode is implemented with low-cost modifications of the camera lenses of  the three Fabry-P\'erot systems that will be used to cover the wavelength range 380--860~nm. Switching between the full-aperture and multi-aperture modes can be done quickly and independently for the three FPI systems. This allows flexible optimisation of EST, taking into account the daily variability of seeing, and that the impact of seeing is much stronger at short wavelengths than at long wavelengths, and when the Sun is observed at low elevation. Providing EST with a multi-aperture option represents a value-added aspect without any negative impact on the EST primary or secondary optical systems or on the actual FPI systems. Additionally, by tilting the high-resolution etalon of the FPI system in multi-aperture mode, the number of wavelength tunings in a spectral line can also optionally be reduced by a factor two to support high-cadence observations. The multi-aperture addition to EST provides a powerful and flexible option that has the potential of significantly improving the quality and amount of its science data before MCAO is operational, and perhaps even after installation of MCAO, when the Sun is observed at low elevation. In this publication, we perform simulations and image reconstructions of simulated data to demonstrate the benefits of the multi-aperture option, and provide a simple optical design to demonstrate its feasibility.
}

\maketitle
\section{Introduction}\label{Introduction}

 The construction of 4-m class solar telescopes -- the Daniel K. Inouye Solar Telescope \citep[DKIST;][]{2018SPIE10700E..0VW,2020SoPh..295..172R} and the planned European Solar Telescope \citep[EST;][]{2022A&A...666A..21Q} -- is motivated in roughly equal parts by the need to resolve the fundamental scales in the solar atmosphere, and the need to improve signal-to-noise ratio (SNR) in spectropolarimetry. One major science goal is to observe energetic and dynamic transient events and their associated relatively weak magnetic fields in the chromosphere and corona at the highest possible spatial resolution, and with a cadence that matches their temporal evolution. 
However, measuring weak magnetic fields will require prioritising SNR by sacrificing spatial resolution, either in software or in hardware. Therefore, the design of EST and its instrumentation must be simultaneously optimised with respect to both high spatial resolution and high SNR. A crucial aspect of that instrumentation is the use of short exposures\footnote{ That is, exposures that are short enough to freeze the point spread function (PSF). In principle, this requires integration times of  a few ms, but in practice exposure times of 10-15~ms, or even longer, are used. When calculating averaged PSFs, individual PSFs are here aligned either by removal of the individual wavefront tip-tilt terms, or by aligning the PSFs with respect to their brightest speckle, as described in Sects. \ref{sec:seeing_degraded_psf} and \ref{sec:psf_analysis}} and image reconstruction techniques to dramatically improve image quality.

Whereas there are thus strong arguments supporting the need for the 4.2-m aperture of EST, this leads to the awkward challenge of dealing with high-altitude seeing, in particular that originating close to the tropopause some 8--10~km above the telescope. During observations in the morning, with the Sun at low elevations, this seeing layer is at a distance of 20~km or more from the telescope. To reach the science goals of EST, a large field of view (FOV) of about 1\arcmin{} is required. As explained below, this requires a complex multi-conjugate adaptive optics (MCAO) system that compensates for seeing from several layers in the atmosphere. However, for technical reasons, MCAO will not be available initially, and this will strongly limit the useful part of ESTs science FOV. The proposal made here aims at mitigating the absence of such a system with a focus on observations made in the morning, with the Sun at low elevation. This is when the ground-layer seeing often is excellent, but also is at the peak of the high-altitude seeing. The proposal we will make adds flexibility and improves the science potential of EST quite significantly, without any negative impact on its overall performance or complexity.

Below, we first explain why an MCAO system is particularly important for a telescope like EST but not for smaller telescopes, such as the Swedish 1-m Solar Telescope (SST). To simplify the discussion, we will ignore ground-layer seeing completely, and only consider the impact of high-altitude seeing. This corresponds to assuming that EST operates with a ground-layer AO system that is 100\% efficient, which obviously is unrealistic, but that helps to isolate the particular challenges of the high-altitude seeing.

The impact of high-altitude seeing can be compensated for with a 1-m solar telescope such as SST, simply by using short exposures and advanced image reconstruction techniques \citep{1994A&AS..107..243L, 2002SPIE.4792..146L, 2005SoPh..228..191V, 2021A&A...653A..68L}. However, for a 4.2~m telescope this is much more challenging because the impact of seeing worsens dramatically when increasing the telescope diameter from 1~m to 4~m.

Seeing is often characterised in terms of Fried's parameter $r_0$ \citep{1966JOSA...56.1372F}, which is related to the average wavefront variance and corresponding Strehl ratio of a short exposure through \citep[][Eq. 2.17]{2004aoa..book.....R}
\begin{equation}
\sigma^2 = 0.134 (D/r_0)^{5/3},
\end{equation}
where $\sigma$ is the wavefront rms (in radians), $D$ is the telescope diameter,  and the Strehl ratio $S$ is given by
 \begin{equation}
S = e^{-\sigma^2},
\end{equation}
when the aberrations are small such that the Strehl is higher than about 0.3. At present, we cannot provide a rigorous estimate of the range of $r_0$ values to be expected from high-altitude seeing above La Palma, but based on experience with image restoration of data from SST and seeing measurements made with the SST AO system, we can make rough but meaningful estimates -- this is discussed in Appendix \ref {SST_AO_measurements}. In the following, we adopt values of  $r_0$ based on that discussion. 

For observations made in the morning with the Sun at low elevation, which is when the daytime seeing often peaks, assuming $r_0=0.35$~m must be considered optimistic. If we adopt this value and furthermore assume $D=0.98$~m, corresponding to the diameter of SST, we obtain $\sigma=0.86$~radian or 0.14~wave, and $S=0.47$. This corresponds to a moderate impact on the image quality, which can be further improved with multi-frame blind deconvolution (MFBD), multi-object multi-frame blind deconvolution (MOMFBD), and phase diversity (PD) image reconstruction techniques \citep{1994A&AS..107..243L, 2002SPIE.4792..146L, 2005SoPh..228..191V, 2021A&A...653A..68L}. But if we instead assume $D=4.2$~m, corresponding to the diameter of EST, we obtain $\sigma=2.90$~radian or 0.46~wave, and $S=0.0002$ (although Eq. 2 should not be used to calculate Strehl values that are less than about 0.3). Though the increase of wavefront rms ``only'' is a factor 3.4, this obviously has a devastating effect on the image quality. This is why a large telescope such as EST needs an  MCAO system that compensates seeing from both the ground layer and high altitude \citep{2010SPIE.7736E..0US}, whereas the 1-m SST only needs single-conjugate adaptive optics (SCAO). 

The problem we are addressing here, is that according to the present Project Management Plan \citep{managementplan2025} and Operations Plan \citep{operationsplan2025}, EST will only have SCAO, but not MCAO, operating during its first year(s) of operation. This will strongly limit its imaging performance. With a wavefront sensor (WFS) that has a small FOV, the SCAO system can compensate also for the high-altitude seeing, but because of the large distance to that seeing layer, this compensation will only be good over a very small angular region, corresponding to the isoplanatic patch \citep[][Eq. 2.45]{2004aoa..book.....R}, which is typically $0.3 r_0/L$.  With $L=20$~km, this isoplanatic patch is less than a few arc secs, which is too small. 

The challenges of reaching high image quality {\gs over a large FOV without MCAO is then that the modulation transfer function (MTF) suffers severe degradation at mid-to-high spatial frequencies, such that recovering the full spatial resolution in restored images will require (much) higher SNR than with a highly performing MCAO system. This problem is} aggravated by the generic difficulty of achieving near diffraction limited images of solar fine structure when the telescope diameter is increased. This is a direct consequence of power spectra of solar fine structure decreasing in amplitude at higher and higher spatial frequencies, such that achieving near diffraction limited performance with a large telescope requires higher SNR than with a smaller telescope. {\gs In this way, the achievable spatial resolution is highly object- and noise-dependent, in addition to dependending on the strength of the high-altitude seeing. This aspect of the performance of EST is of particular importance for the use with narrow-band FPI systems, with which the SNR is strongly limited by the required cadence of recording spectropolarimetric data. In Sects. \ref{sec:seeing_degraded_psf}, \ref{sec:artificial_data} and the subsequent sections, we therefore investigate the combined effects of high-altitude seeing and noise on the image quality, by means of simulations.}

The above discussion provides the main motivation for the present proposal. The key to the solution proposed is that solar imaging, whether made with wideband imaging or narrowband imaging with FPI-based spectropolarimeters, is made with short exposures (typically 10--15~ms). It is well-known that by using short exposures, sufficiently small telescopes can reach diffraction limited resolution without AO \citep{1966JOSA...56.1372F}, (see Roddier, Ch. 3.3, in particular his Fig. 3.2). The optimum gain in spatial resolution is obtained when the telescope diameter is about 4--5 times $r_0$, which in the case of $r_0$ being 0.35~m, corresponds to a diameter of about 1.4--1.8~m. For a smaller value of $r_0$ of 0.20~m, the optimum diameter would be slightly less than 1~m. 

\begin{figure}[!t]
\centering
\includegraphics[angle=0, width=0.7\linewidth,clip] {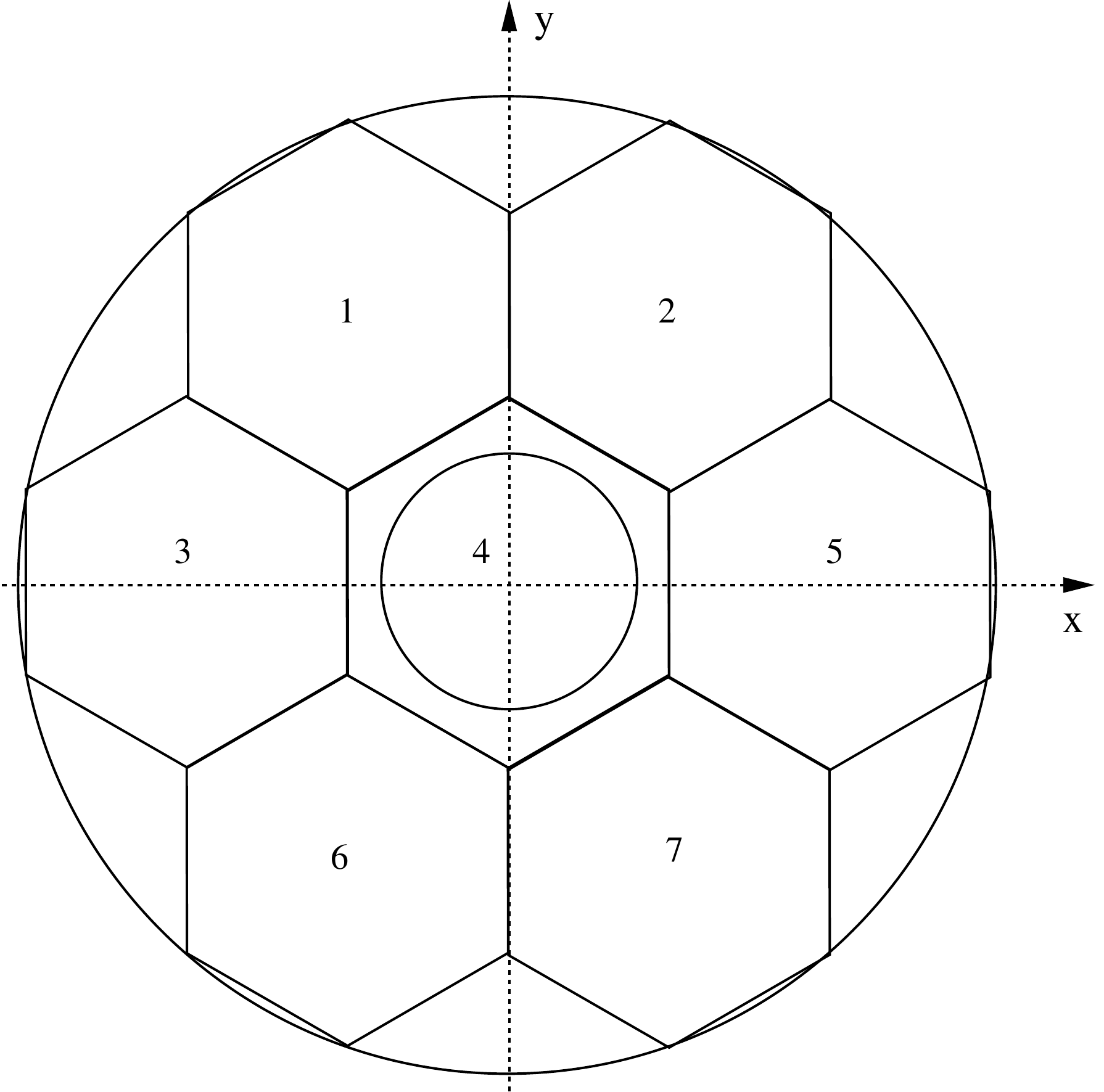}
 \caption{Schematic layout of the EST aperture (large circle) with its 1.1~m central obscuration (small circle), and the proposed seven (six useful) 1.4~m hexagonal subapertures. The numbering of the subapertures shown is used in Tables \ref{table_EST-V}, \ref{table_EST-B}, and \ref{table_EST-R} discussing layout and predicted performance. The $x$ and $y$ coordinate system shown, normalised to the diameter of a sub-aperture, is used to define the center coordinates of the sub-apertures. Tilting of the etalon is implemented as a rotation around the x-axis.}
 \label{fig:multi-aperture_layout}
 \end{figure}

In this paper, we therefore propose to provide a multi-aperture option for EST, which effectively converts its 4.2~m aperture into six 1.4~m subapertures, and to record images from these separately. Schematically, this is shown in Fig.~\ref{fig:multi-aperture_layout}\footnote{Hexagonal subapertures are preferred over circular subapertures because of their higher fill factor. Note that the orientation angle of the hexagons is given by the layout of the subimages on the detector.}. Repeating the previous calculations with $D=1.4$~m, we obtain $\sigma=1.16$~rad or 0.18~wave, and $S=0.26$. This Strehl value is more than an order of magnitude higher than obtained with the full 4.2~m aperture, and corresponds to an image quality that stands a good chance of being further improved by image restoration. We argue that a choice of 1.4~m sub-apertures represents a well balanced compromise in the sense of trading the diffraction limited resolution of EST over a very small FOV for a lower spatial resolution, but with excellent image quality and outstanding signal-to-noise, over a large (1\arcmin{} diameter) FOV. 

\begin{figure}[!t]
 \centering
     \includegraphics[width=0.49\linewidth]{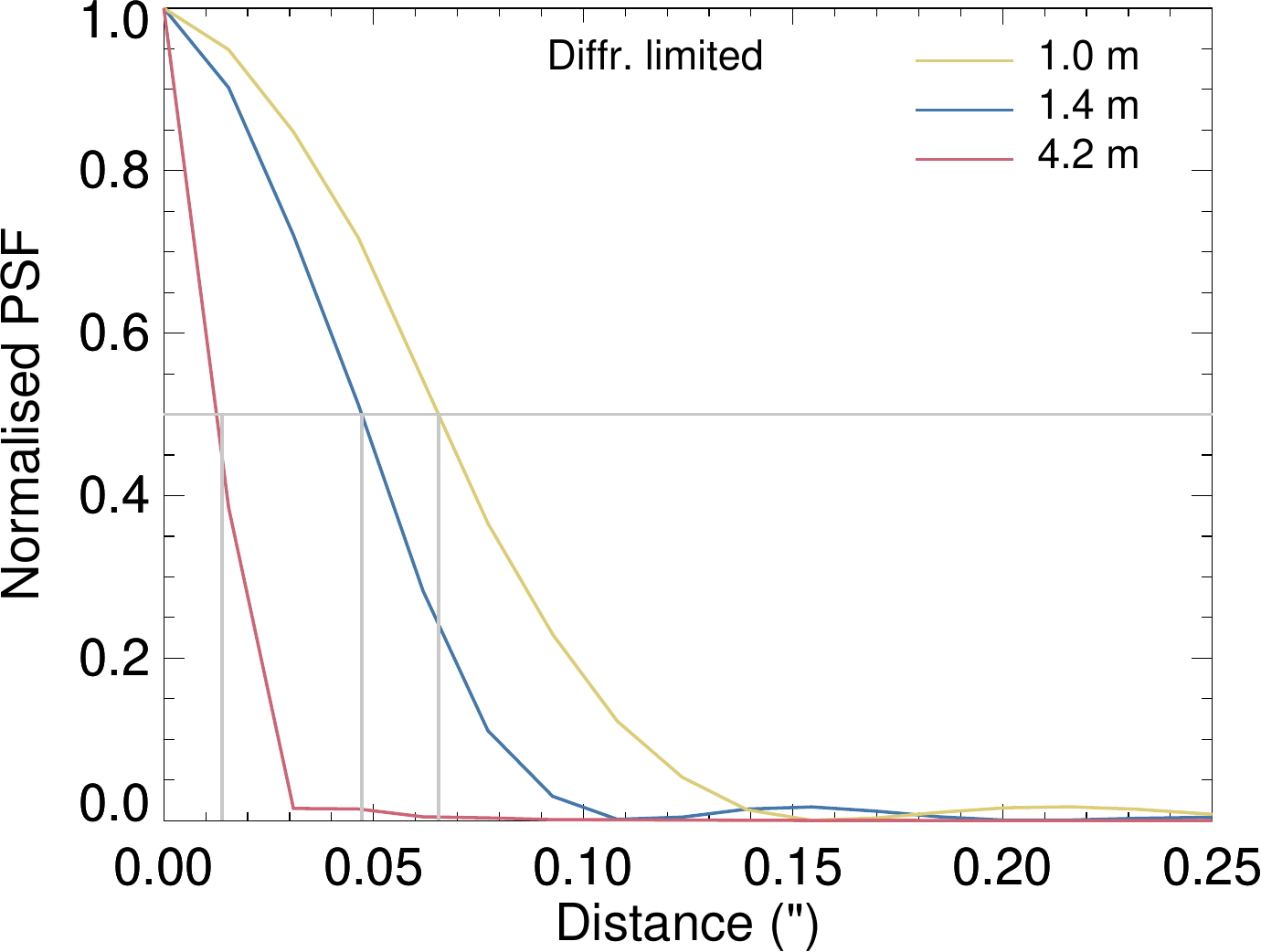} 
     \includegraphics[width=0.49\linewidth]{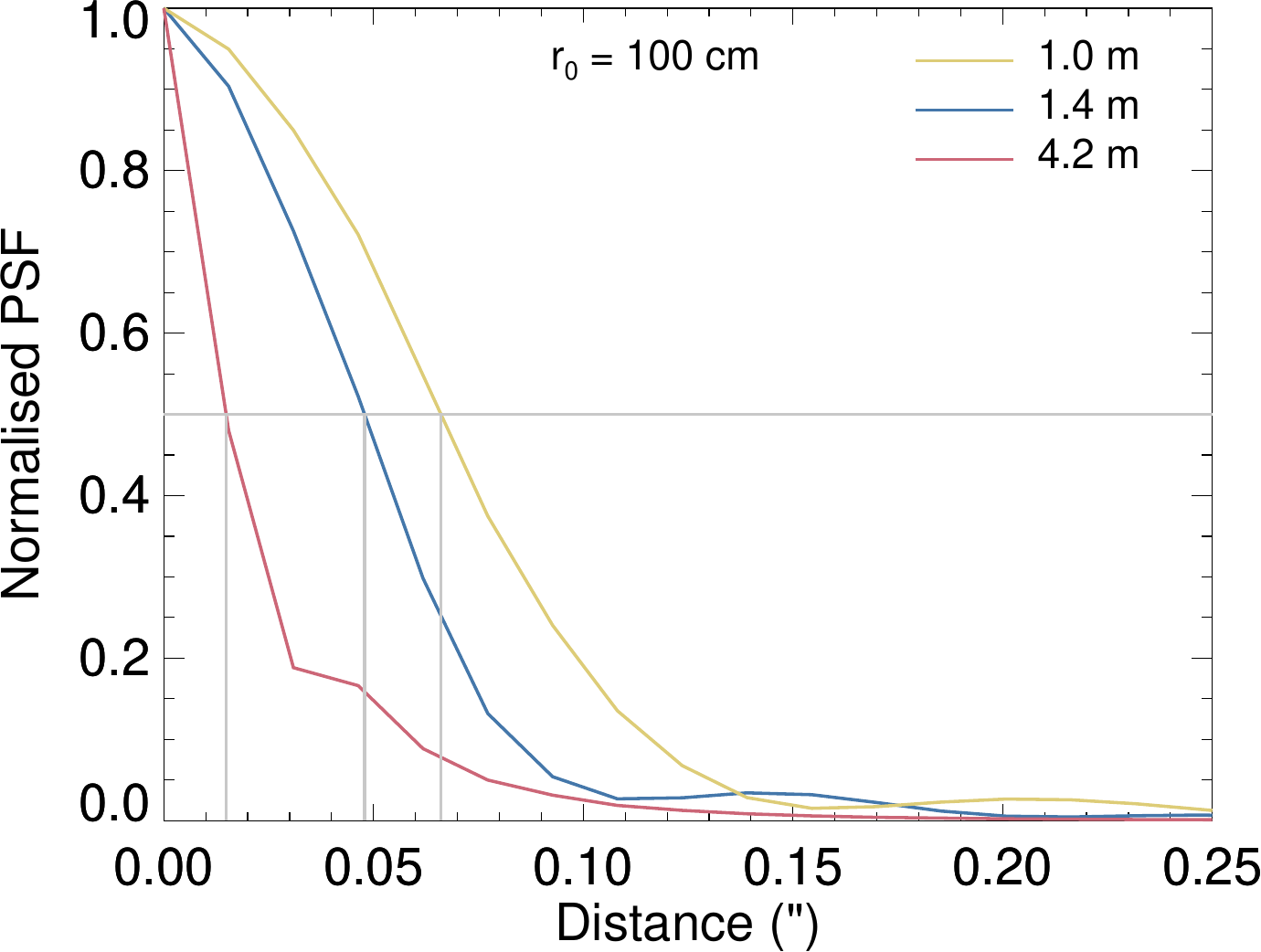} \\[1.5mm]
     \includegraphics[width=0.49\linewidth]{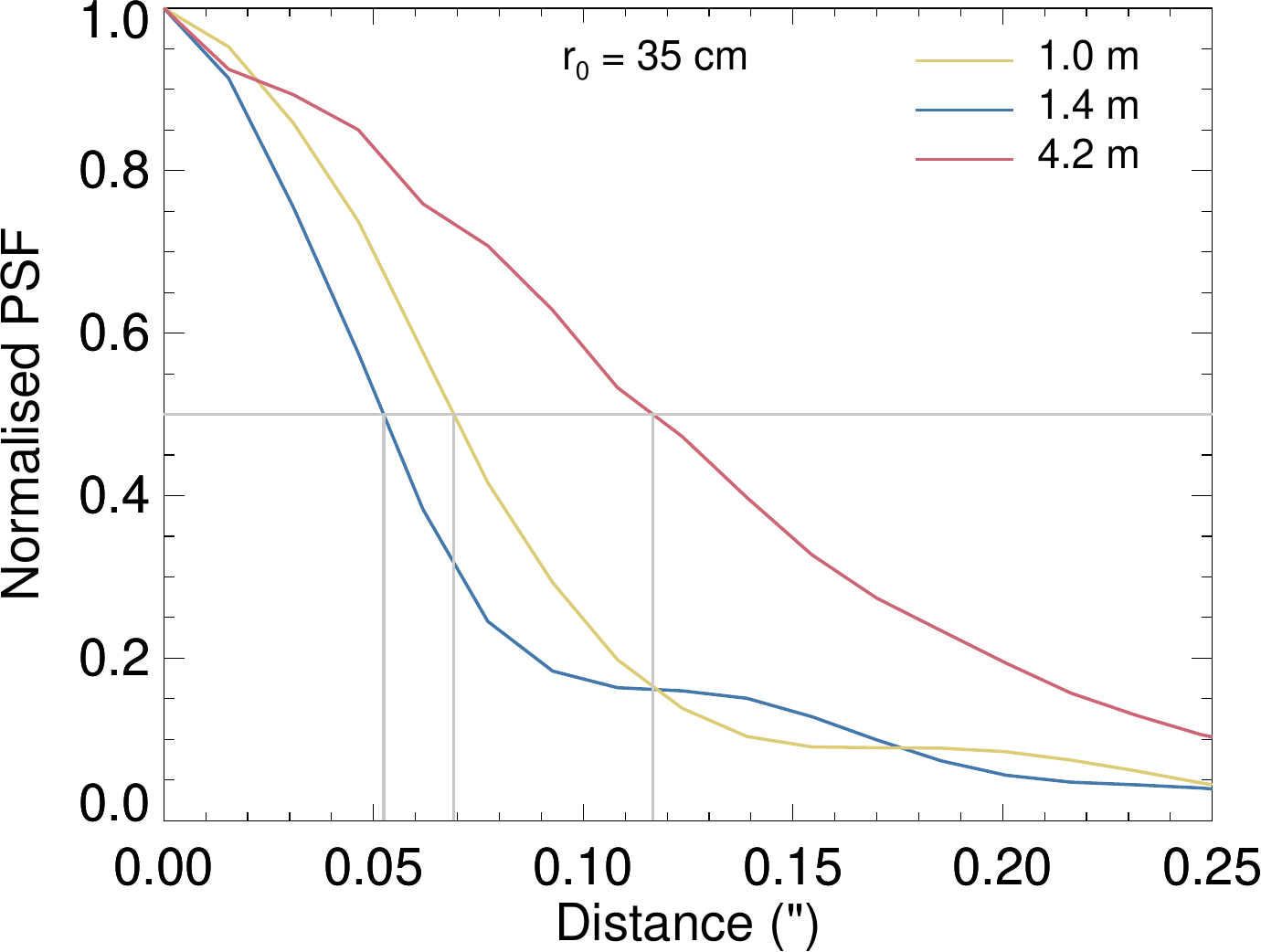}
     \includegraphics[width=0.49\linewidth]{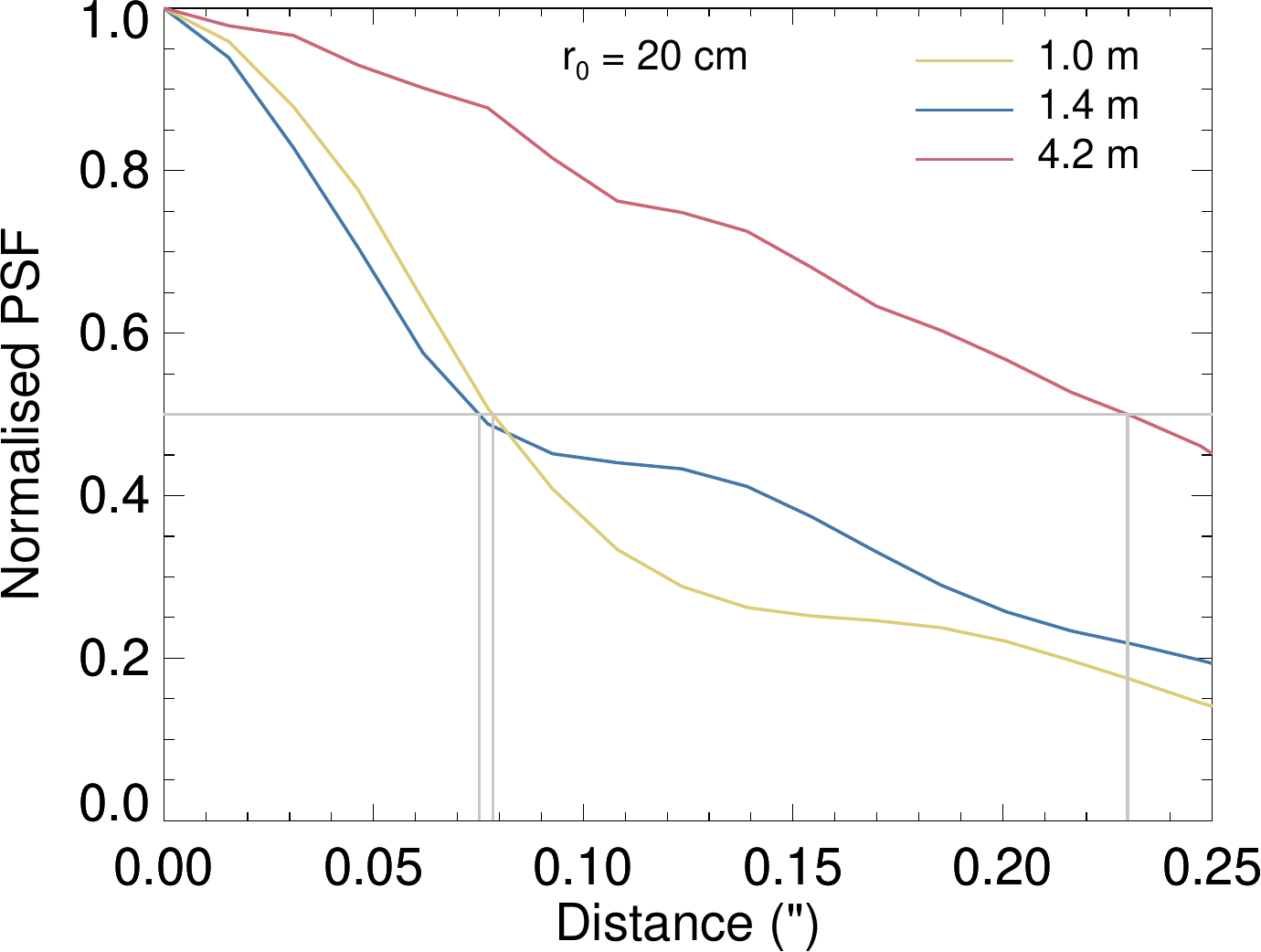} 
  \caption{The short-exposure PSFs of 1~m, 1.4~m, and 4.2~m apertures, normalised to unity at their peaks. Note that the FWHM (indicated with vertical gray lines) of the seeing degraded PSFs is larger for the 4.2~m aperture than for the smaller apertures when $r_0$ is 0.35~m or smaller. }
  \label{fig:psfs}
\end{figure}

This multi-aperture option is intended to enable science requiring a large FOV using SCAO alone (without MCAO), thereby optimising the scientific output of EST in prevailing seeing conditions. However, SST statistics of the 25\% best high-altitude seeing implies $r_0$ values higher than 0.45~m when the Sun is observed at small zenith distances (Appendix \ref{SST_AO_measurements}, Fig. \ref{fig:SST_seeing}). Based on experience with SST, such days with excellent ground layer seeing close to noon are not commonplace but when they do occur, observations with the full 4.2~m aperture can deliver more highly resolved images of excellent quality. This is then the preferred configuration of EST. 

The $r_0$ values discussed above and in Appendix \ref{SST_AO_measurements} refer to seeing quality at a wavelength of 500~nm. However, $r_0$ scales with wavelength as $\lambda^{6/5}$ \citep[][Table 2.1]{2004aoa..book.....R}. When $r_0$ is 0.3~m at 500~nm, then $r_0$ equals 0.22~m at 390~nm and 0.58~m at 860~nm, which are wavelengths of particular interest for diagnosing dynamics and magnetic fields in the chromosphere. This example, which would suggest using the multi-aperture mode at 390~nm and the 4.2~m full-aperture mode at 860~nm, clearly illustrates that the choice of  observing mode of EST must be made on the basis of  $r_0$ at the wavelength observed. The solution we present here entails multi-aperture configurations that are implemented individually for the three FPI systems of EST, covering the wavelength ranges 380--500~nm, 500--630~nm, and 630--860~nm \citep{2026A&A...705A..55S, 2025arXiv250521053S}. This provides the desired flexibility in optimising EST for each FPI system independently. 

The actual implementation of the multi-aperture configuration of EST is discussed in Sect.~\ref{sect:multiaperture_implementation}. First, we validate the above proposal by means of simulations of seeing and noise degraded solar images (Sects. 2 and 3) and image reconstruction of these images (Sect. 4), using the same MFBD techniques that have successfully been used for processing of data from the SST FPI systems CRISP, CRISP2 and CHROMIS \citep{2026A&A...705A..55S}. We demonstrate with simulations the advantages of the proposed multi-aperture approach to observing with EST.

\section{Simulations}
\subsection{Seeing simulations}
\label{sec:seeing}

The seeing simulations are simplified by making the assumption that the EST SCAO system perfectly compensates for the ground layer seeing. The simulations then represent the remaining seeing from the high-altitude layers (tropopause) that remains uncompensated for in the absence of an MCAO system. However, we relocate this seeing layer to the entrance pupil, such that we represent the strength of that seeing layer, but ignore the (strong) angular variation of that seeing.

To generate random phase screens with a given $r_0$, we use the power spectrum method, as described by \citet{2012ApOpt..51.7953K}. The phase screen $\phi_i(x)$ is given by its Fourier spectrum times a complex array of random numbers, $R_{N,N}$, where the real and imaginary parts are independent realizations of a Normal distribution with unit variance,
\begin{equation}
  \label{eq:2}
  \phi_i(x) = \Fourier^{-1}\left\{
    \sqrt{0.023} \left(\frac{Ns}{r_0}\right)^{5/6}  
    \abs{u}^{-11/6} R_{N,N} 
  \right\}.
\end{equation}
Here $u$ is the 2D integer coordinates in the Fourier domain, $N$ is the size of the square array, $s$ is the pupil sampling in meters per pixel, and $\Fourier^{-1}$ is the inverse Fourier transform operator.

The power method is known to produce phase screens for which the power at low spatial frequencies is underrepresented. We correct for this deficiency by replacing a low-order component of $\phi_i$ with one based on a sum of Karhunen--Loeve (KL) modes. The low-order component to be subtracted is made by least-squares fitting a number of KL modes to $\phi_i$. The added replacement component is made by summing the same KL modes with coefficients that have normal distributions, and variances as expected from Kolmogorov turbulence. In this way, we obtain a correct combination of low-order random KL modes at large spatial scales and high-order modes from the power method, and phase screens that include all scales from the pixel scale up to the scales of the 4.2~m pupil of EST.

A fundamentally important feature of our simulations is the goal to predict the image quality obtained with short exposures. To implement that, we remove the Zernike tip and tilt modes from all simulated phase screens, before calculating averages of seeing degraded point spread functions and images. 

\subsection{Seeing degraded PSFs and MTFs}
\label{sec:seeing_degraded_psf}
Figure \ref{fig:psfs} shows the diffraction limited PSFs and the corresponding seeing degraded short-exposure PSFs, based on averages from 100 phase screens and removing their tip-tilt modes. The PSFs are normalised to unity at their peaks. The plots in the upper row of panels show the diffraction limited PSF and seeing averaged PSF with $r_0=1.0$~m. This large value of $r_0$ is not realistic but provides us with a baseline in the sense of allowing high quality image restorations using MFBD applied to EST images. The plots in the lower row shows the PSFs with  $r_0=0.35$~m  and 0.20~m. As can be seen, the PSF of the 4.2~m aperture broadens considerably when $r_0$ is small, such that it is wider than the PSFs of the 1.0~m and 1.4~m apertures. Calculations demonstrate that, when $r_0$ is less than about 0.5~m, the FWHM of  the 1.4~m aperture PSF is narrower than for a 4.2~m aperture, when short exposures are used. 
\begin{figure}[!t]
 \centering
     \includegraphics[height=0.375\linewidth]{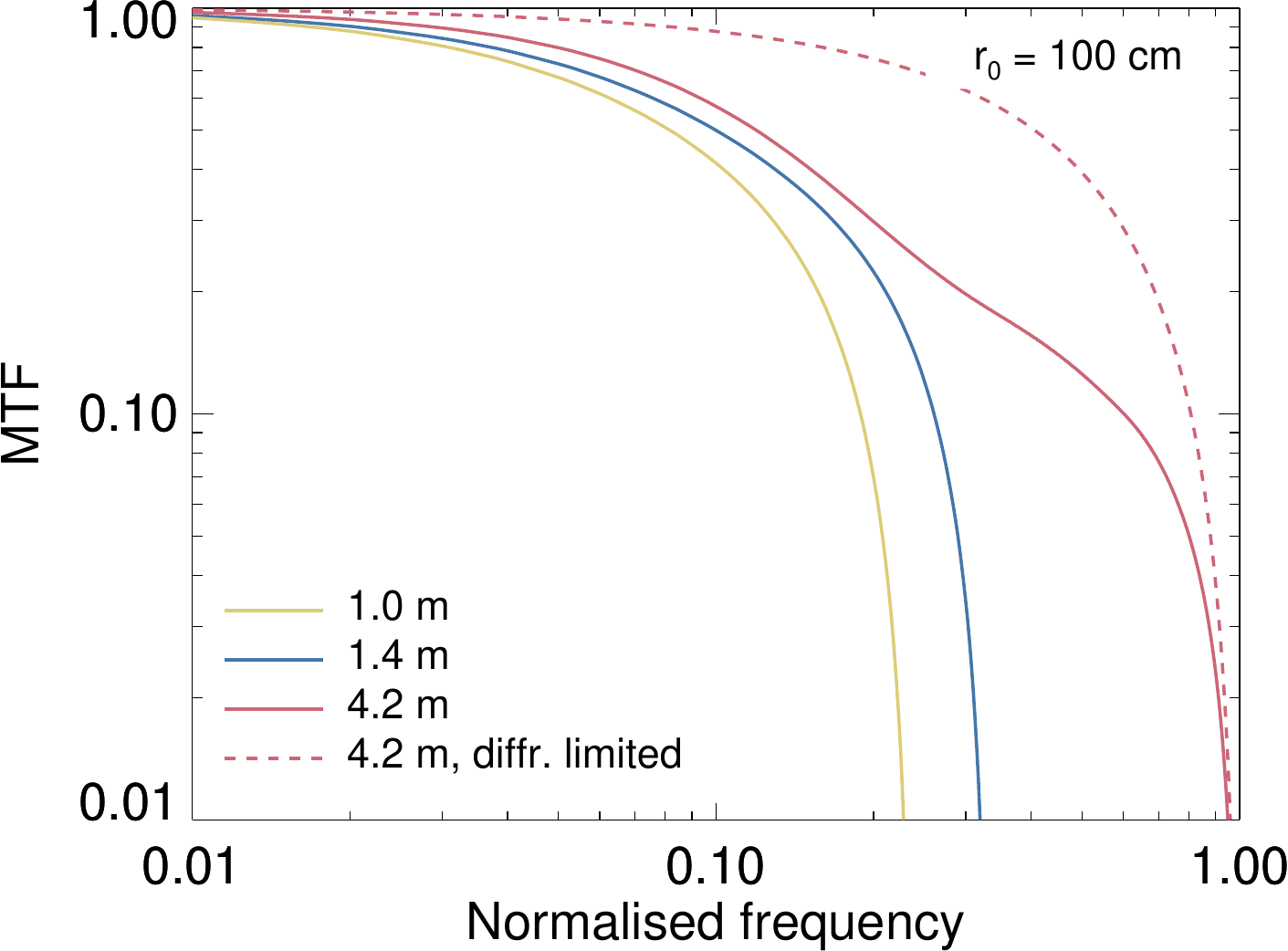 }
     \includegraphics[height=0.375\linewidth]{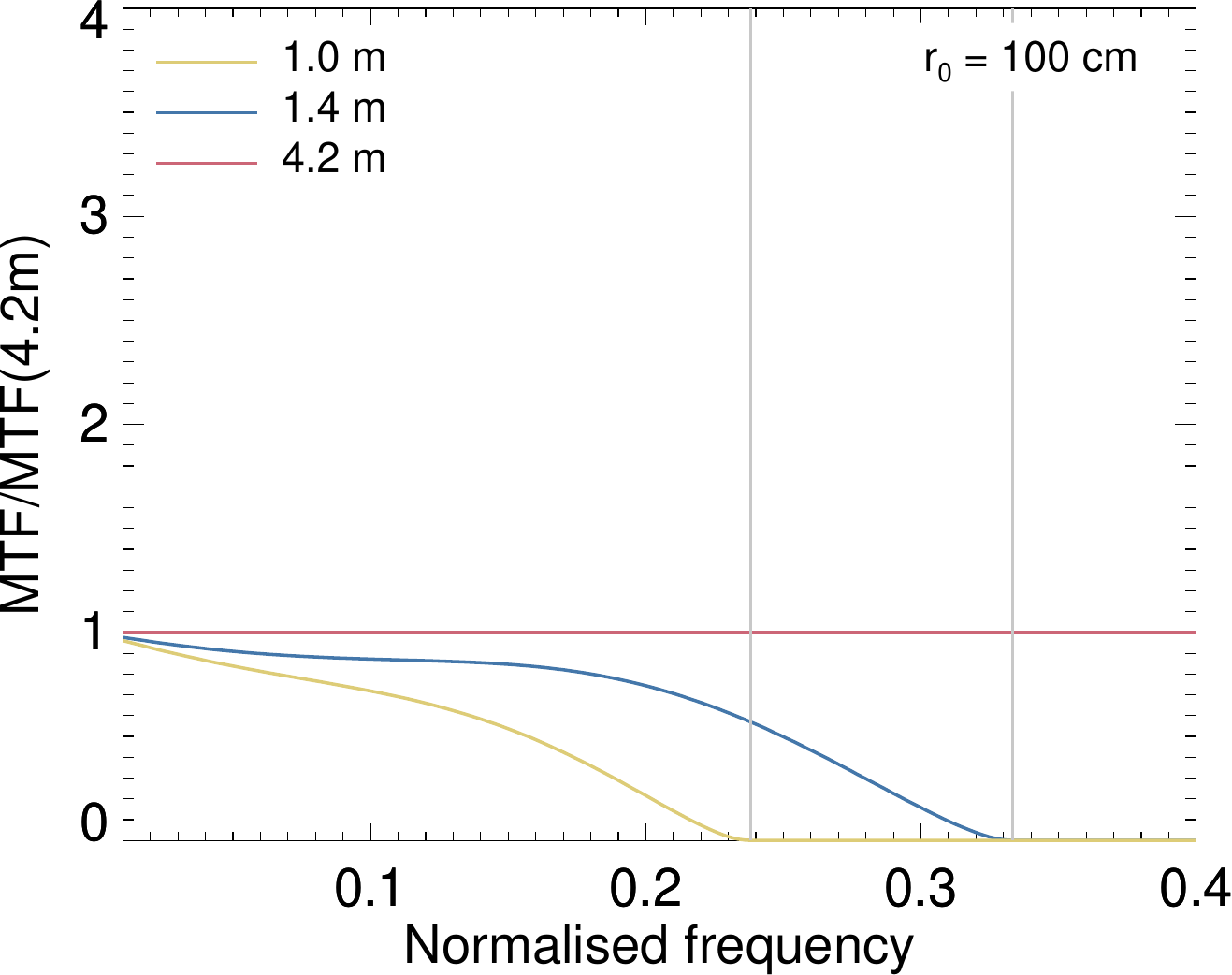 }\\[1.5mm]
     \includegraphics[height=0.375\linewidth]{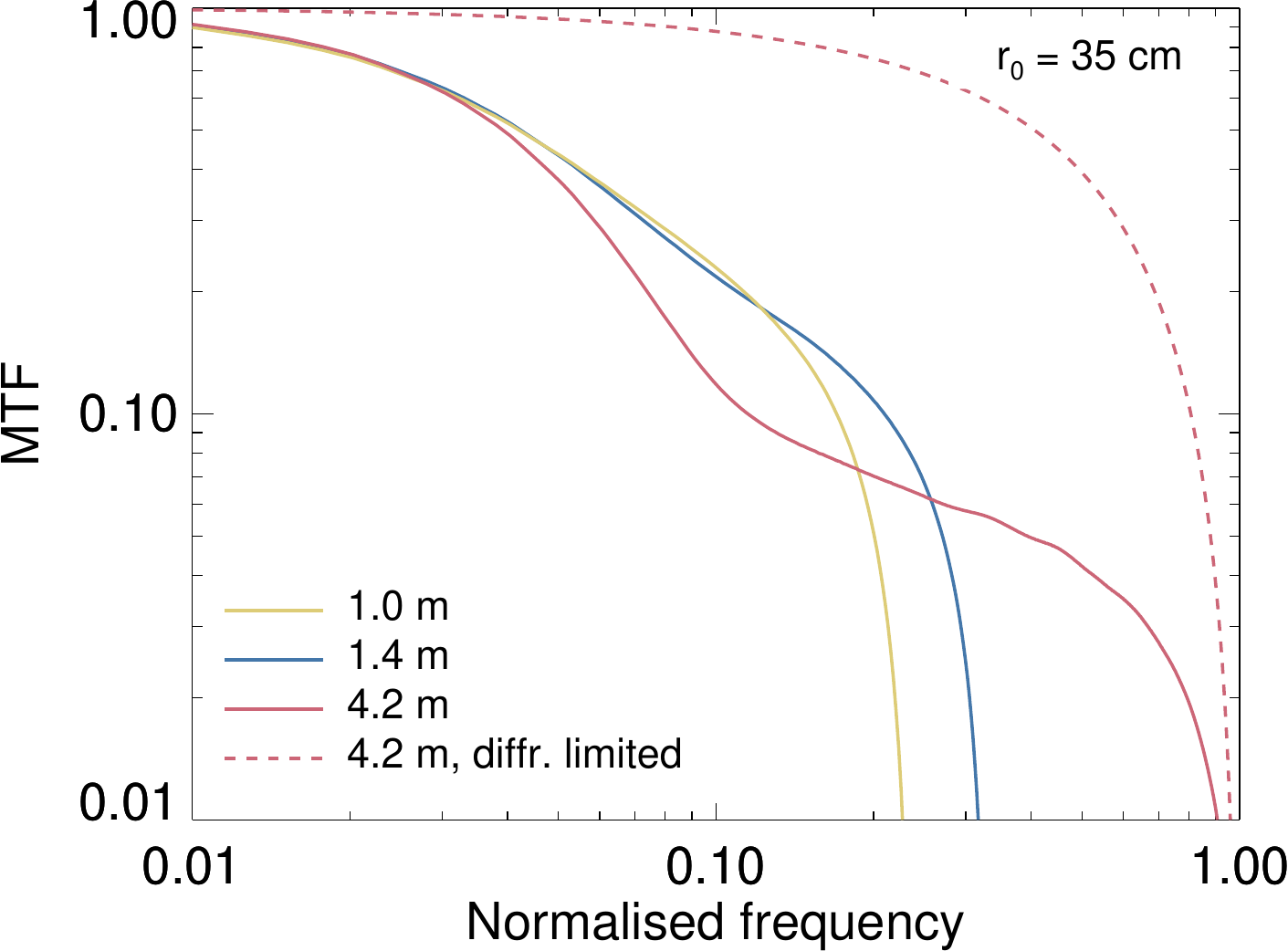 }
     \includegraphics[height=0.375\linewidth]{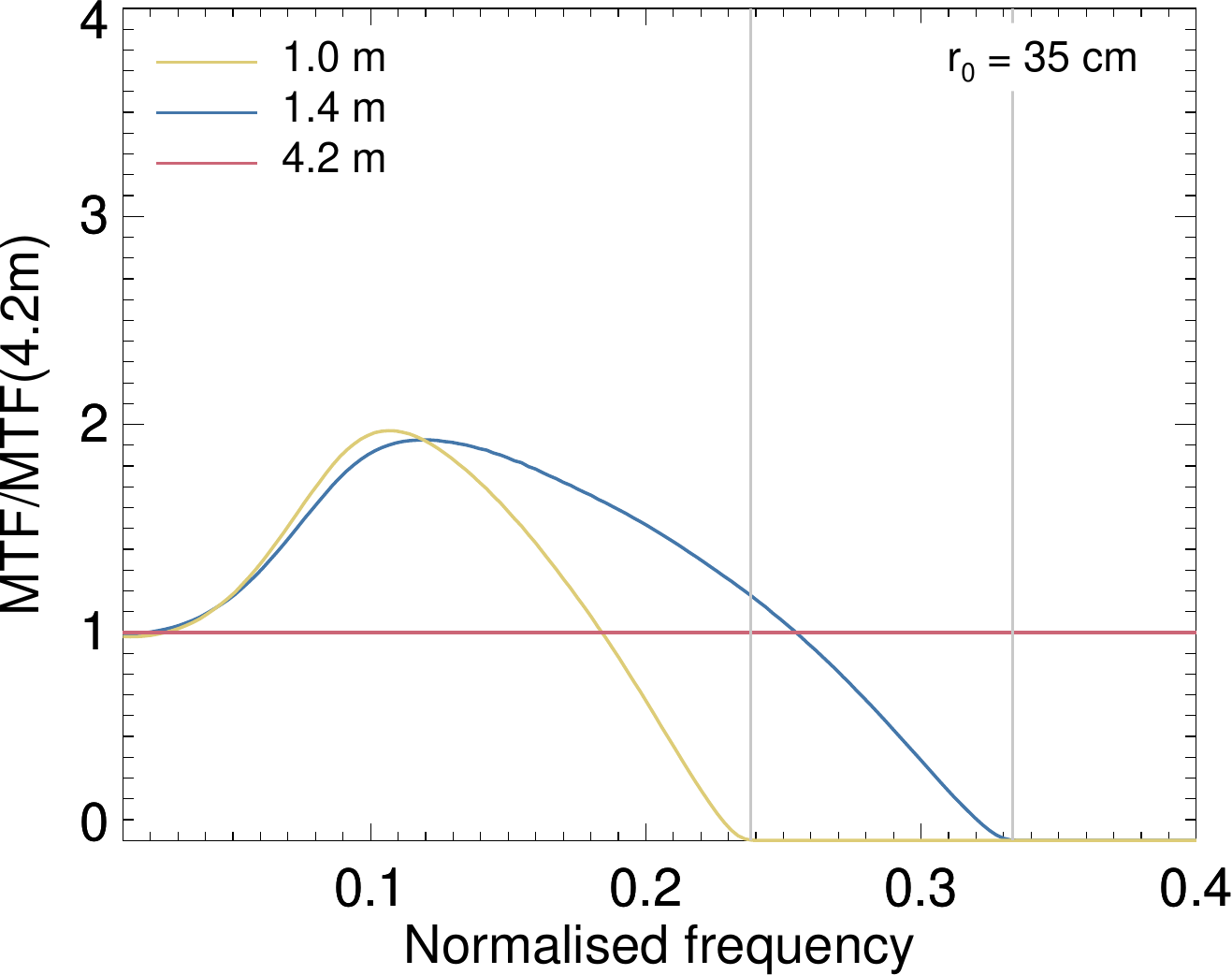 }\\[1.5mm]
     \includegraphics[height=0.375\linewidth]{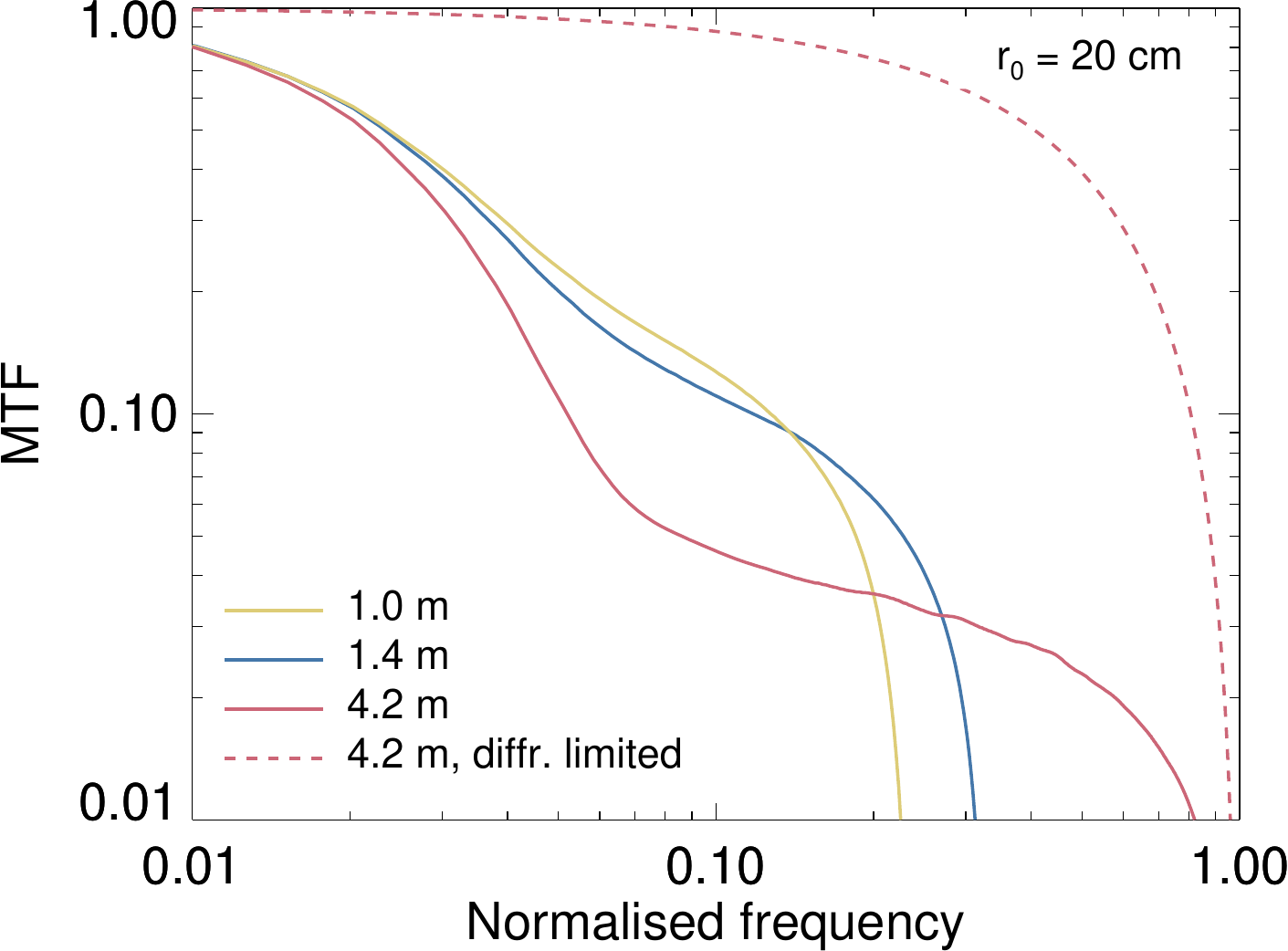 }
     \includegraphics[height=0.375\linewidth]{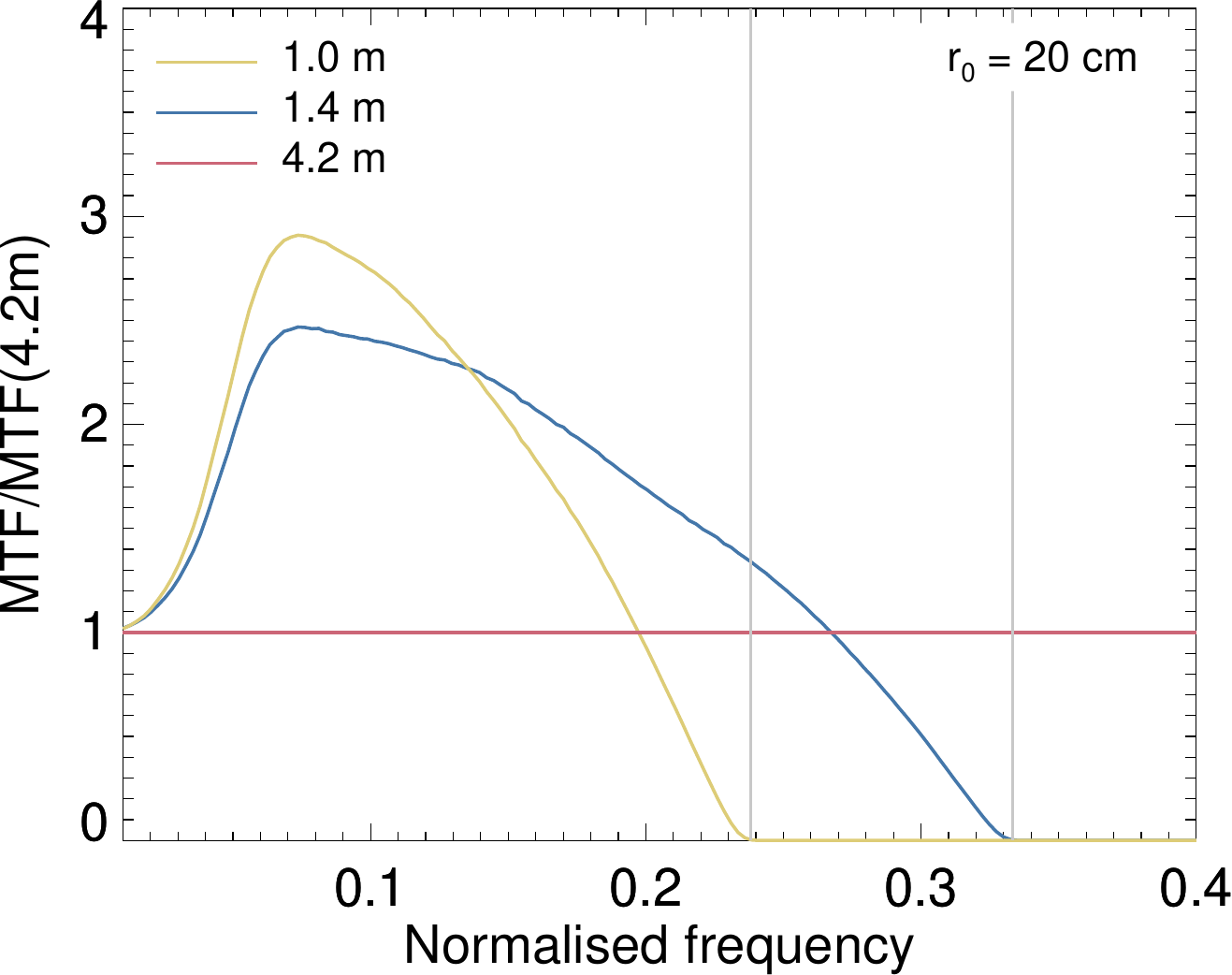 }          
  \caption{Left: MTFs of  1.0~m, 1.4~m, and 4.2~m apertures. Right: the corresponding MTFs normalised to that of a 4.2~m aperture. The vertical gray lines mark the diffraction limits of the 1.0~m and 1.4~m apertures. The spatial frequency scale is in units of the limiting spatial frequency of a 4.2~m telescope.}
  \label{fig:mtfs}
\end{figure}

Figure \ref{fig:mtfs} shows in the left column {\gs the averaged MTFs} corresponding to the PSFs shown in Fig.~\ref{fig:psfs} for $r_0=1$~m (top), $r_0=0.35$~m (middle) and $r_0=0.20$~m (bottom). The uppermost long-dashed curve corresponds to the MTF of a diffraction limited 4.2~m aperture, emphasising that the seeing degraded MTFs\footnote{The use of MTFs to compare the performance is justified, because the total area of the six 1.4~m subapertures is comparable to that of the 4.2~m aperture.} with $r_0=0.35$~m or smaller reduce the SNR of the images with the 4.2~m aperture by over an order of magnitude over a wide range of spatial frequencies. Notably, the MTF of the 4.2~m aperture is so strongly attenuated that the MTFs of the 1.0~m and 1.4~m telescopes deliver better SNR over a wide range of spatial frequencies. The right column shows the ratios of the MTFs of the 1.0~m and 1.4~m apertures, normalised to the MTF of the 4.2~m aperture for the $r_0$ corresponding to the left column. These plots demonstrate that the SNR improves up to a factor 2.5 for a 1.4~m  aperture, relative to the 4.2~m aperture, when $r_0=0.20$~m. We also note, that the MTF of the 1.4~m apertures drops below that of the 4.2~m telescope somewhat above the diffraction limit of the 1.0~m aperture, suggesting that the 4.2~m aperture could be the preferred choice. However, this happens where all MTFs drop strongly with increasing spatial frequency, clearly also implying that the level of noise will have a decisive role -- this important aspect is investigated in Sects.~\ref{sect:seeing_degraded} and \ref{sect:mfbd_restored}.

\subsection{Artificial data}
\label{sec:artificial_data}
For simulations of the imaging performance of EST, up to the diffraction limit of its full 4.2~m aperture and its proposed 1.4~m subapertures, we use a snapshot from a 3D MHD simulation of the solar atmosphere, representing a weak network region with a net magnetic flux corresponding to 200G flux density. The simulation was computed in a domain with 9~Mm horizontal and 3.24 Mm vertical extent and a grid spacing of 7$\times$7$\times$5~km. This setup is based on simulations from \citet{2020ApJ...894..140R}. The simulation uses the FreeEOS equation of state \citep{2012ascl.soft11002I} with  \citet{2009ARA&A..47..481A} abundances and non-grey radiative transfer with 12 opacity bins. 

\begin{figure}[!t]
  \centering
    \includegraphics[width=0.8\linewidth]{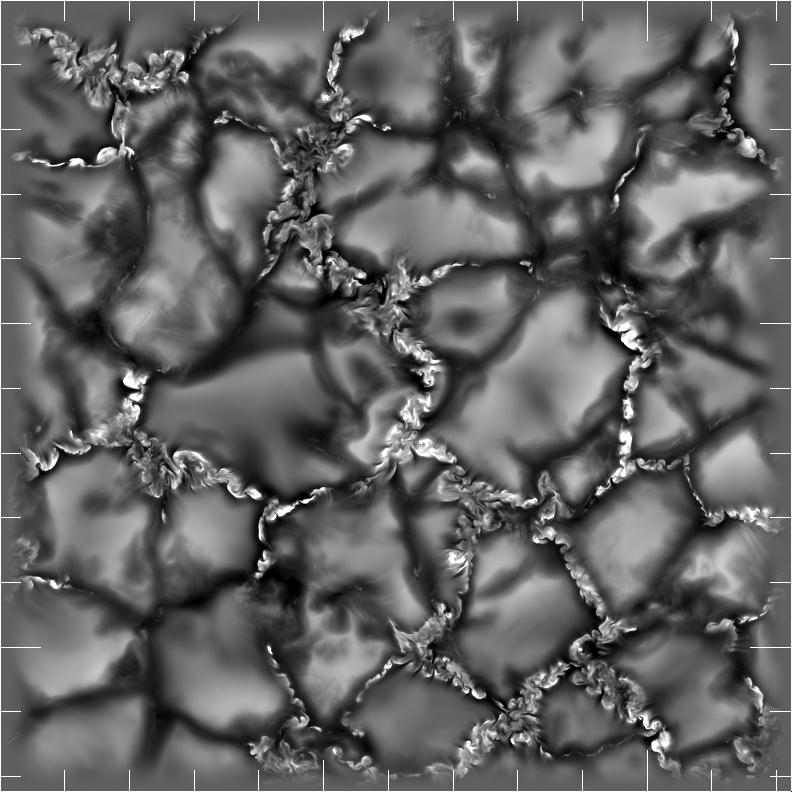}
  \caption{The ``true'' 12\farcs2$\times$12\farcs2 image that the
    artificial ``observed'' images are based on. This is based on a a simulation snapshot calculated by \citet{2020ApJ...894..140R}. From this, we calculated a synthetic image as seen through the 0.4~nm FWHM 630.2~nm prefilter of CRISP2, centered on a pair of magnetically sensitive FeI lines.  1\arcsec{}
    tickmarks.}
  \label{fig:truth_image}
\end{figure}

\begin{figure}[!t]
  \centering
  \includegraphics[width=0.8\linewidth]{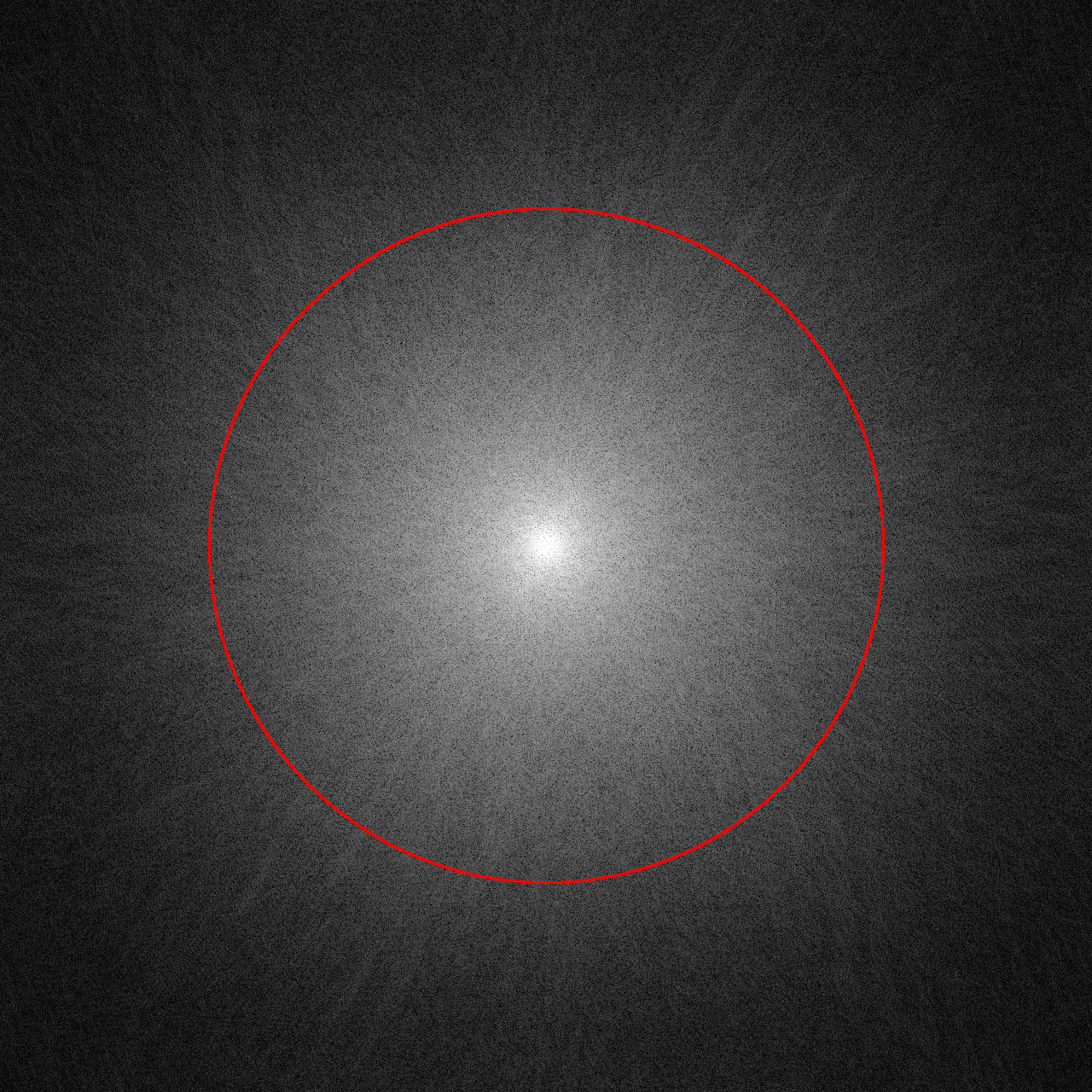}
  \caption{Logaritm of 2D power spectrum of the image in
    Fig.~\ref{fig:truth_image}. The circle corresponds to the
    diffraction limit of a 4.2~m telescope.}
  \label{fig:truth_power2d}
\end{figure}

\begin{figure}[!t]
 \centering
     \includegraphics[width=0.8\linewidth]{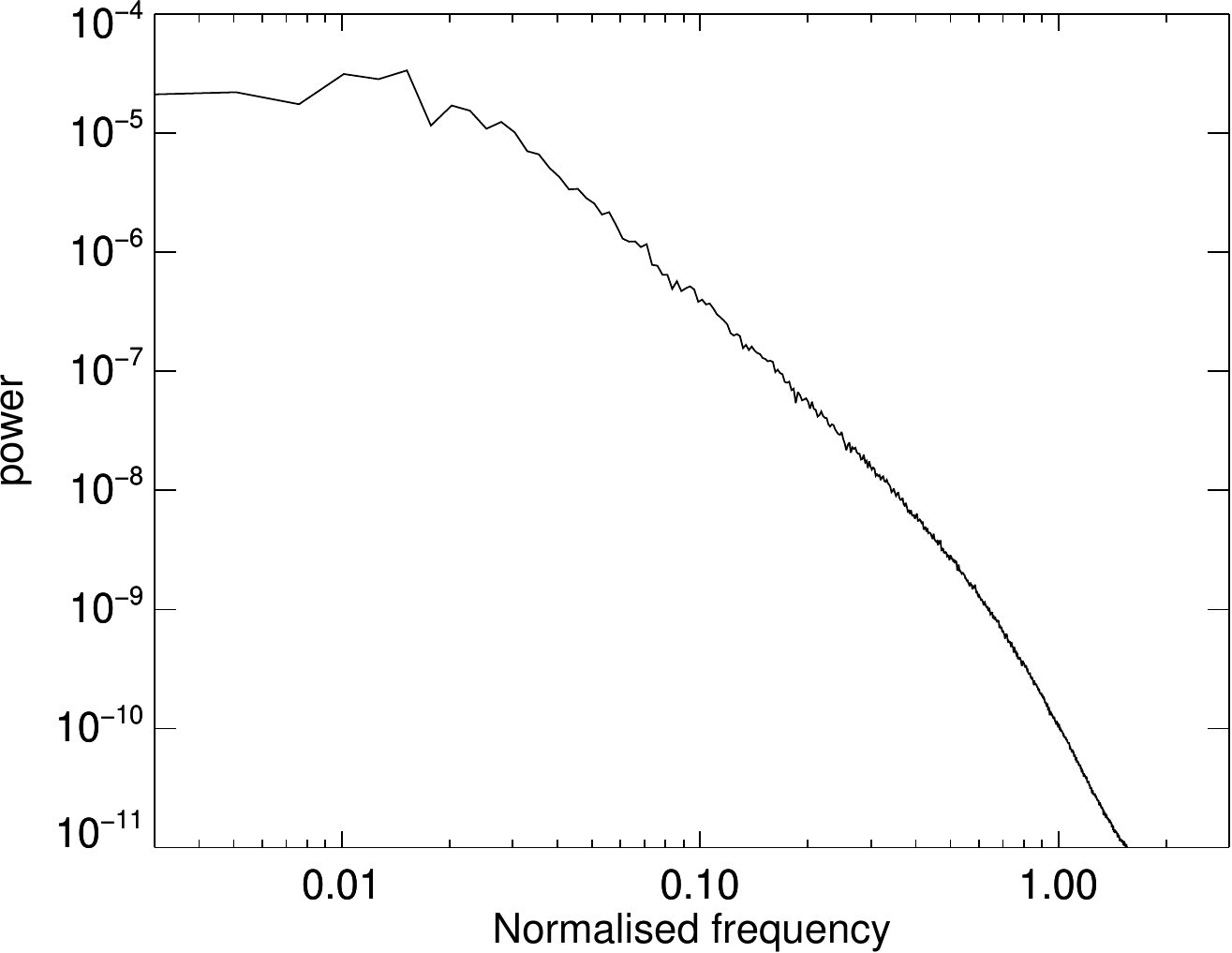}
  \caption{1D power spectrum, based on the 2D power spectrum shown in Fig.~\ref{fig:truth_power2d}. The angular averages are calculated after taking the logarithm. The spatial frequency coordinate is normalised to the diffraction limit of a 4.2~m telescope. The power spectrum of the synthetic image continues beyond this diffraction limit.}
  \label{fig:truth_power1d}
\end{figure}

This snapshot was used to calculate  synthetic spectra from the entire FOV within a 0.8~nm spectral region centered on the two Fe~I lines at 630.2~nm, but also including a dozen weaker spectral lines. The snapshots at different wavelengths were then integrated with a synthetic 0.4~nm  FWHM 2-cavity filter, which corresponds to the FPI prefilter of CRISP2 \citep{2026A&A...705A..55S} for this wavelength. The resulting image is shown in Fig.~\ref{fig:truth_image}. We use this image for emulating images observed with the wide-band channel of an FPI system that is similar to CRISP2 and the future EST-V, with three different telescope aperture diameters and in a wide range of seeing conditions. The 2D power spectrum shown in Fig.~\ref{fig:truth_power2d} reveals that the input image has significant power beyond the diffraction limit of a 4.2~m telescope, which is demonstrated more clearly in the 1D power spectrum shown in Fig.~\ref{fig:truth_power1d}. 
 
\subsubsection{Noise levels of  simulated images}\label{sect:gband_noise}
During the initial phases of this project, we based our simulations on restorations of a G-band image recorded with DKIST and kindly provided by Friedrich Wöger. This reached a spatial resolution of about 40\% of the DKIST diffraction limit, which prompted us to investigate the impact of noise on the achievable spatial resolution. We found that this image, which is based on a burst of 80 observed images, corresponds to a recorded total of about \expten{3}{6} photons per pixel, and that approximately \expten{20}{6} photons would have been needed to achieve a spatial resolution of 80\% of the DKIST diffraction limit.  Similarly, MFBD image restoration of an outstanding three minute long time sequence of  DKIST images recorded at a wavelength of  416~nm \citep{Kuridze2026}, required approximately  \expten{10-20}{6} photons per pixel to reach approximately 90\% of the diffraction limit of DKIST (M. van Noort, private communication). We note that both of these data sets were recorded with the Sun at small zenith distance ($z$<35\degr). This is of relevance since $r_0\propto (\cos z)^{3/5}$, which implies that $r_0$ would be 1.7 times smaller if these observations were instead made with the Sun at a zenith distance of 70\degr. 

The extremely high SNR that is needed to reach the diffraction limit of a 4-m telescope is a consequence of both the absence of MCAO and the aforementioned decrease of power of solar fine structure at high spatial frequencies (Fig. \ref{fig:truth_power1d}), as will be further discussed in the following sections. We therefore settled for simulations with the following number of photons: \expten{1}{8} (which seems virtually unachievable with any narrowband imaging spectro-polarimeter)\footnote{Assuming a total efficiency of EST and the visual FPI system with its cameras of 5-10\%, we estimate a count rate of \expten{1.5-3}{6} photons per second integration time in the continuum at 600~nm. However, such long integration times would in most cases lead to polarimetric line scans with unacceptably low cadence. }, \expten{1}{6} (which is an ambitious but achievable goal), and \expten{1}{5} (which has been achieved with CRISP on numerous occasions).  

We remark, that here and in the following, we define the resolution as given by the spatial frequency for which the power spectrum of an image reaches the noise level. This takes into account the spatial content of the entire FOV, and furthermore that image restoration techniques (speckle or MFBD) involve a low-pass filter that effectively reduces the spatial resolution to that of a similar spatial frequency. It is also common to estimate the spatial resolution by identifying the smallest spatial structures in an image and measuring their widths, but this is not our preferred method.

\begin{figure}[!t]
 \centering
      \includegraphics[width=0.98\linewidth]{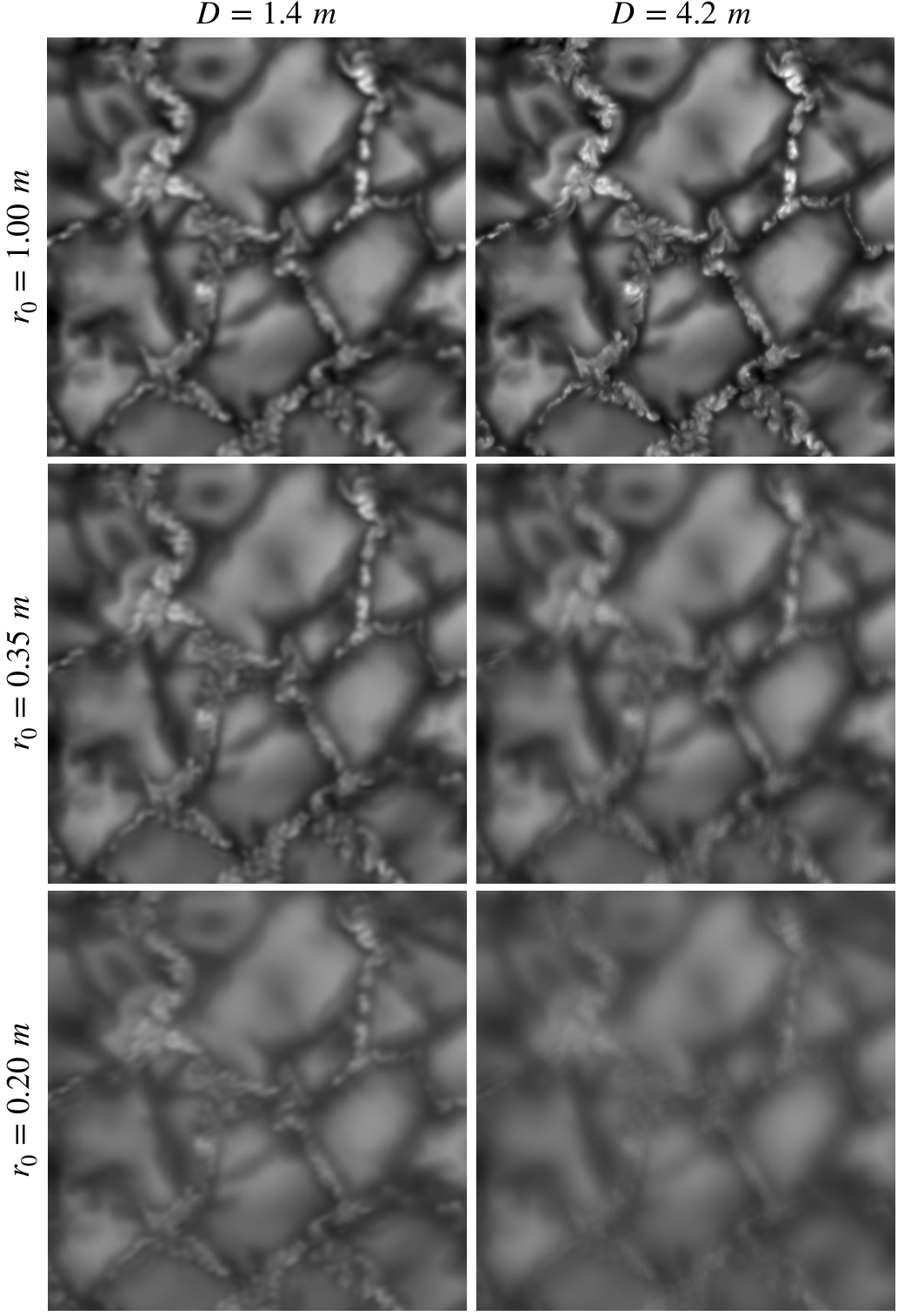}                
      \caption{Part of the FOV of the truth image in Fig.~\ref{fig:truth_image}, showing the image with the highest small-scale contrast out of 100 seeing degraded images with 1.4~m and 4.2~m aperture for  a seeing corresponding to $r_0=1$~m, 0.35~m, and 0.20~m. The intensity is scaled to a range from 0.7 to 1.5 of the average intensity for all panels.}
  \label{fig:best_images}
\end{figure}

\begin{figure}[!t]
 \centering
     \includegraphics[width=0.49\linewidth]{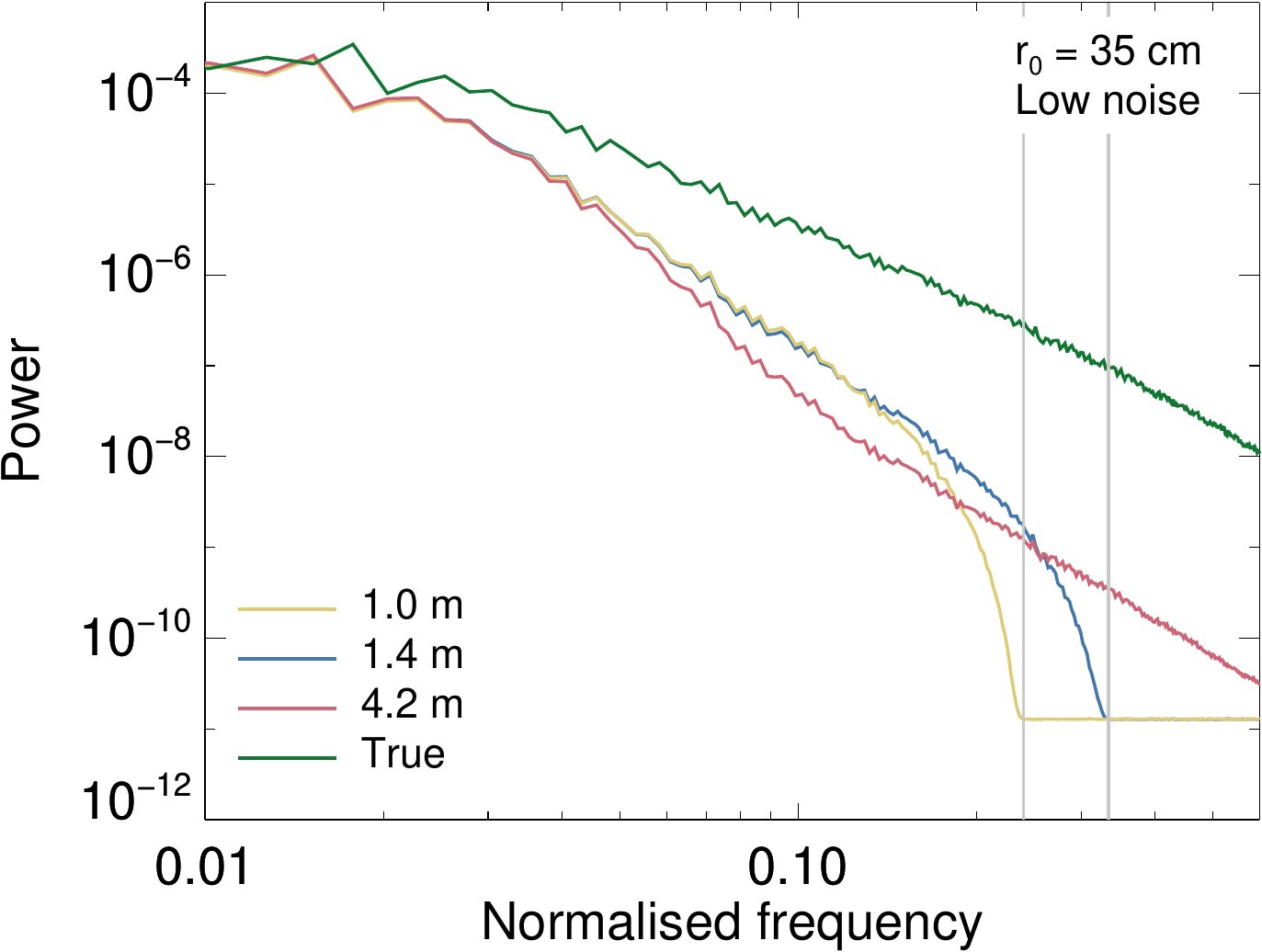}
     \includegraphics[width=0.49\linewidth]{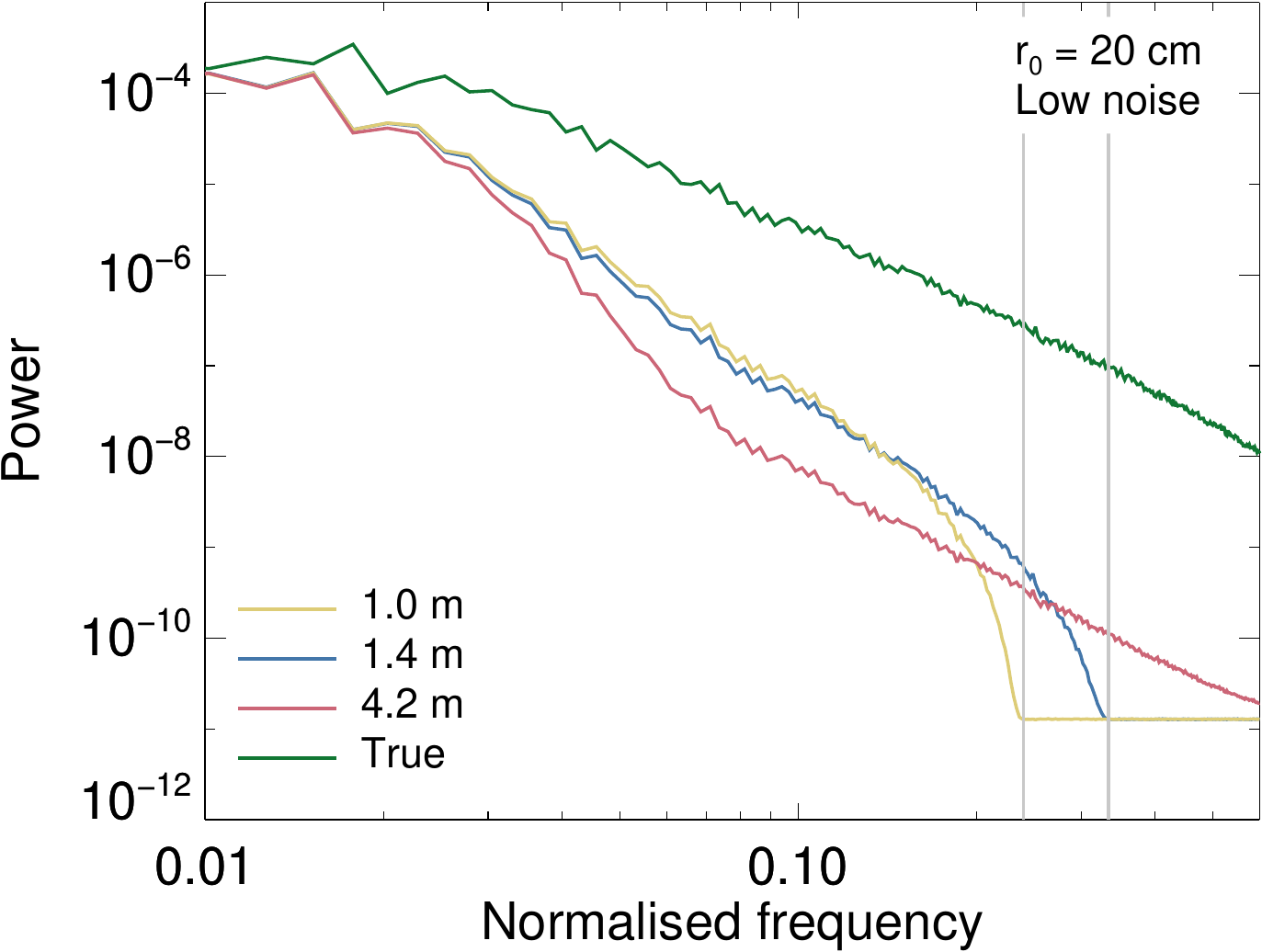}\\[1.5mm]
     \includegraphics[width=0.49\linewidth]{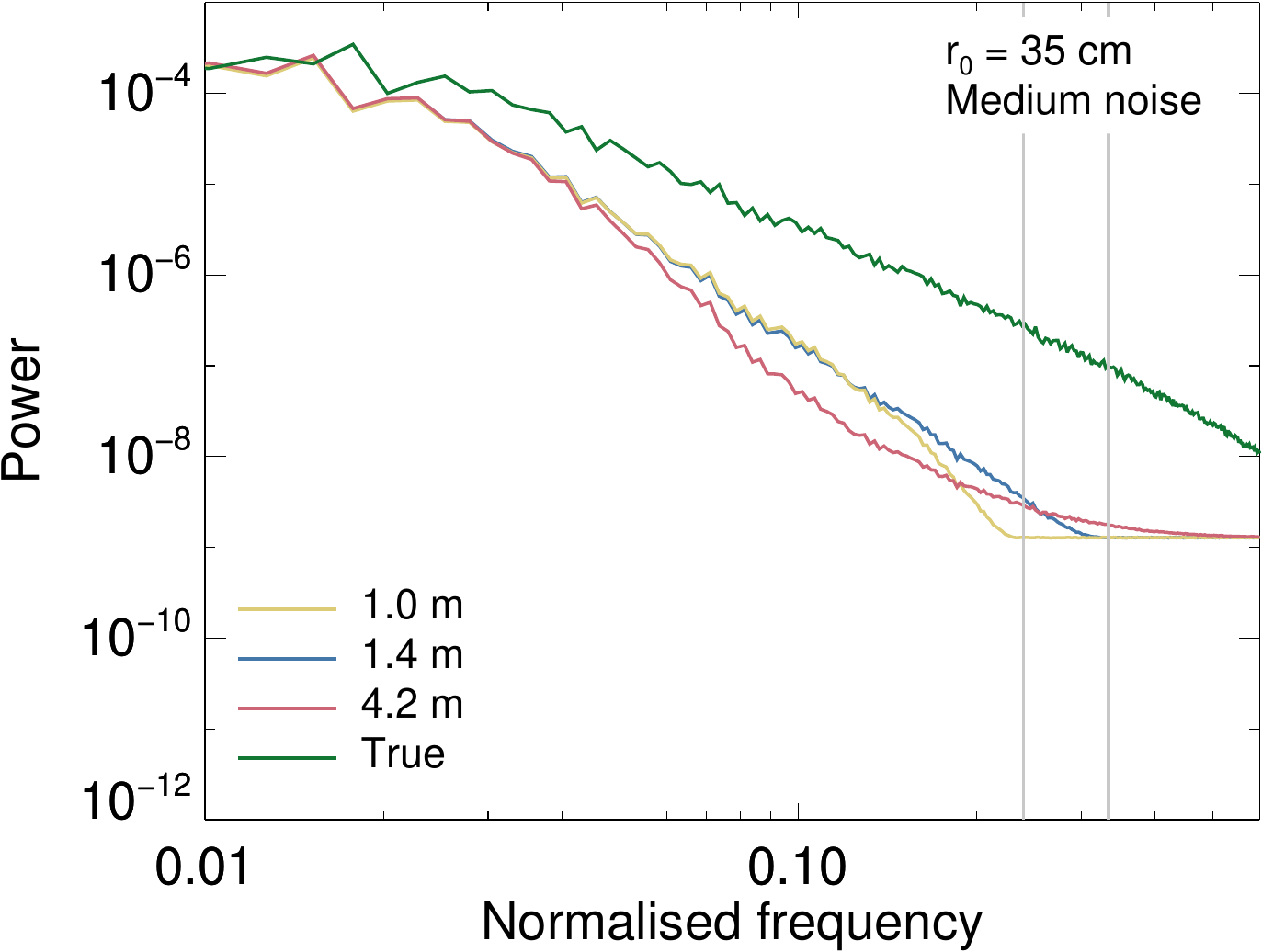}
     \includegraphics[width=0.49\linewidth]{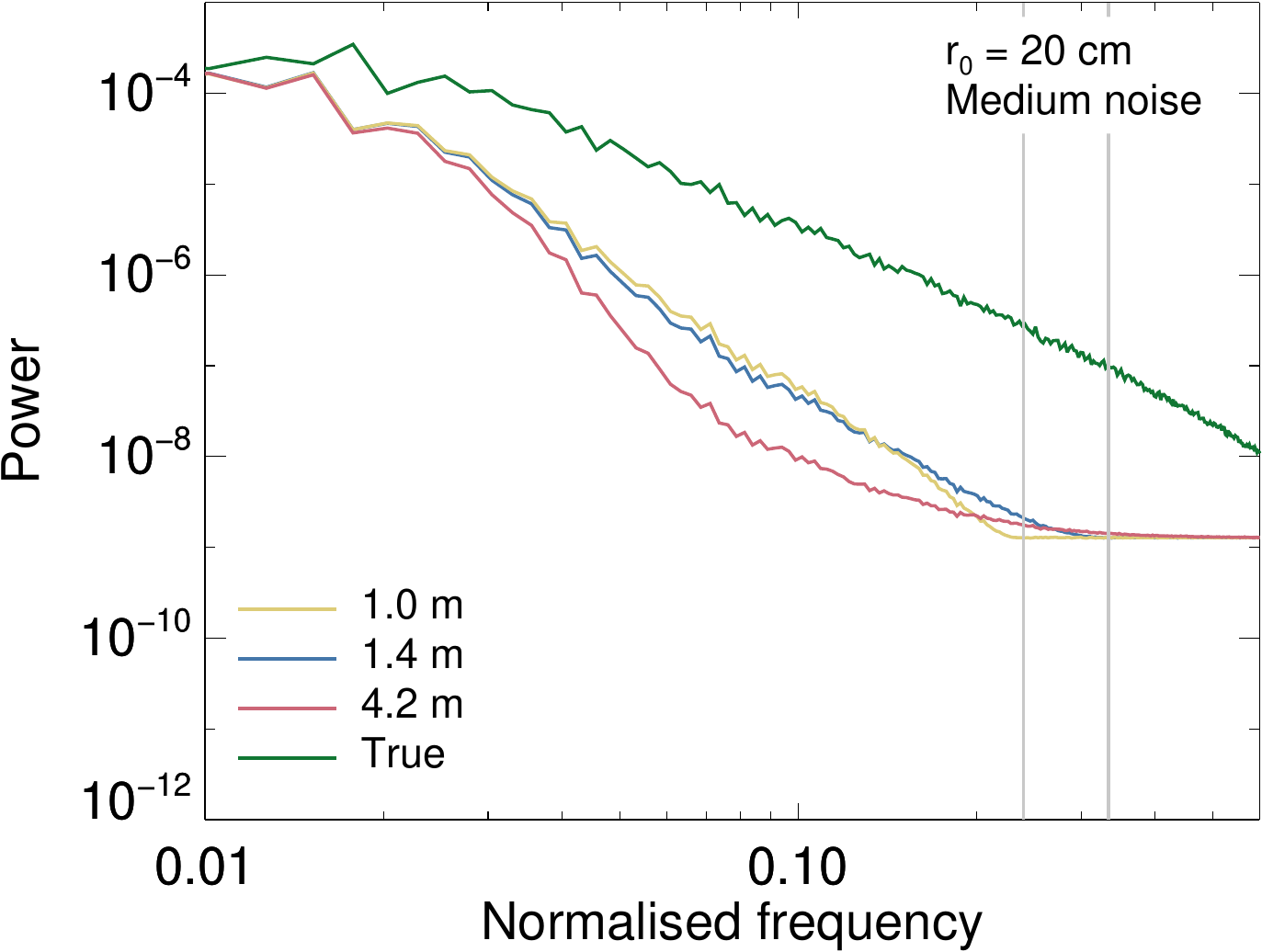}\\[1.5mm]
     \includegraphics[width=0.49\linewidth]{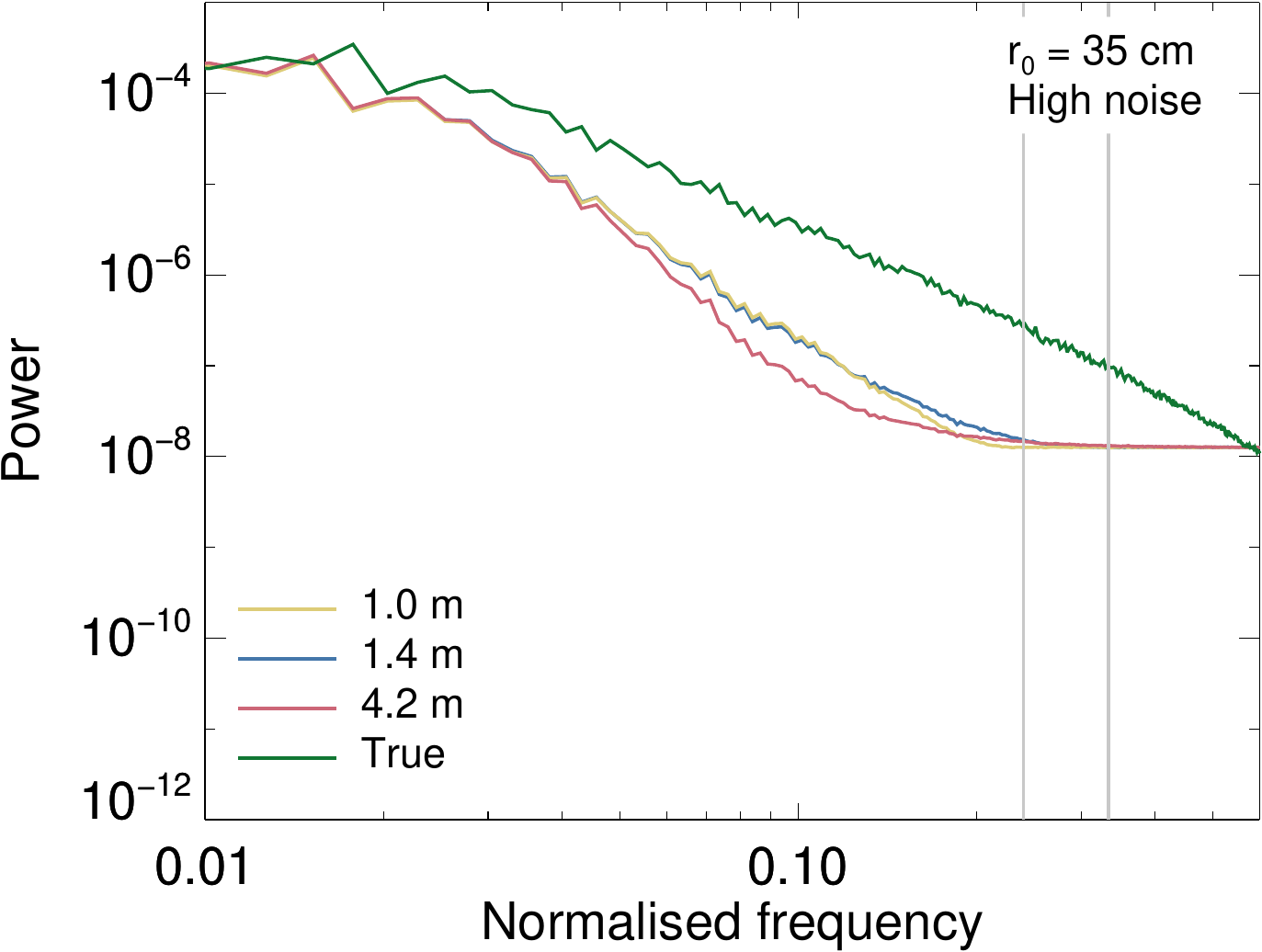}
     \includegraphics[width=0.49\linewidth]{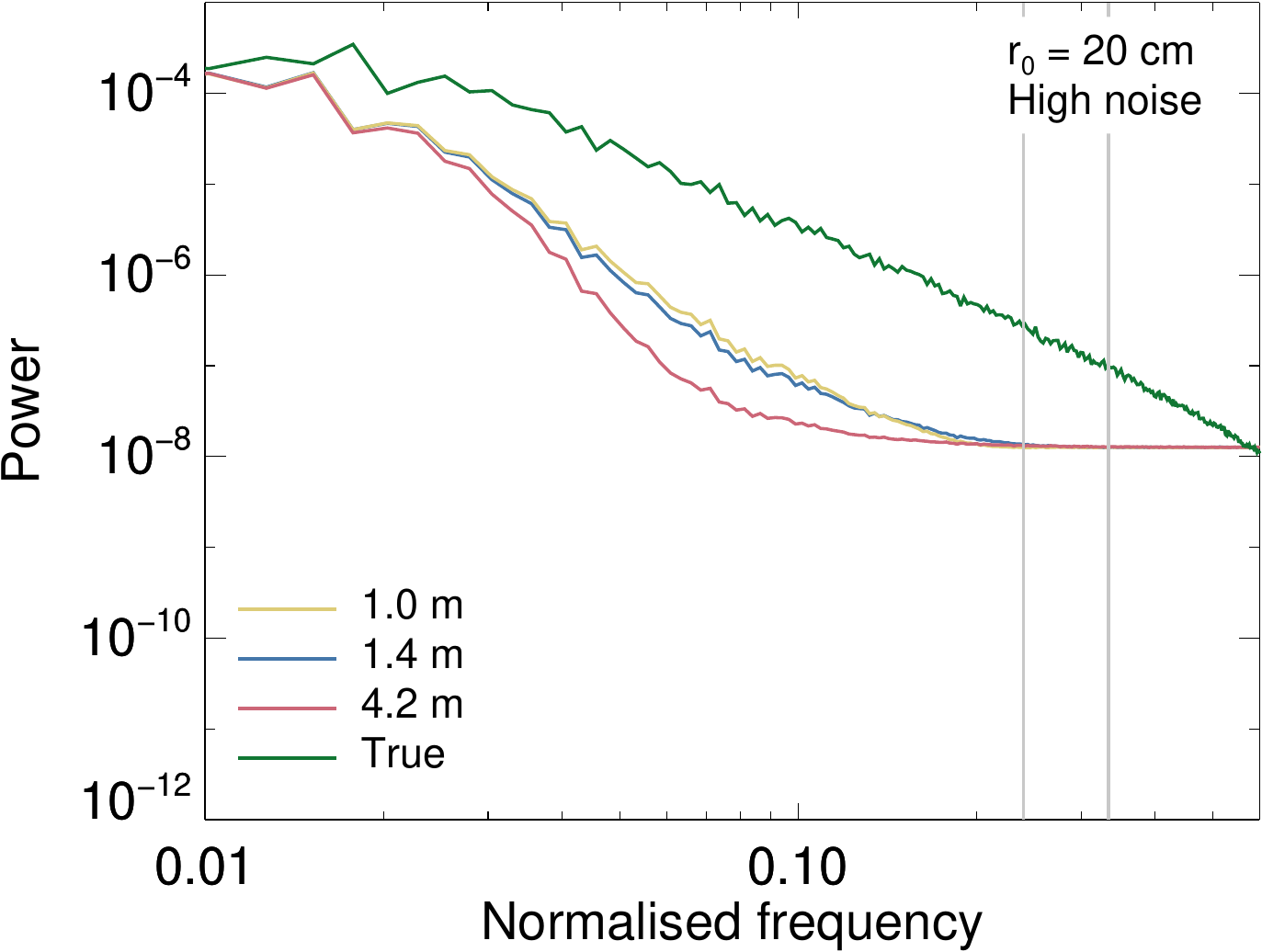}
  \caption{Plots of power spectra for 1.0~m, 1.4~m, and 4.2~m apertures obtained from images degraded by seeing corresponding to $r_0=0.35$~m and 0.20~m, and different noise levels. The vertical gray lines correspond to the diffraction limits of  the 1~m and 1.4~m apertures.}
  \label{fig:raw_power}
\end{figure}

\begin{figure}[!t]
 \centering
     \includegraphics[width=0.49\linewidth]{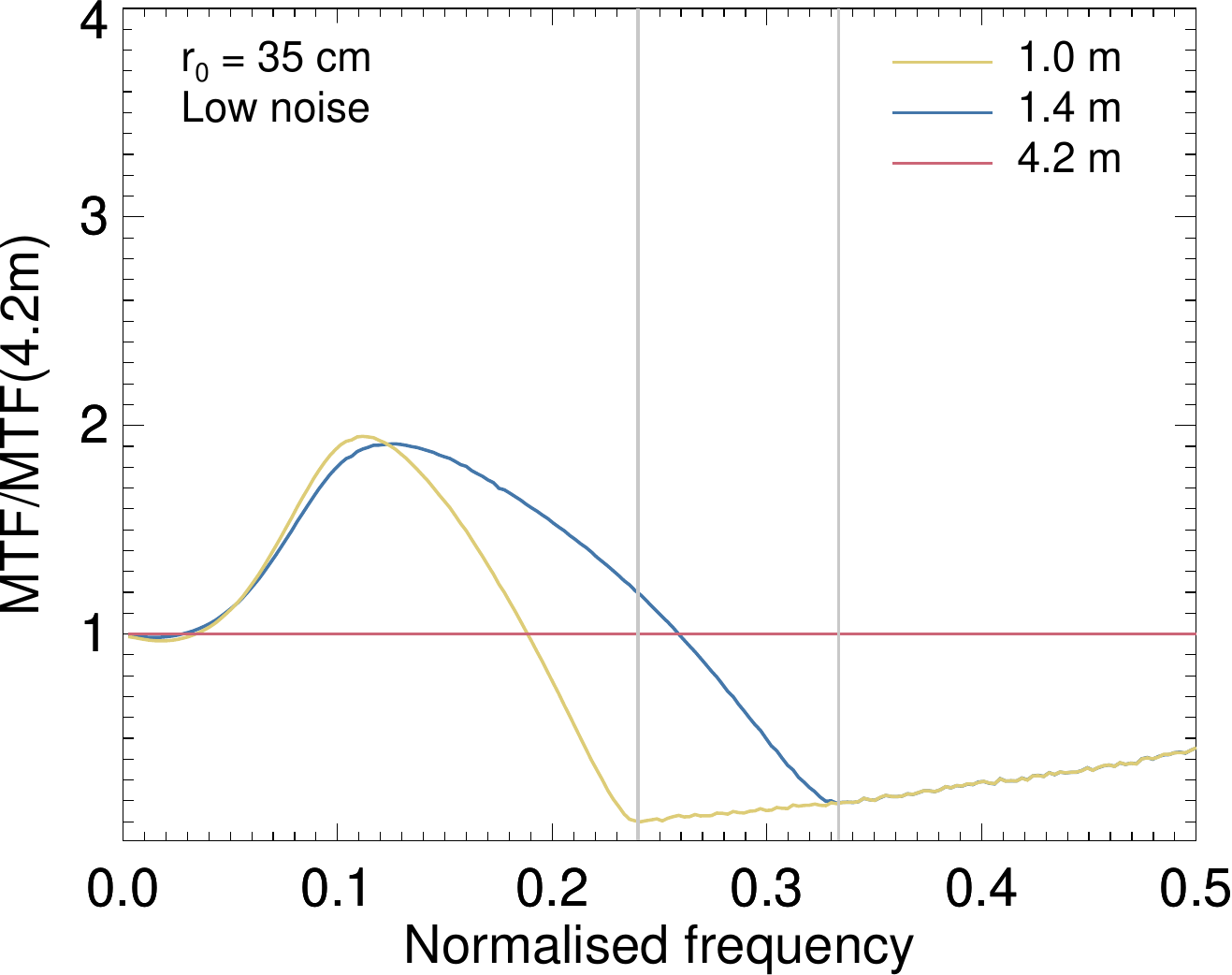}
     \includegraphics[width=0.49\linewidth]{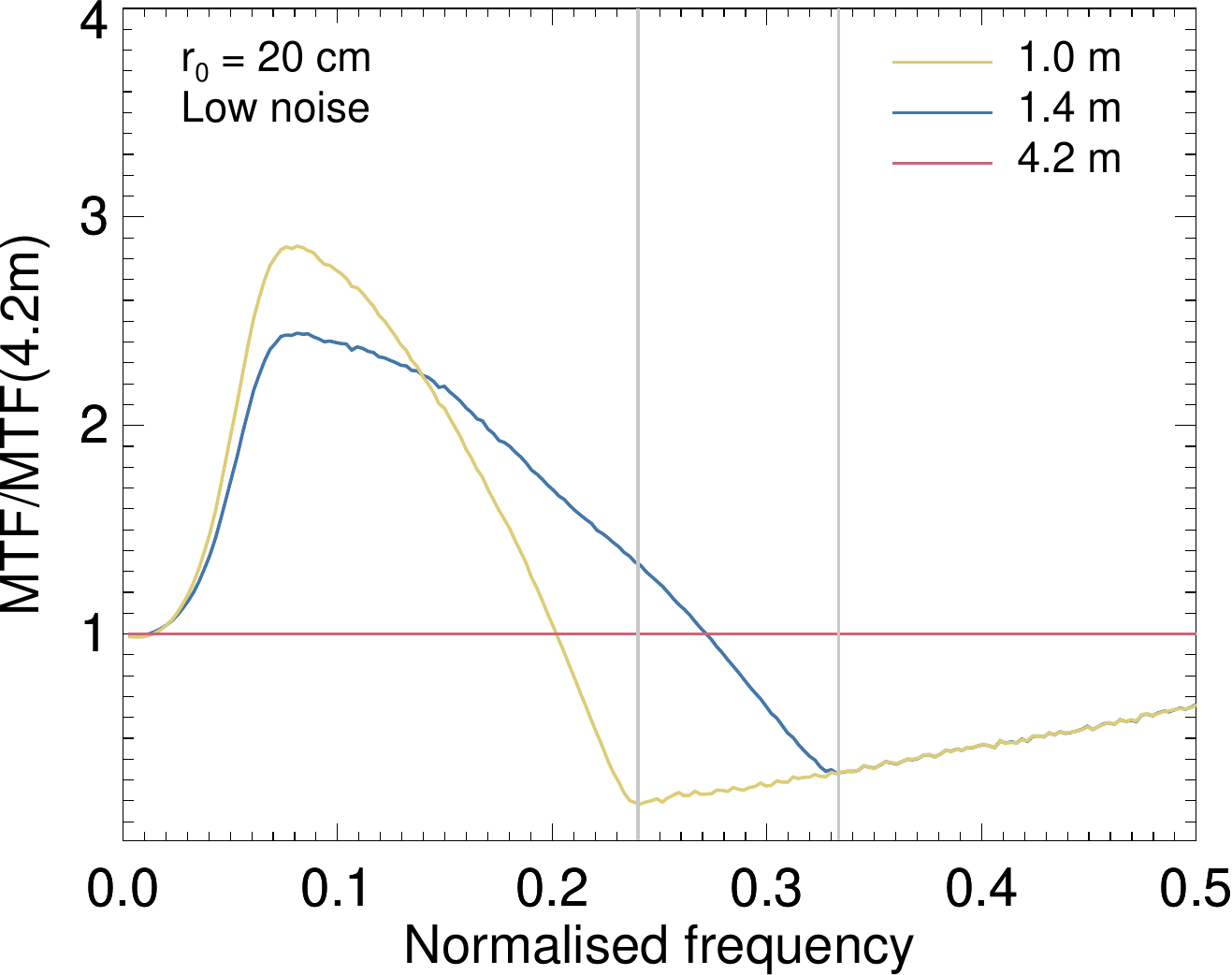}\\[1.5mm]
     \includegraphics[width=0.49\linewidth]{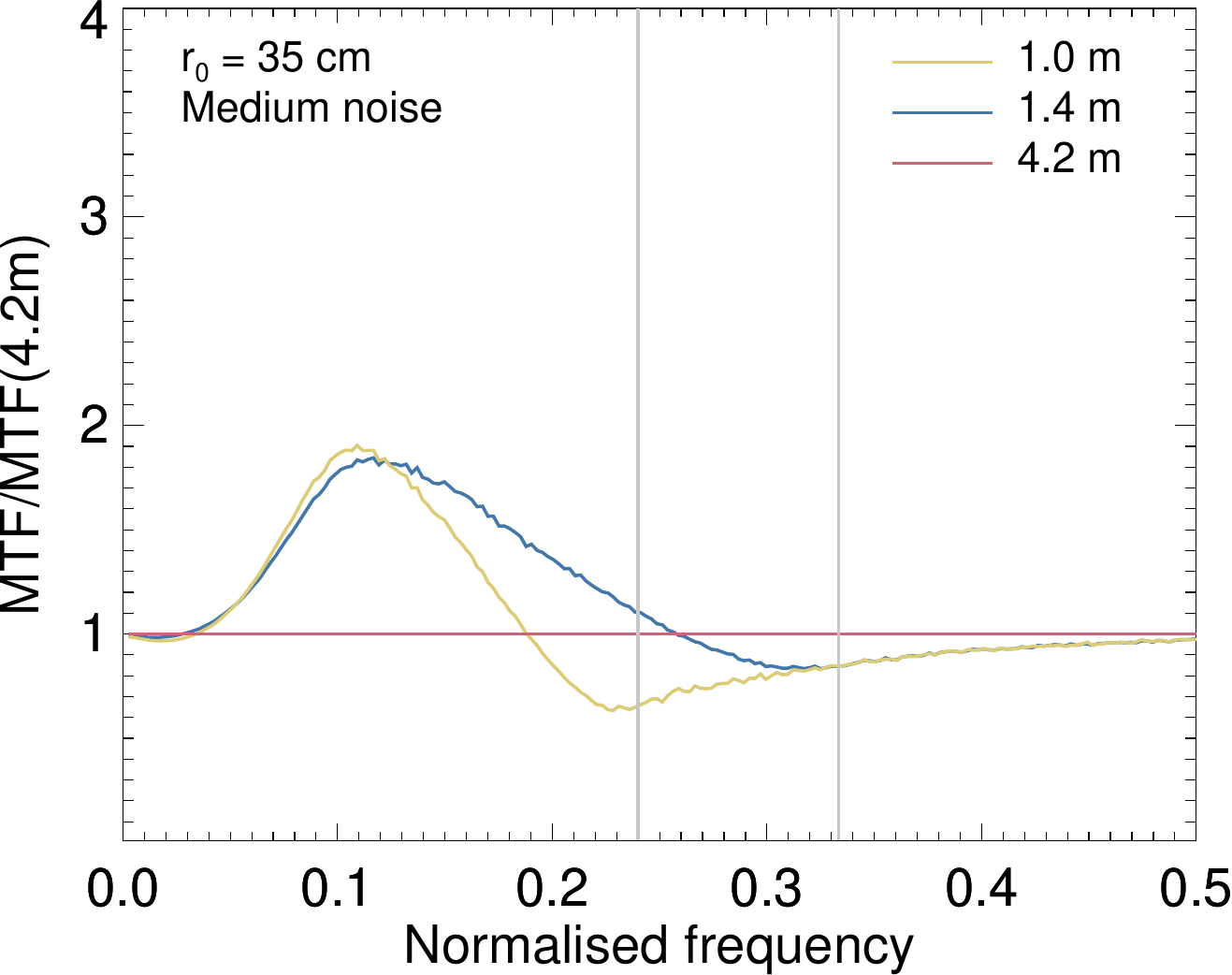}
     \includegraphics[width=0.49\linewidth]{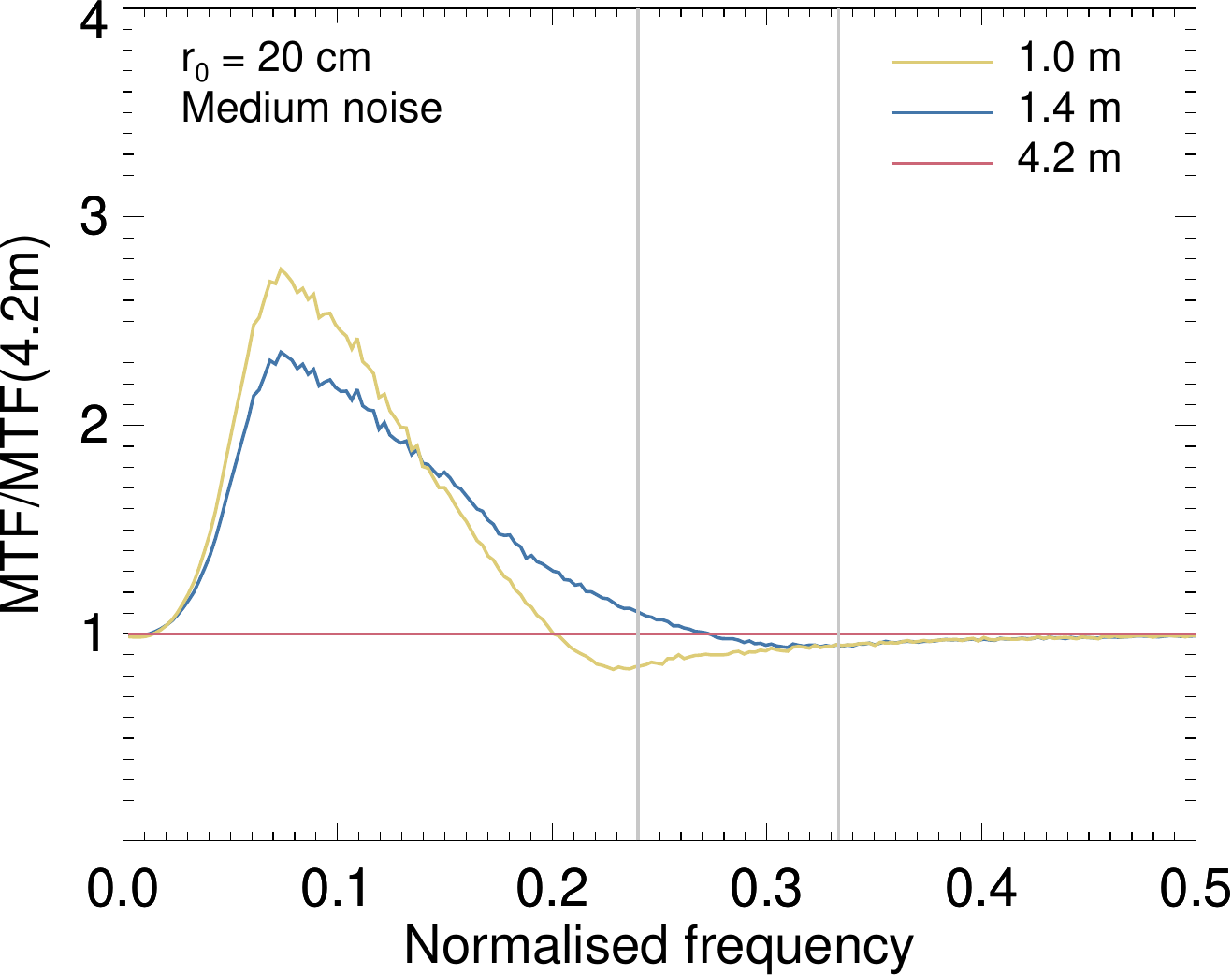}\\[1.5mm]
     \includegraphics[width=0.49\linewidth]{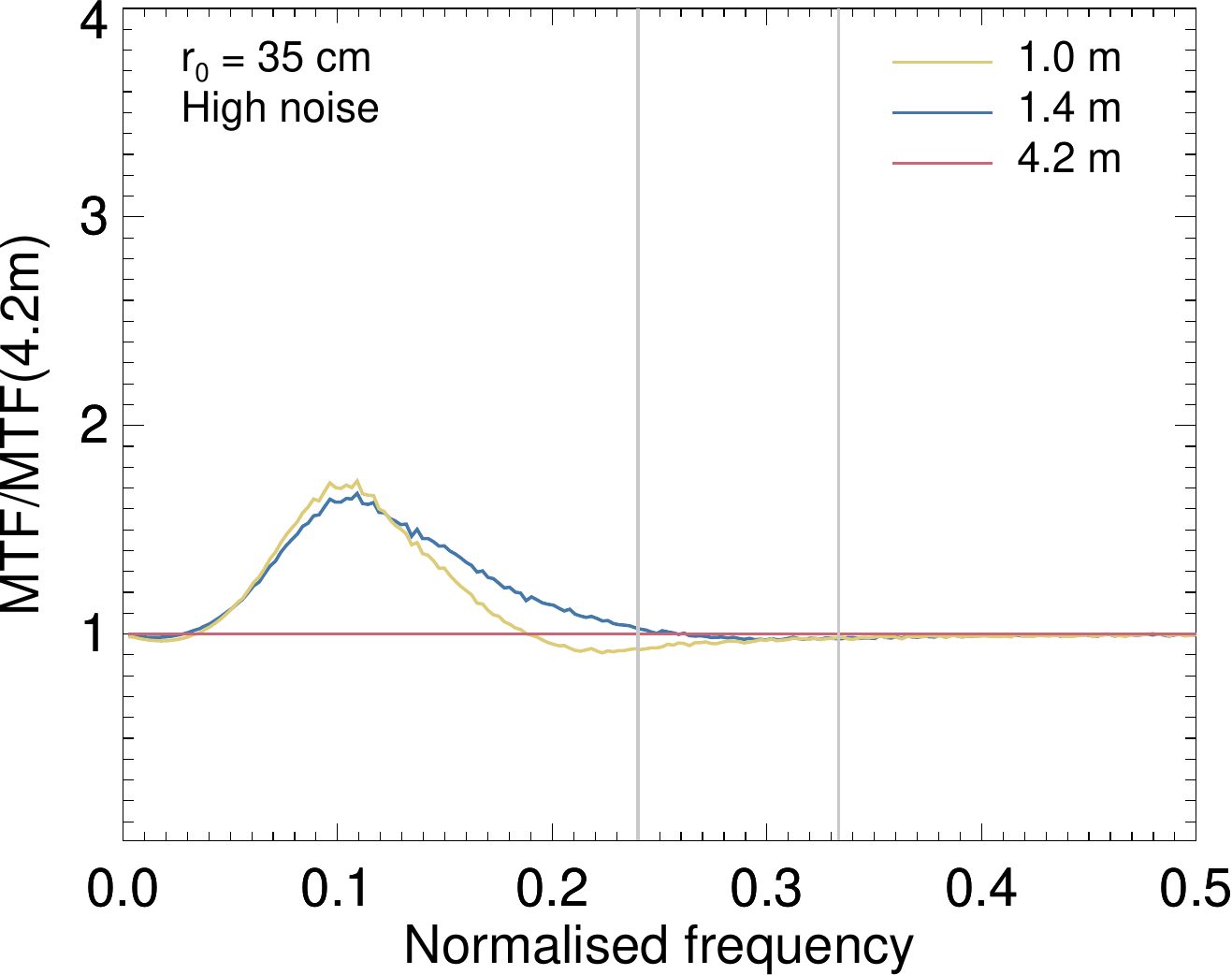}
     \includegraphics[width=0.49\linewidth]{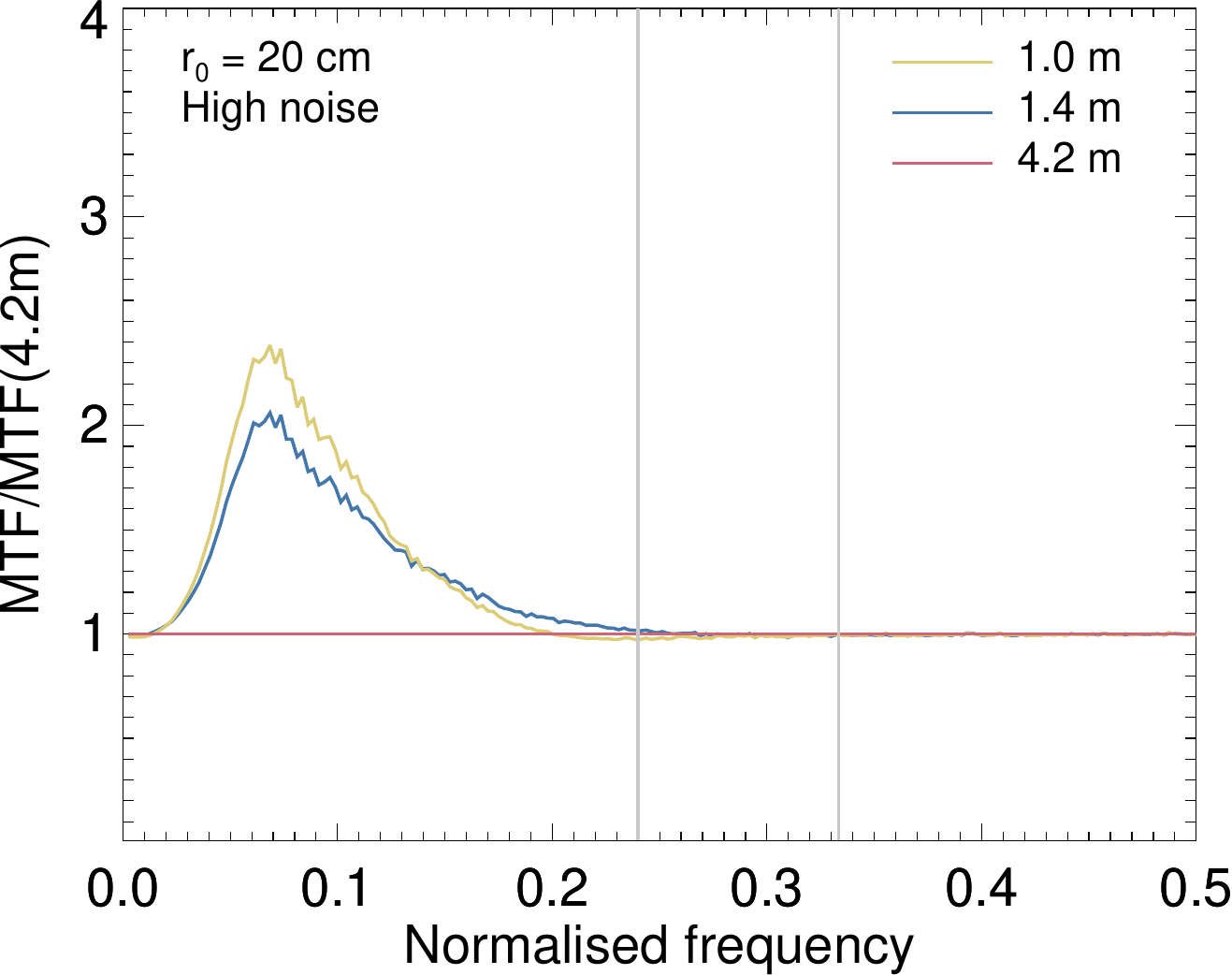}
  \caption{Plots of the square root of the power spectra of the input images for 1.0~m and 1.4~m apertures, degraded with seeing characterised by $r_0=0.35$~m and 0.20~m and noise, and normalised to that of a 4.2~m aperture. Except for the added noise, these plots are very similar to the normalised MTFs shown in the right column of  Fig.~\ref{fig:mtfs}. The vertical gray lines correspond to the diffraction limits of  the 1~m and 1.4~m apertures.}
  \label{fig:raw_power_ratio}
\end{figure}

\begin{table}[tbh]
\caption{\label{Strehl_summary} Simulation parameters and expected Strehl values} 
  \centering
  \small
  \begin {tabular} {lccccccc}
    \hline
        \noalign{\smallskip}
    \mathstrut
    $r_0$ & $N_{KL}$ & 1.0 & 1.4 & 4.2 \\
     (m) &  & (m) & (m) & (m)\\
    \hline
    \noalign{\smallskip}
    1.00 & 3 & 0.87 & 0.79 & 0.23 & tip-tilt only\\
    1.00 & 50 & 0.99 & 0.98 & 0.87\\
    1.00 & 100 & 0.99 & 0.99 & 0.92\\
    1.00 & 200 & 1.00 & 0.99 & 0.96\\
     \hline
     \noalign{\smallskip}
    0.35 & 3 & 0.46 & 0.26 & - & tip-tilt only\\
    0.35 & 50 & 0.93 & 0.88 & 0.45\\
    0.35 & 100 & 0.96 & 0.93 & 0.64\\
    0.35 & 200 & 0.98 & 0.96 & 0.78\\
  \hline
  \noalign{\smallskip}
    0.20 & 3 & 0.14 & 0.03 & - & tip-tilt only\\
    0.20 & 50 & 0.83 & 0.72 & 0.13\\
    0.20 & 100 & 0.90 & 0.83 & 0.32\\
    0.20 & 200 & 0.94 & 0.90 & 0.52\\
  \hline
      \noalign{\smallskip}
  \end{tabular}
  \vspace{1mm}
  \tablefoot{ Predicted Strehl values for aperture diameters of 1.0, 1.4 and 4.2~m,  seeing characterised by Fried's parameter $r_0$, and the number of Karhunen--Loeve modes $N_{KL}$ used with MFBD image reconstruction. Strehl values below 0.3 are only indicative.}
  \label{Strehl_summary}
\end{table}

\section{Noise and seeing degraded images and PSFs}
\subsection{Simulation parameters}
The simulations primarily target a situation where the absence of MCAO leaves the high-altitude seeing layer around the tropopause uncompensated for. As discussed in Appendix \ref {SST_AO_measurements}, our best estimate is that this layer has a strength corresponding to an $r_0$ that is typically somewhere in the range 0.20--0.35~m. Furthermore, the impact of noise on the reconstructed images is clearly of the highest significance, as discussed in Sect.~\ref{sect:gband_noise}. Adding the 1~m aperture diameter of SST as reference to the 1.4~m and 4.2~m aperture diameters of EST, we perform simulations with the following set of parameters:
\begin{itemize}
\item Three $r_0$ values: 0.20, 0.35, and 1.0~m
\item Three total noise levels: \expten{1.0}{-4}, \expten{1.0}{-3}, and \expten{3.0}{-3}, assumed to correspond to the resulting noise after accumulating 50 observed images. For simplicity, we will refer to these noise levels as ``low", ``medium", and ``high" in the following, though it would be more correct to refer to the noise level \expten{1.0}{-4} as ``extremely low"
\item Three aperture diameters: 1.0~m, 1.4~m and 4.2~m
\item For each of the above 27 combinations of parameters, we calculated 100 seeing degraded images, using a FOV of 1024$\times$1024 pixels
\end{itemize}

\subsection{Analysis of seeing degraded images, PSFs and MTFs}\label{sect:seeing_degraded} 

In the following, we evaluate performance of EST with 1.4~m subapertures by simulating both the image quality degradation by seeing and noise, and the reconstruction of the degraded images with MFBD techniques  \citep{2002SPIE.4792..146L, 2005SoPh..228..191V, 2021A&A...653A..68L} using a part of the FOV shown in Fig.~\ref{fig:truth_image}. Figure \ref{fig:best_images} shows  the best image out of 100 seeing degraded, but noise-free, images with the 1.4~m and 4.2~m apertures with seeing quality corresponding to $r_0=$1~m, 0.35~m, and 0.20~m. The identification of the best image is made by calculating the rms contrast of the high-pass filtered image (obtained from the difference between the image and the same image smeared by a Gaussian with a full width at half maximum of 1\farcs95), and represents only a rough quality indicator. However, this suffices to illustrate that when $r_0$ is small, the quality of the images with a 1.4~m aperture is better than those from the 4.2~m aperture. Even $r_0$=1.0~m leads to significant reduction of the contrast of small-scale structures with 4.2~m aperture (compare to Fig. \ref{fig:truth_image}). All images shown here and in the following are scaled to intensities in the same range of 0.7 to 1.5 of their average intensity.

Figure \ref{fig:raw_power} shows power spectra obtained from the images degraded by seeing and added noise as described in the Figure. The power spectrum of the ground truth image is shown as a dotted curve for comparison. We emphasise, that only the top two plots, corresponding to a burst of 50 images with exceptionally low noise (\expten{1.0}{-4}) shows clear evidence of power that goes beyond the diffraction limited limit of the 1.4~m aperture. Note that the MTF of the 4.2~m aperture suppresses the power spectrum of the truth image by more than 200 times at high spatial frequencies. The plots in the middle row, with noise levels of \expten{1.0}{-3} clearly illustrate the statement made in Sect.~\ref{sect:gband_noise}, that recording near-diffraction limited images with a 4~m telescope requires very low noise. However, it must be noted that the power spectra shown are averages of power spectra of individual images in the burst -- the noise level of the combination of images would be roughly seven times lower than that of an individual image. 

\begin{figure}[!t]
 \centering
     \includegraphics[bb=0 0 950 450, clip,width=0.88\linewidth]{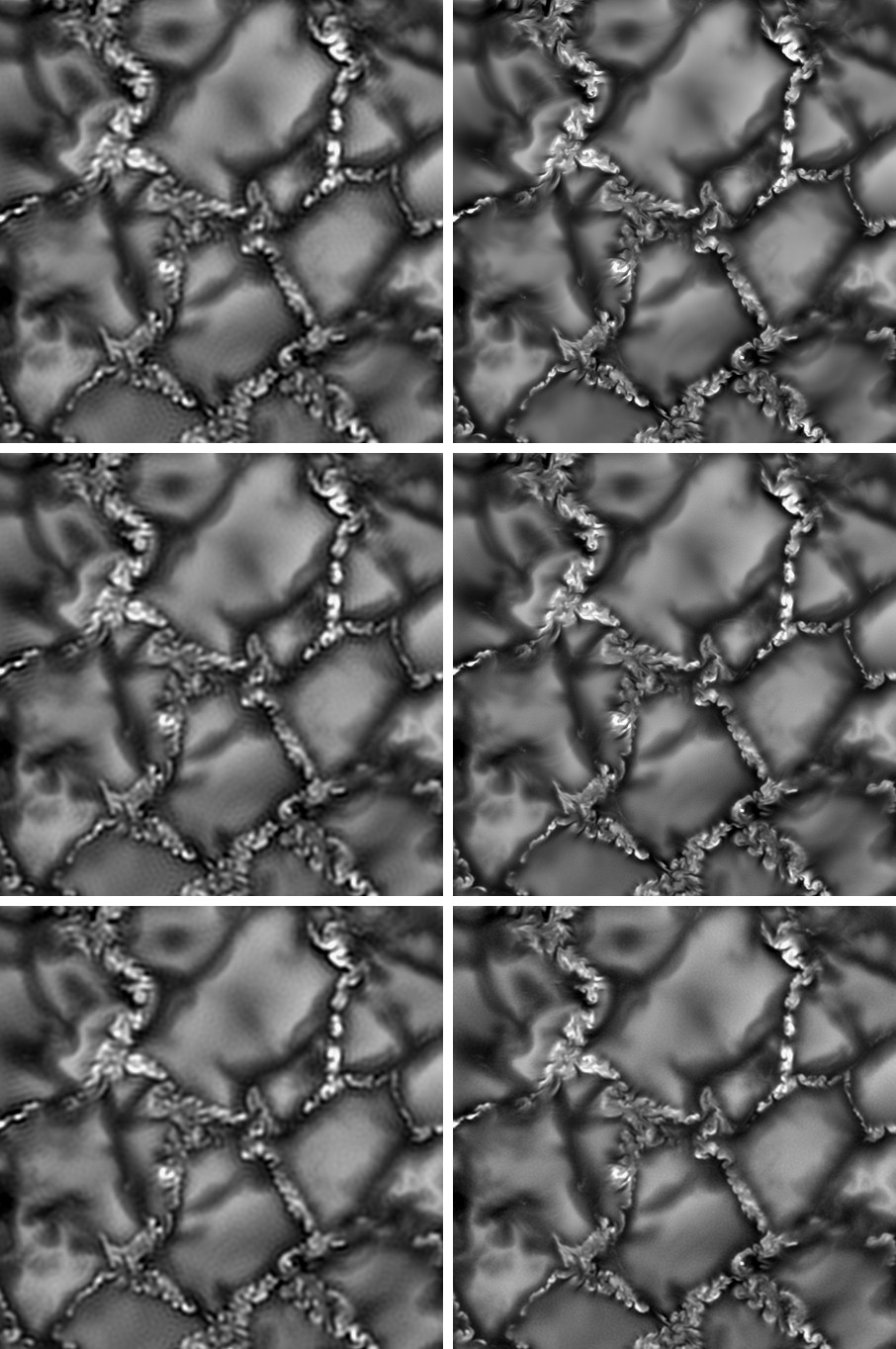}
  \caption{MFBD reconstructed images from the 1.4~m (left) and 4.2~m (right) apertures with outstanding seeing corresponding to $r_0=1$~m, exceptional low noise (\expten{1.0}{-4}), and using 100~Karhunen--Loeve modes for the MFBD reconstruction.}
  \label{fig:mfbd_r0_100cm}
\end{figure}

\begin{figure*}[!t]
 \centering
     \includegraphics[width=0.44\linewidth]{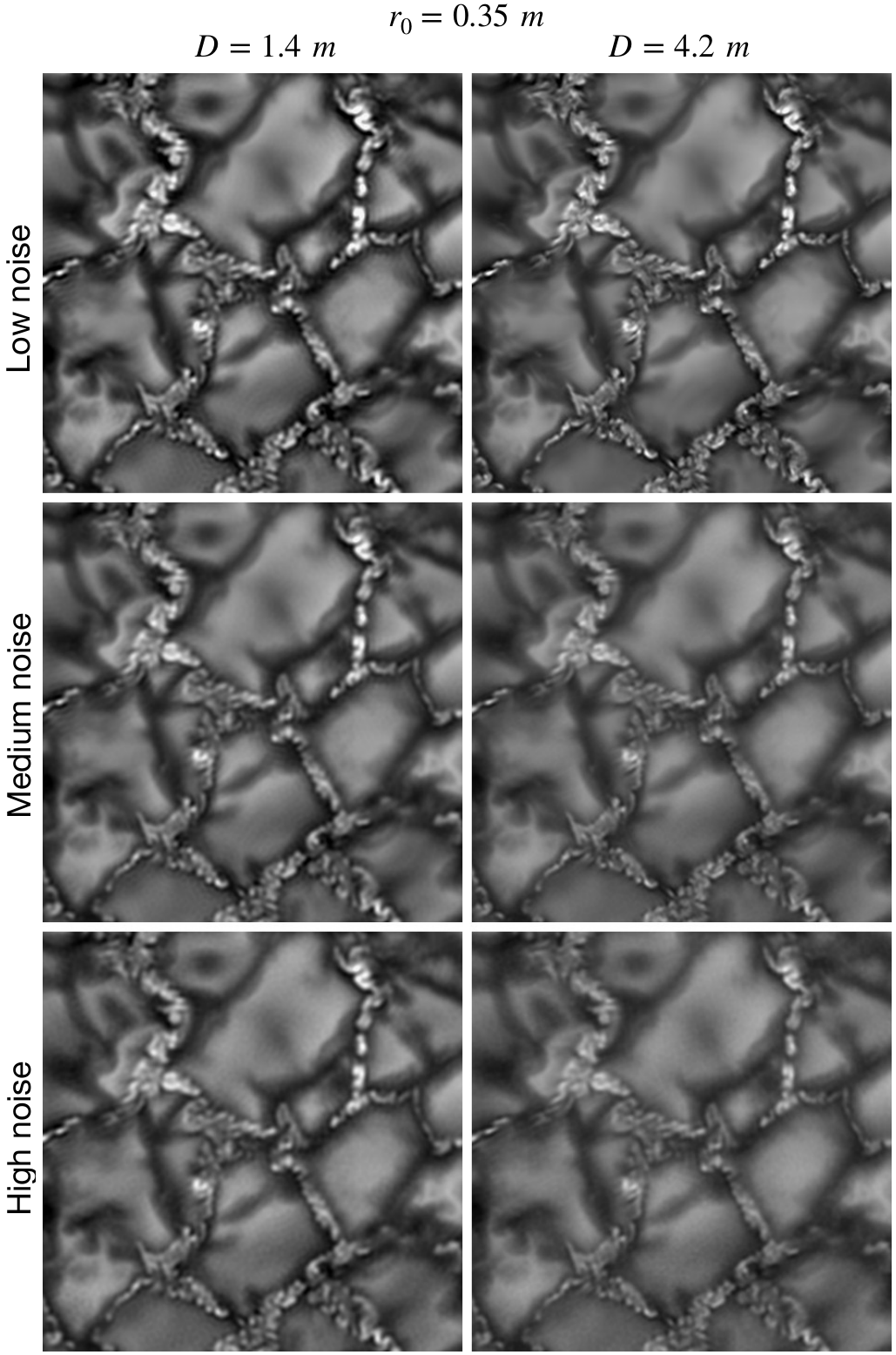}
     \hspace{2mm}
     \includegraphics[width=0.44\linewidth]{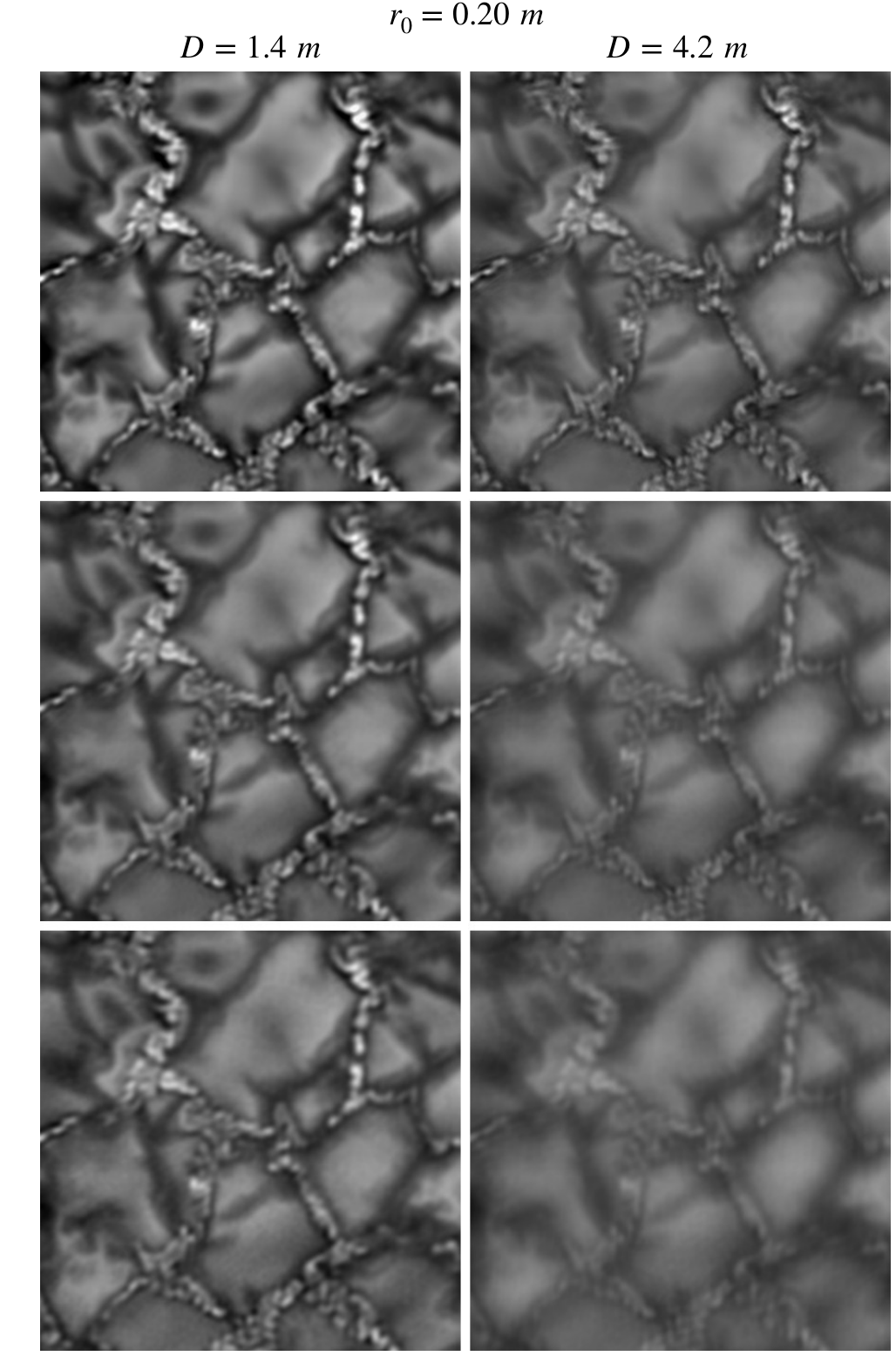}
  \caption{MFBD reconstructed images with 50~Karhunen--Loeve modes from the 1.4~m and the 4.2~m apertures, with seeing corresponding to $r_0=0.35$~m and 0.20~m, and different levels of noise.}
  \label{fig:mfbd_50KL}
\end{figure*}

\begin{figure}[!t]
 \centering
     \includegraphics[width=0.88\linewidth]{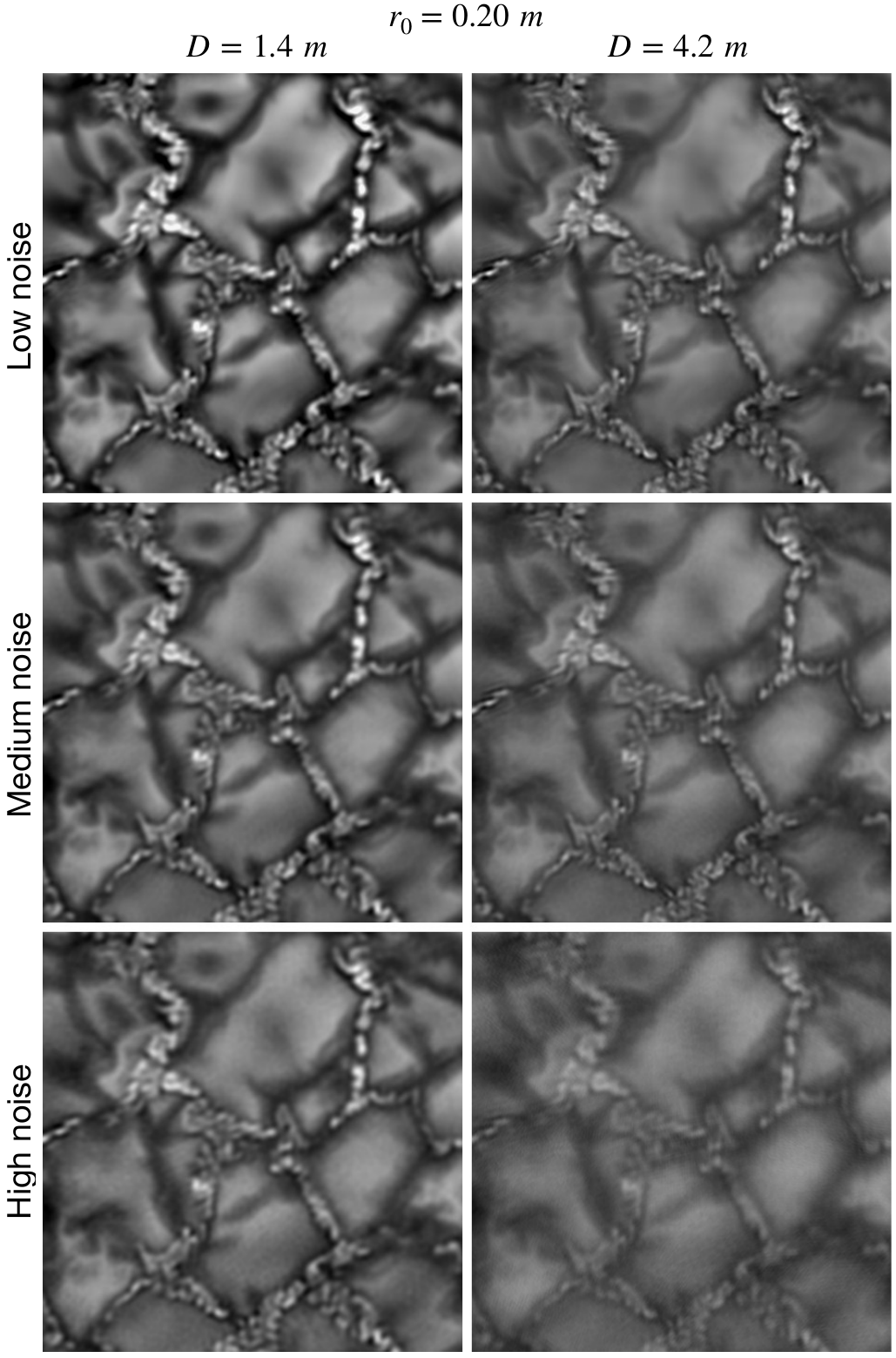}     
  \caption{MFBD reconstructed images using 100~Karhunen--Loeve modes from the 1.4~m and the 4.2~m apertures, with seeing corresponding to $r_0=0.20$~m, and different levels of noise.}
  \label{fig:mfbd_100-200KL}
\end{figure}

\section{Analysis of MFBD restored images and PSFs}
\subsection{ MFBD restored images \label{sect:mfbd_restored}}
The preceding analysis clearly indicates significant advantages of operating EST with six 1.4~m subapertures, instead of with its full 4.2~m aperture, in the absence of MCAO but using short exposures. However, what matters at the end of the day is the quality of the final science images. This quality to a very significant extent relies on post-processing and image reconstruction techniques, and is ultimately limited by noise. 

We have during several decades developed and repeatedly improved PD and MFBD techniques for use at SST \citep{1994A&AS..107..243L, 2002SPIE.4792..146L, 2005SoPh..228..191V, 2021A&A...653A..68L}, and here we employ MFBD to the simulated seeing and noise degraded images. For this processing, we expand the pupil phase errors in KL functions. Our experience involving reconstruction of numerous sets of SST images and also a small set of images from DKIST, suggests that stable reconstructions with excellent reconstructed images are obtained by using on the order of 60--80 KL functions. Another aspect is the importance of performing the MFBD reconstructions over sufficiently small sub-fields in order to take into account anisoplanatism, and to use overlapping sub-fields to smooth out any variations from one sub-field to another. Typically, $4\arcsec\times4\arcsec$ sub-fields are used for SST data. Note that for the MFBD inversions with small subfields, it makes no difference whether the phase errors originate close to the telescope, or at high altitude, or represent an accumulation of phase errors from many phase screens at various heights. 

MFBD with small subfields is considerably more challenging than with large subfields. We therefore applied MFBD techniques to simulated EST images with 256$\times$256 pixel ($4\arcsec\times4\arcsec$) overlapping sub-fields on a 3$\times$3 grid. We then mosaicked the reconstructed sub-images to form a single 512$\times$512 pixel image (448$\times$448 pixels after removal of the apodised parts). We discuss these simulations below.

Figure \ref{fig:mfbd_r0_100cm} shows the MFBD reconstructed images with 100 KL functions, in outstanding seeing conditions ($r_0=1$~m) and with negligible noise (\expten{1.0}{-4}). These images serve as references to the best possible image quality achievable with MFBD techniques.

Figure \ref{fig:mfbd_50KL} shows the impact of stronger and more realistic seeing ($r_0=0.35$~m and 0.20~m) and varying levels of noise (\expten{1.0}{-4}, \expten{1.0}{-3}, and \expten{3.0}{-3}), on the image quality.  When the noise is negligibly small (\expten{1.0}{-4}) and $r_0=0.35$~m, the 4.2~m image has somewhat more fine structure but also lower contrast than the truth image. Already with a noise level of \expten{1.0}{-3}, the 1.4~m image has a visibly more contrast than the 4.2~m image and shows the same amount of small-scale structure. With stronger, but yet low, noise level (\expten{3.0}{-3}) and poorer seeing ($r_0=0.20$~m), the spatial resolution and contrast of the 1.4~m images is superior to those of  the 4.2~m images.

Figure \ref{fig:mfbd_100-200KL} shows MFBD reconstructed images for $r_0=0.20$~m with 100 KL functions. The reconstructions are all well behaved and deliver superior results for the 1.4 m aperture when compared to the 4.2~m aperture reconstructions. Reconstructions with 200 KL functions show evidence of  problems such as too high contrast with 1.4~m aperture, and artefacts in the form of fringe-like structures with 4.2~m aperture. This indicates that neither 1.4~m images nor 4.2~m images should be reconstructed with more than on the order of 100 KL functions, which is consistent with our experience from reconstruction of SST images and the aforementioned G-band and 416~nm images (M. van Noort, private communication) recorded with DKIST. In the following, we constrain the discussions to inversions made with 50 or 100 KL functions. 

\begin{figure}[!t]
 \centering
      \includegraphics[width=0.49\linewidth]{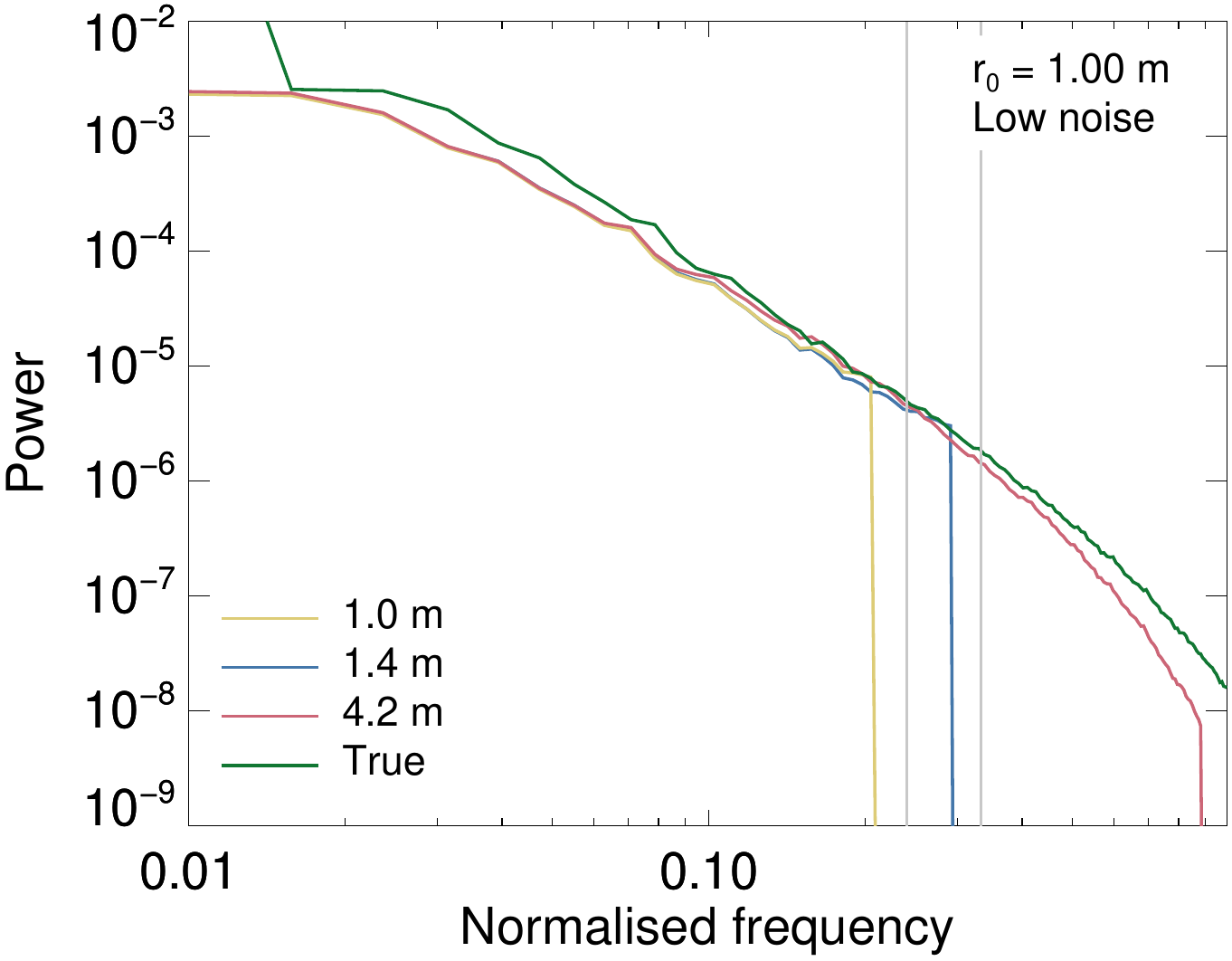}
      \includegraphics[width=0.49\linewidth]{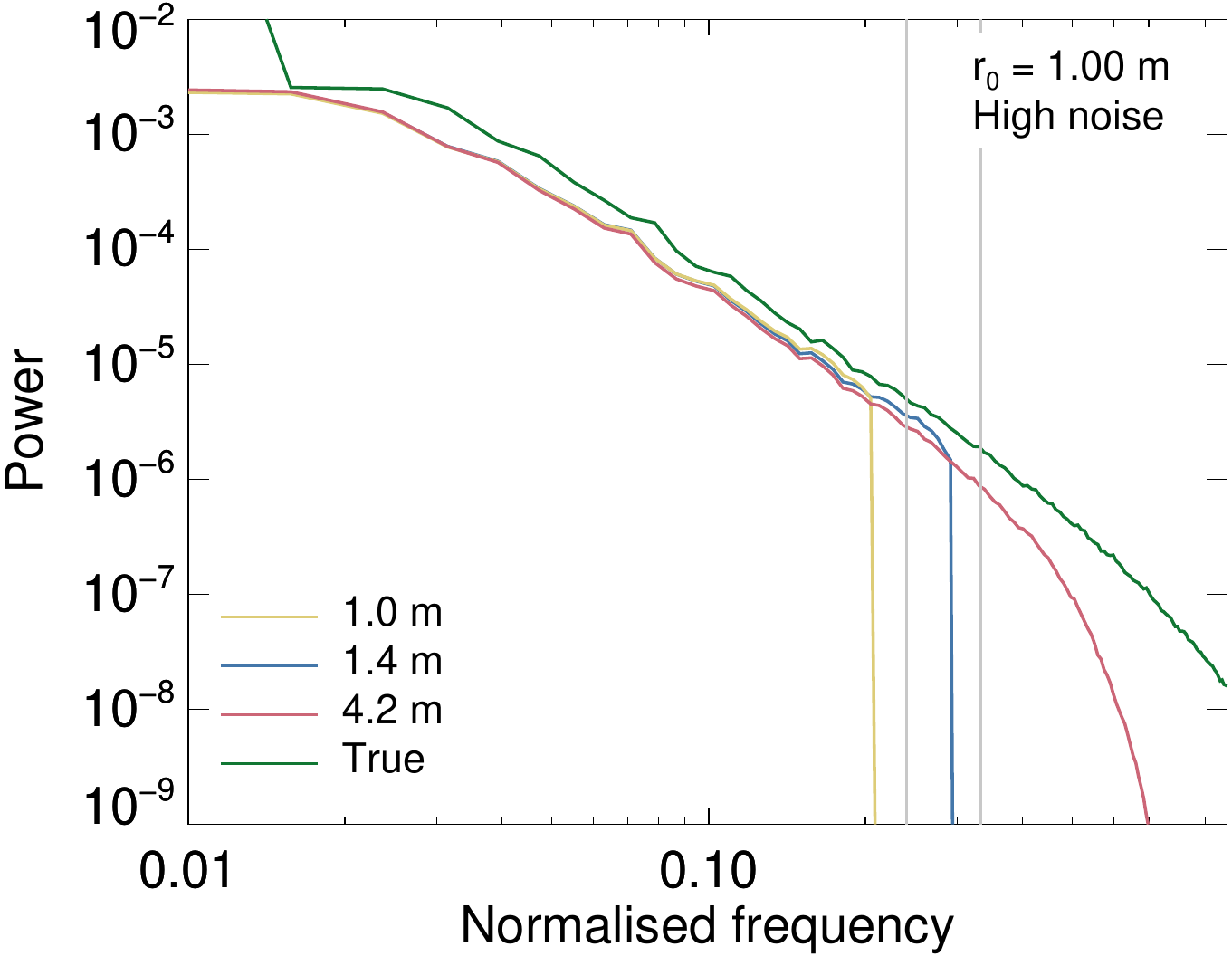}\\[1.5mm]
      \includegraphics[width=0.49\linewidth]{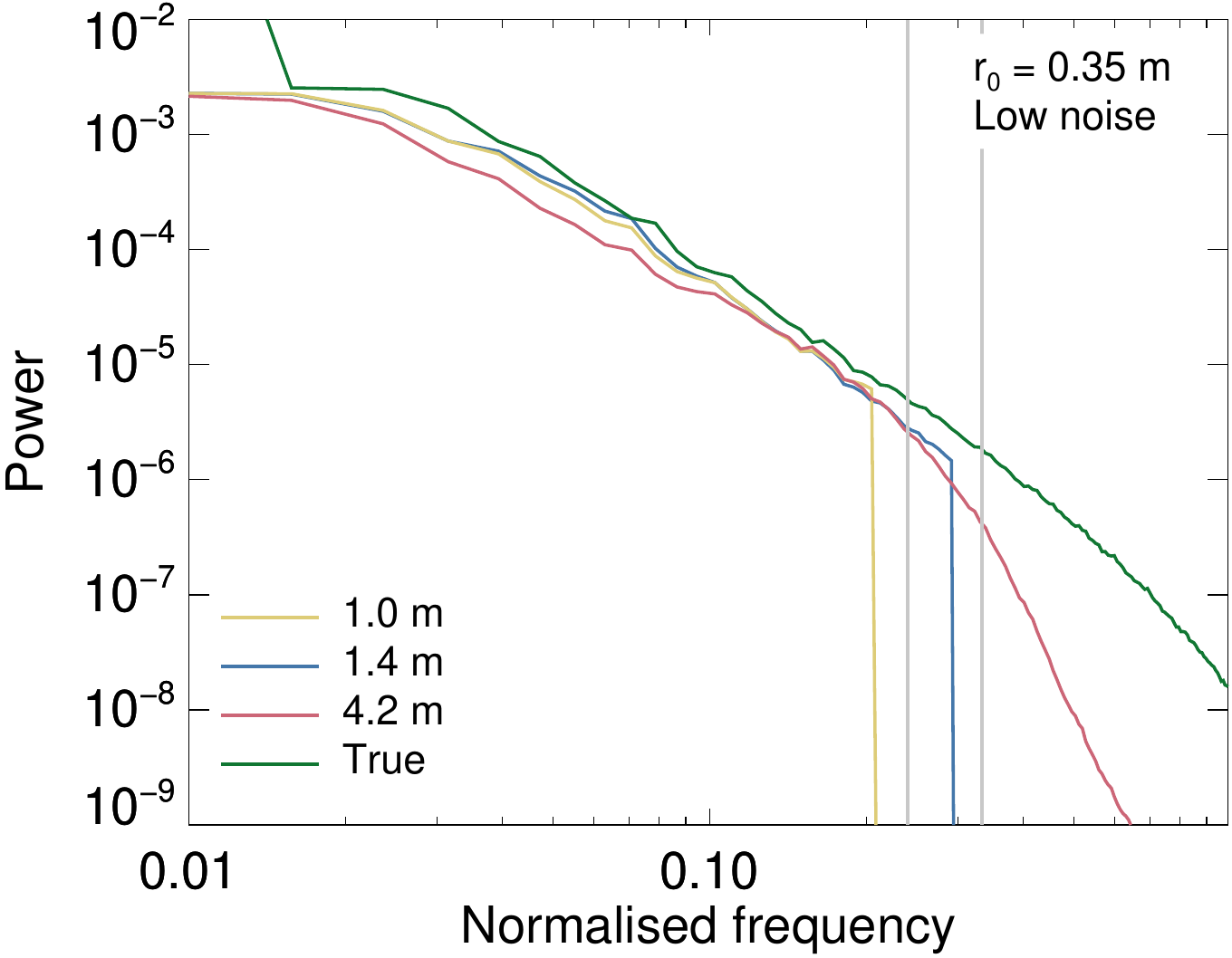}
      \includegraphics[width=0.49\linewidth]{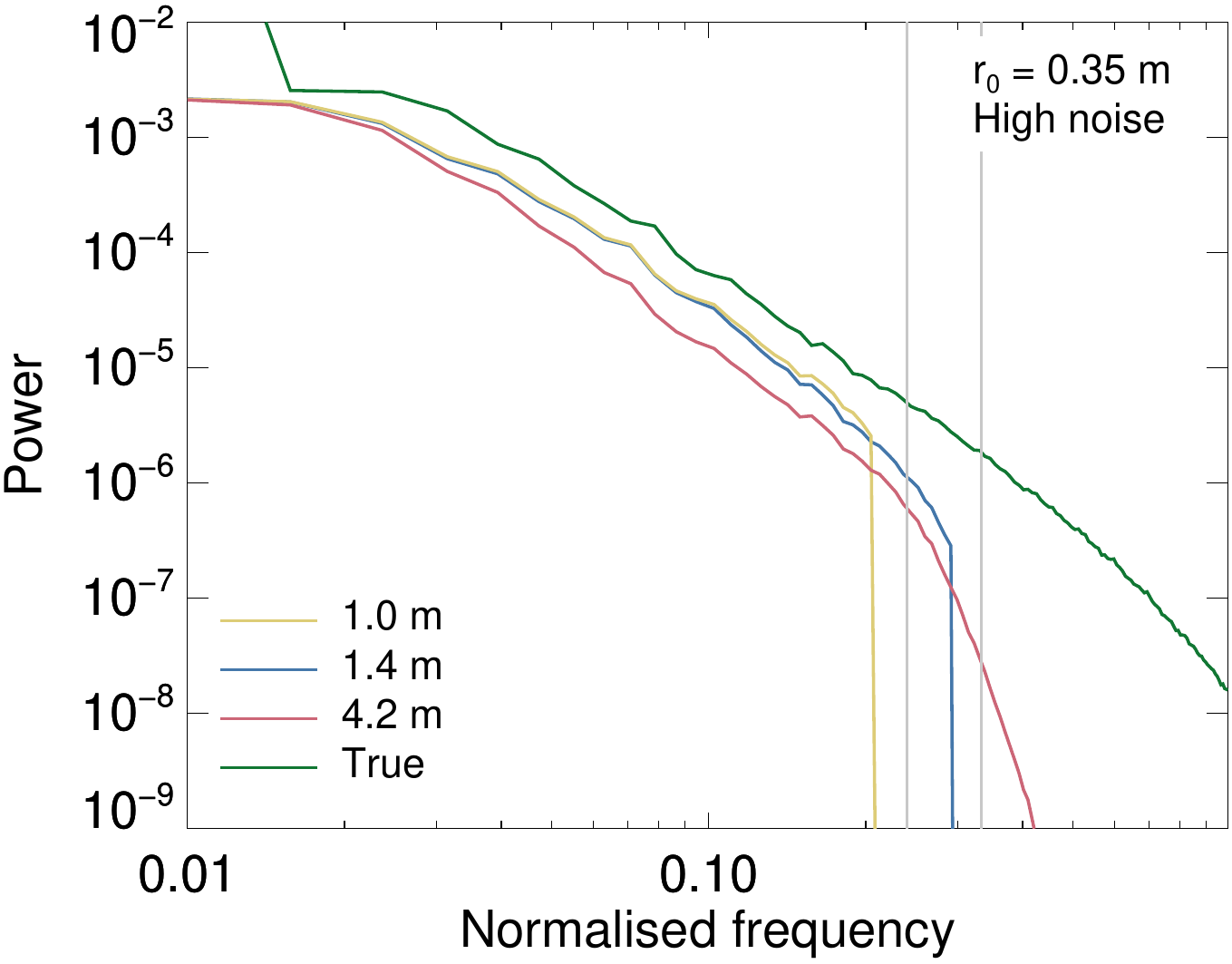} 
      \includegraphics[width=0.49\linewidth]{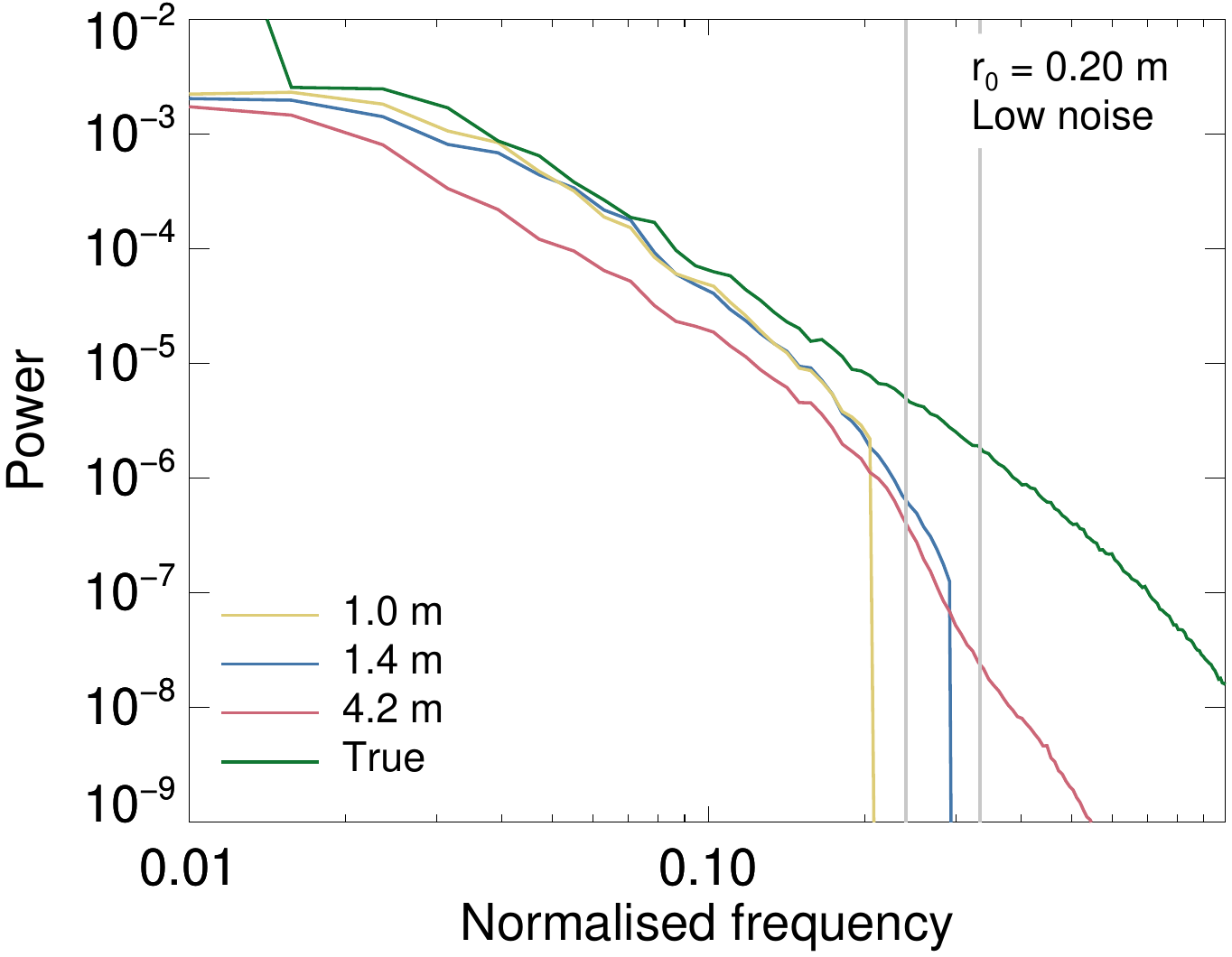}
      \includegraphics[width=0.49\linewidth]{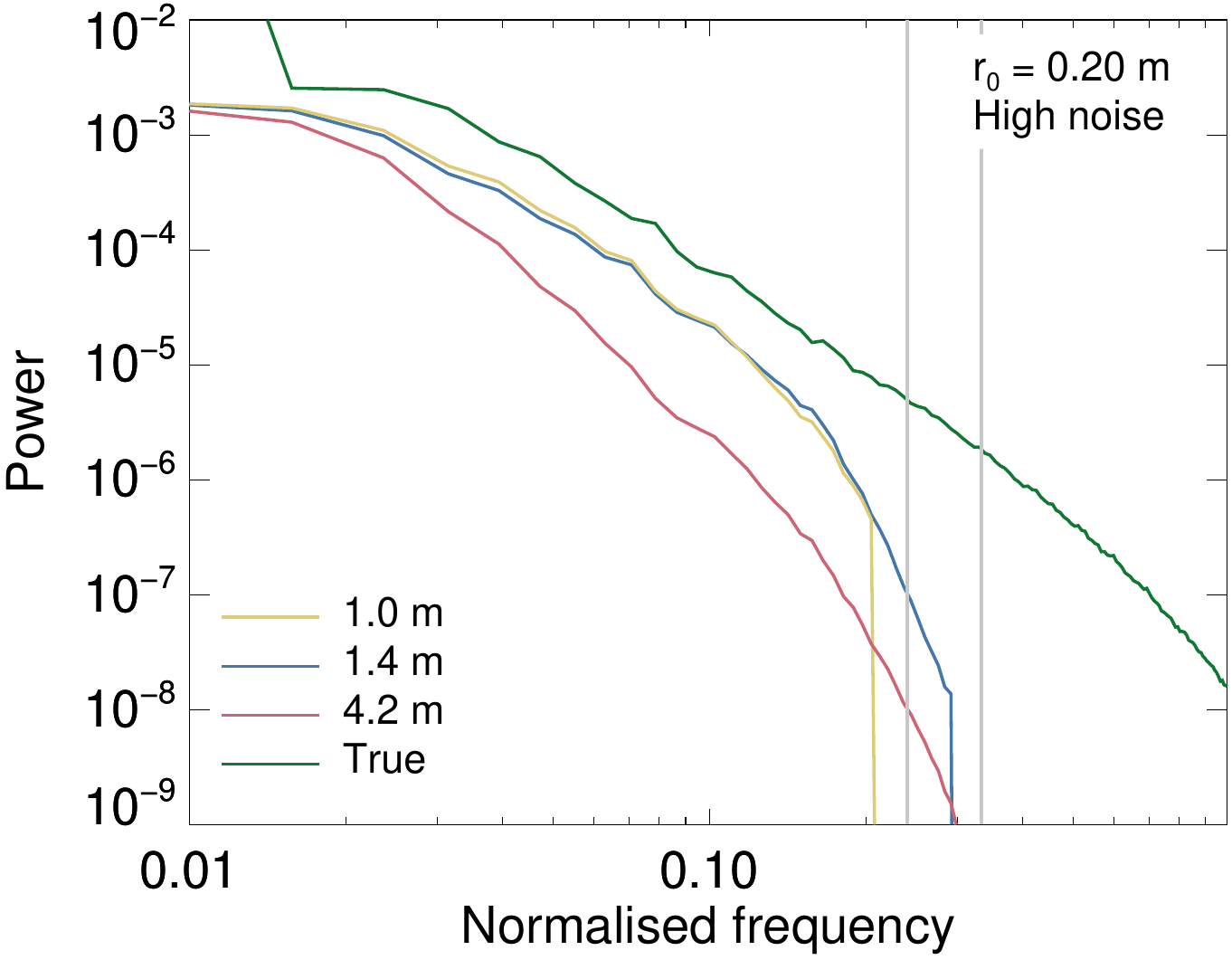}

  \caption{Plots of power spectra of MFBD reconstructed images with 1~m, 1.4~m, and 4.2~m apertures, seeing characterised by $r_0=$1.0~m, 0.35~m, and 0.20~m, and a total noise level from 50 frames corresponding to \expten{1.0}{-4} (low noise), and \expten{3.0}{-3} (high noise). For comparison, the power spectrum of the truth image is included. }
  \label{fig:mfbd_power}
\end{figure}

Figure \ref{fig:mfbd_power} shows power spectra for the truth image and for the MFBD reconstructed images. The power spectra of the 1.4~m aperture are consistently closer to that of the truth image than those of the 4.2~m aperture. This tendency increases with stronger seeing and increasing noise level, such that the power spectrum with a noise level of \expten{3.0}{-3} and $r_0=0.20$~m for the 4.2~m aperture, if anything, looks like a failure. Note that the noise filter applied to these data is set to clip at 90\% of the diffraction limit for each aperture diameter.

Figure \ref{fig:mfbd_power} also demonstrates, that if $r_0=0.35$~m for the high-altitude seeing layer, EST cannot reach the diffraction limit without MCAO even with extremely low noise level (\expten{1.0}{-4}). If we consider the resolution limit to be the spatial frequency for which the power spectrum of the restored image drops one order of magnitude below that of the truth image, then the resolution reached is about 40\% of the diffraction limit in this case. This implies, that to reach the diffraction limit of EST over a large FOV without MCAO, $r_0$ needs to be substantially larger than 0.35~m.

\begin{figure}[!t]
 \centering
      \includegraphics[width=0.49\linewidth]{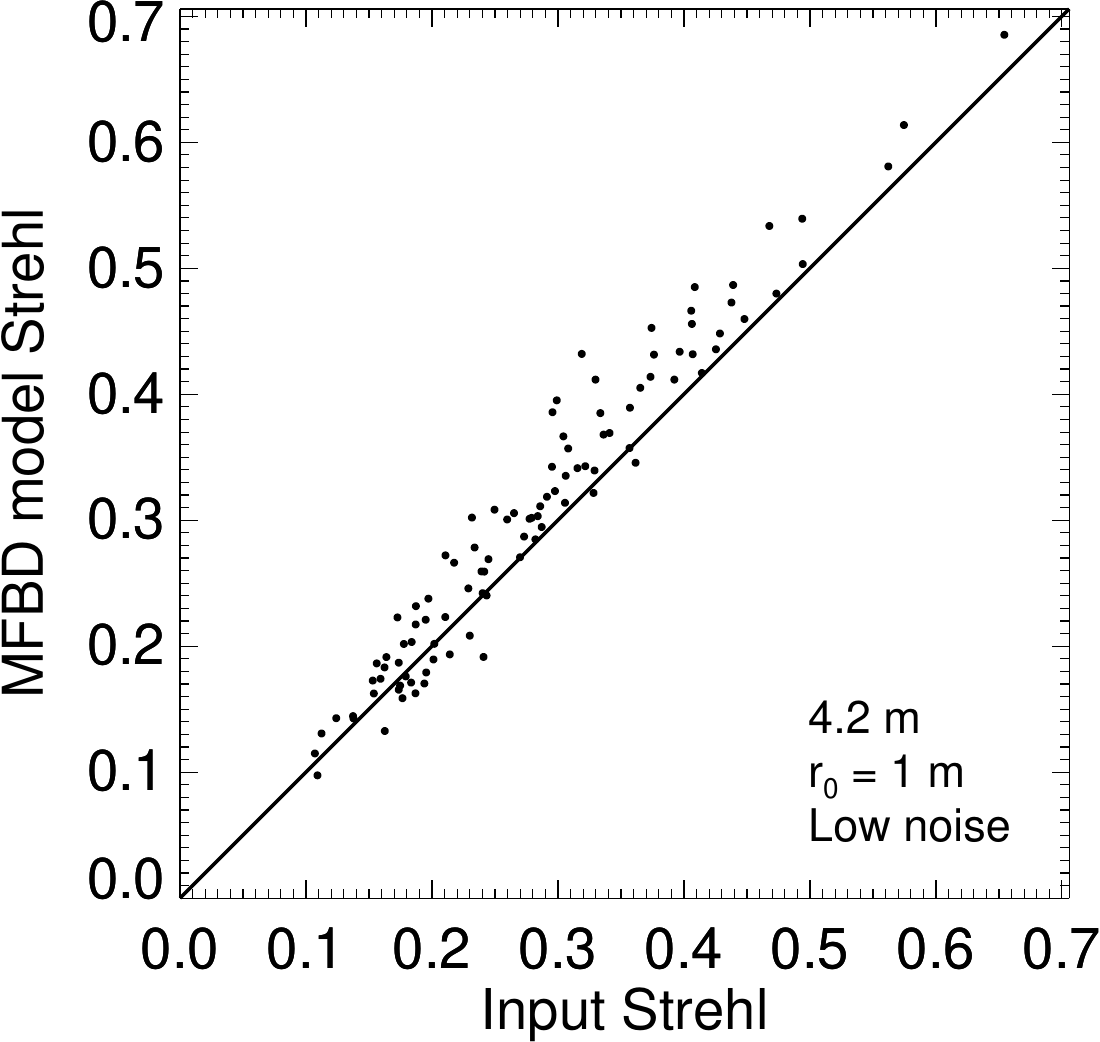}
      \includegraphics[width=0.49\linewidth]{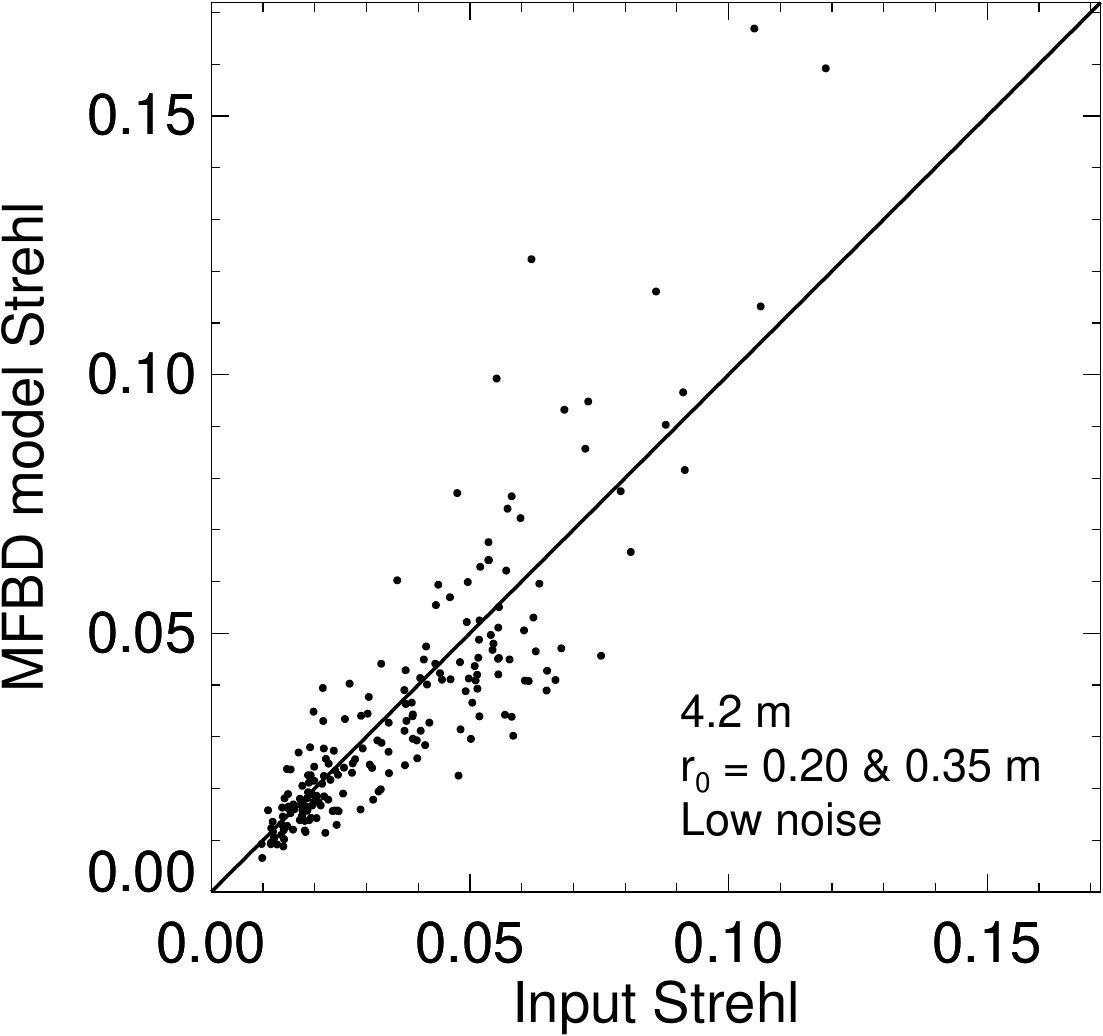}\\[1.5mm]
      \includegraphics[width=0.49\linewidth]{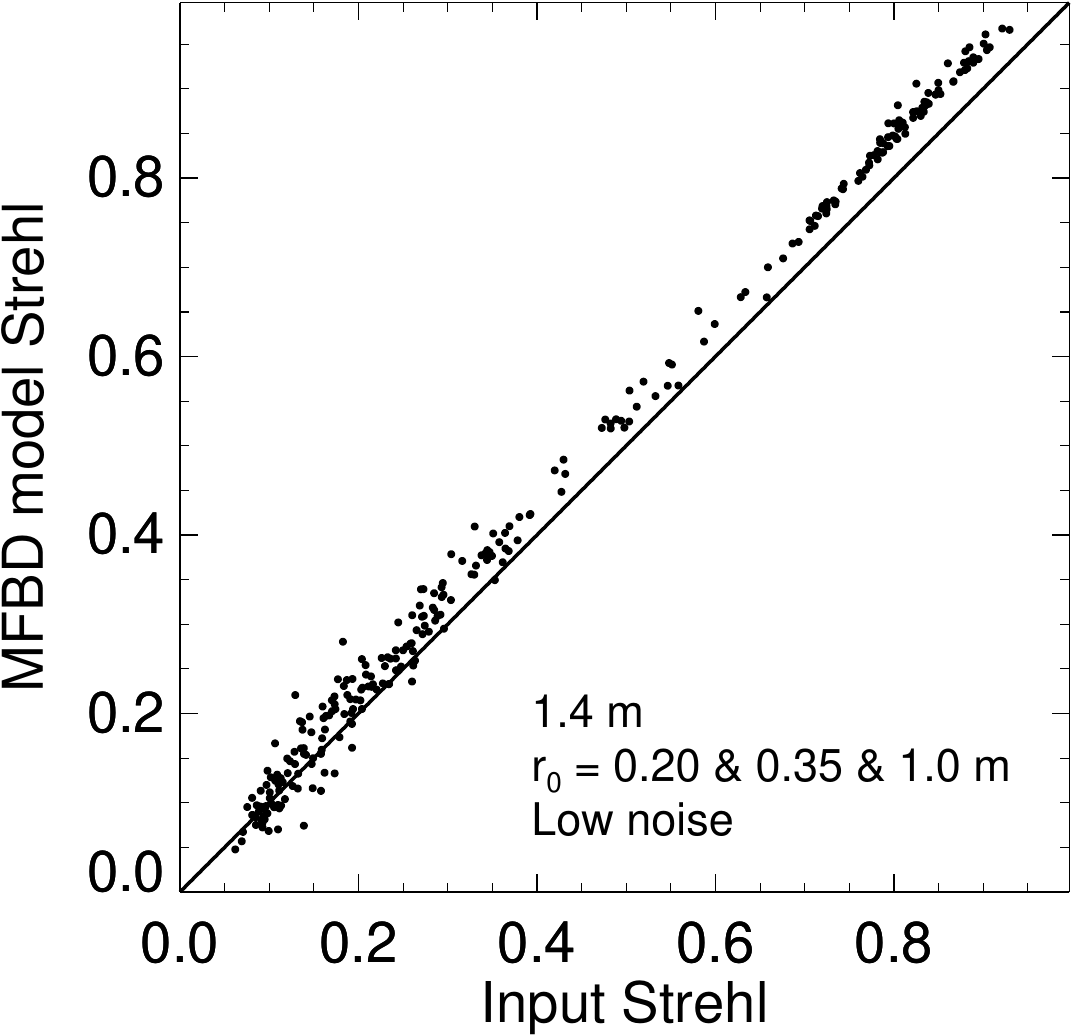}
      \includegraphics[width=0.49\linewidth]{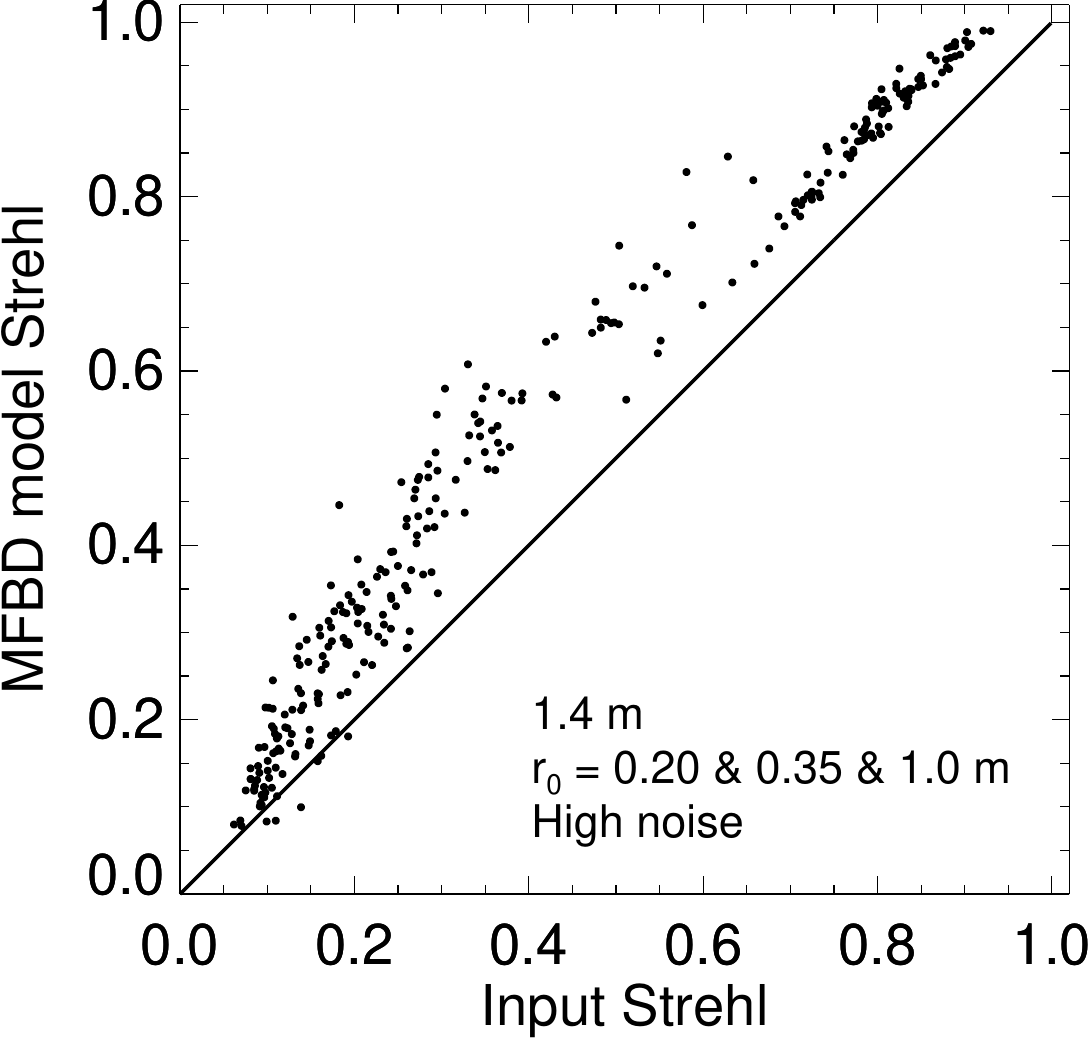}        
  \caption{Comparisons of input and output (from MFBD processing) Strehl values with 4.2~m and 1.4~m apertures. Low noise corresponds to \expten{1.0}{-4} and high noise to \expten{3.0}{-3} with 50 frames.}
  \label{fig:Strehl_comparison}
\end{figure}

\begin{figure}[!t] 
 \centering
  \includegraphics[width=0.49\linewidth]{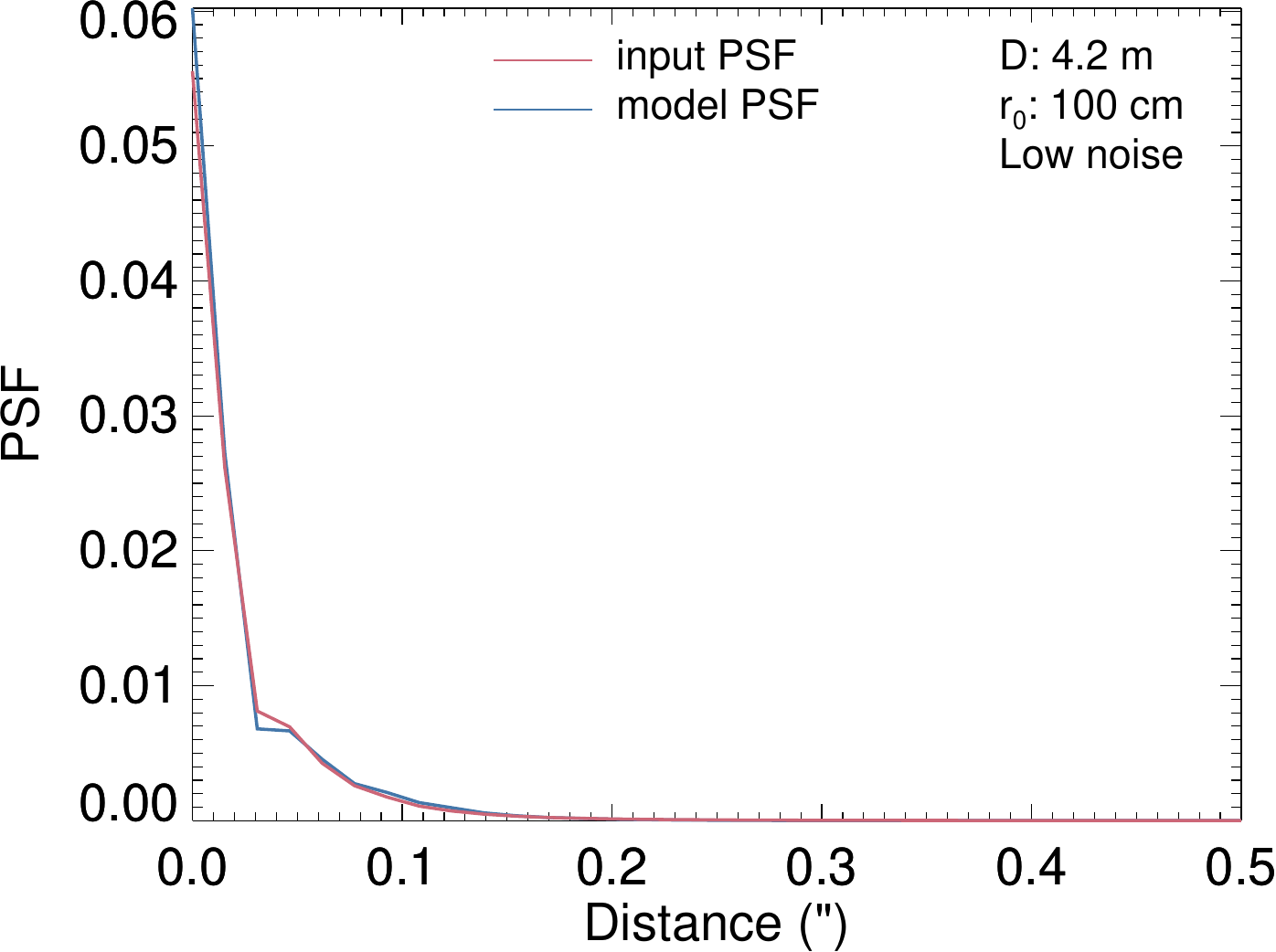}
  \includegraphics[width=0.49\linewidth]{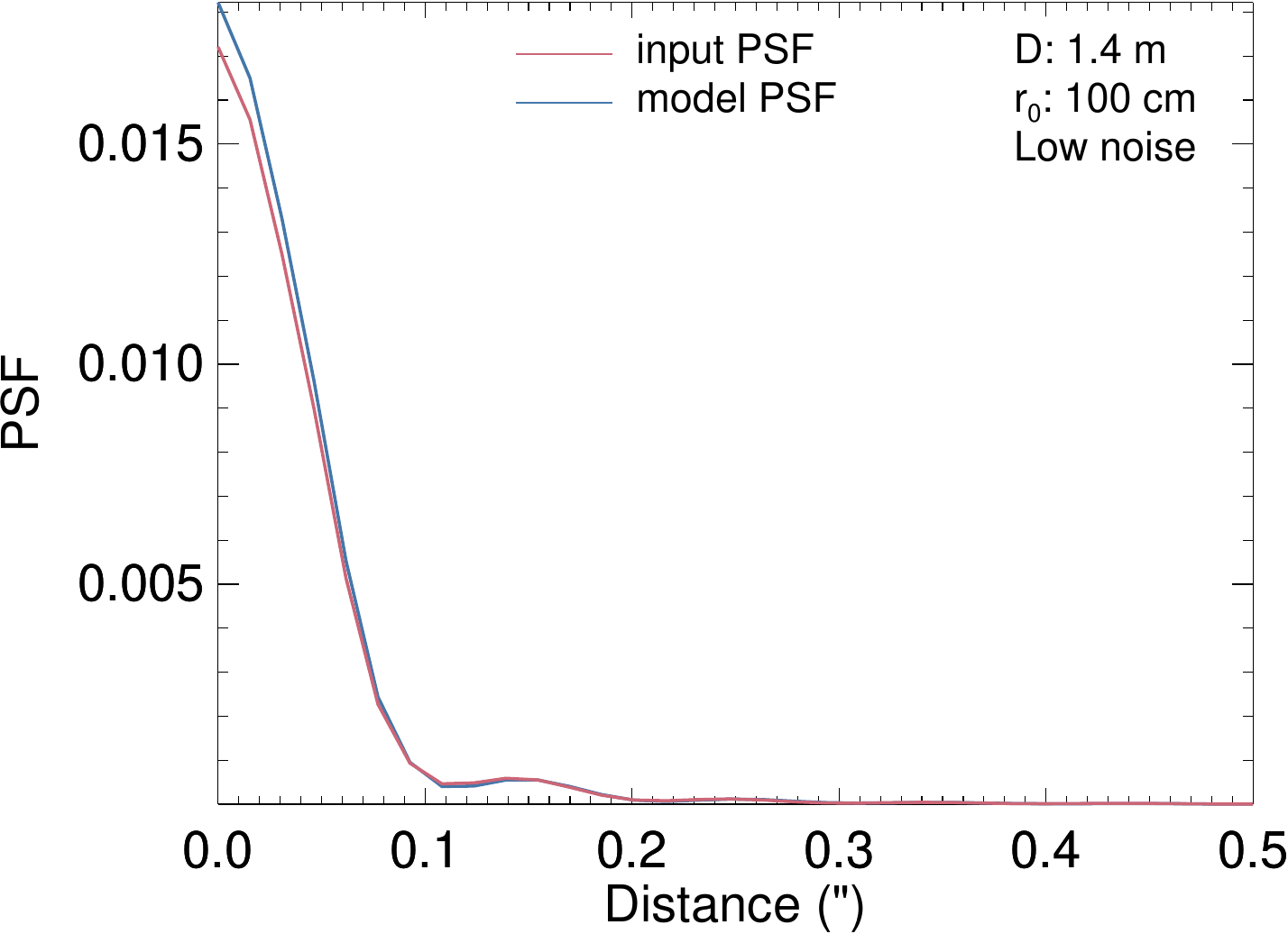}\\[1.5mm]
  \includegraphics[width=0.49\linewidth]{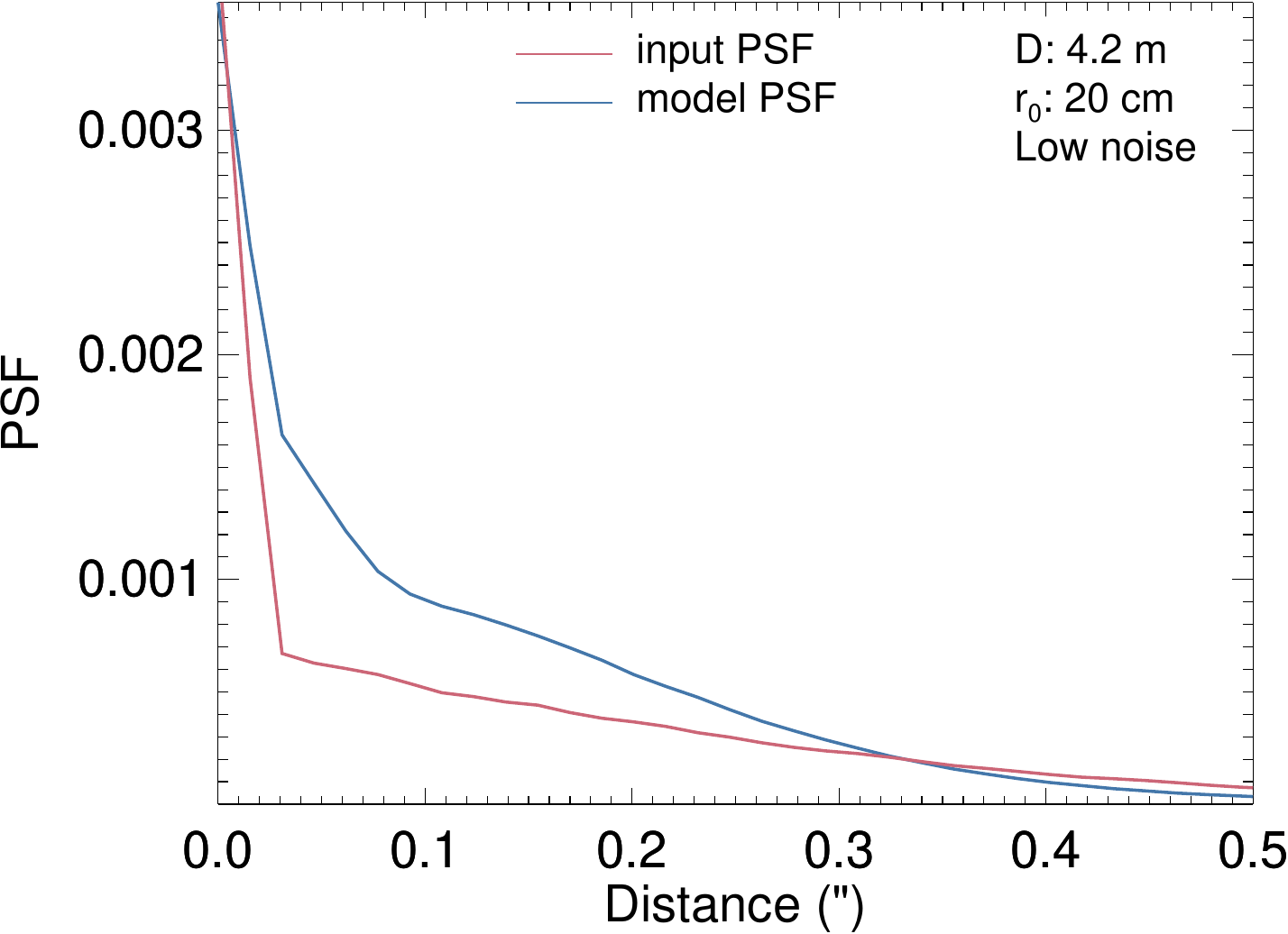}
  \includegraphics[width=0.49\linewidth]{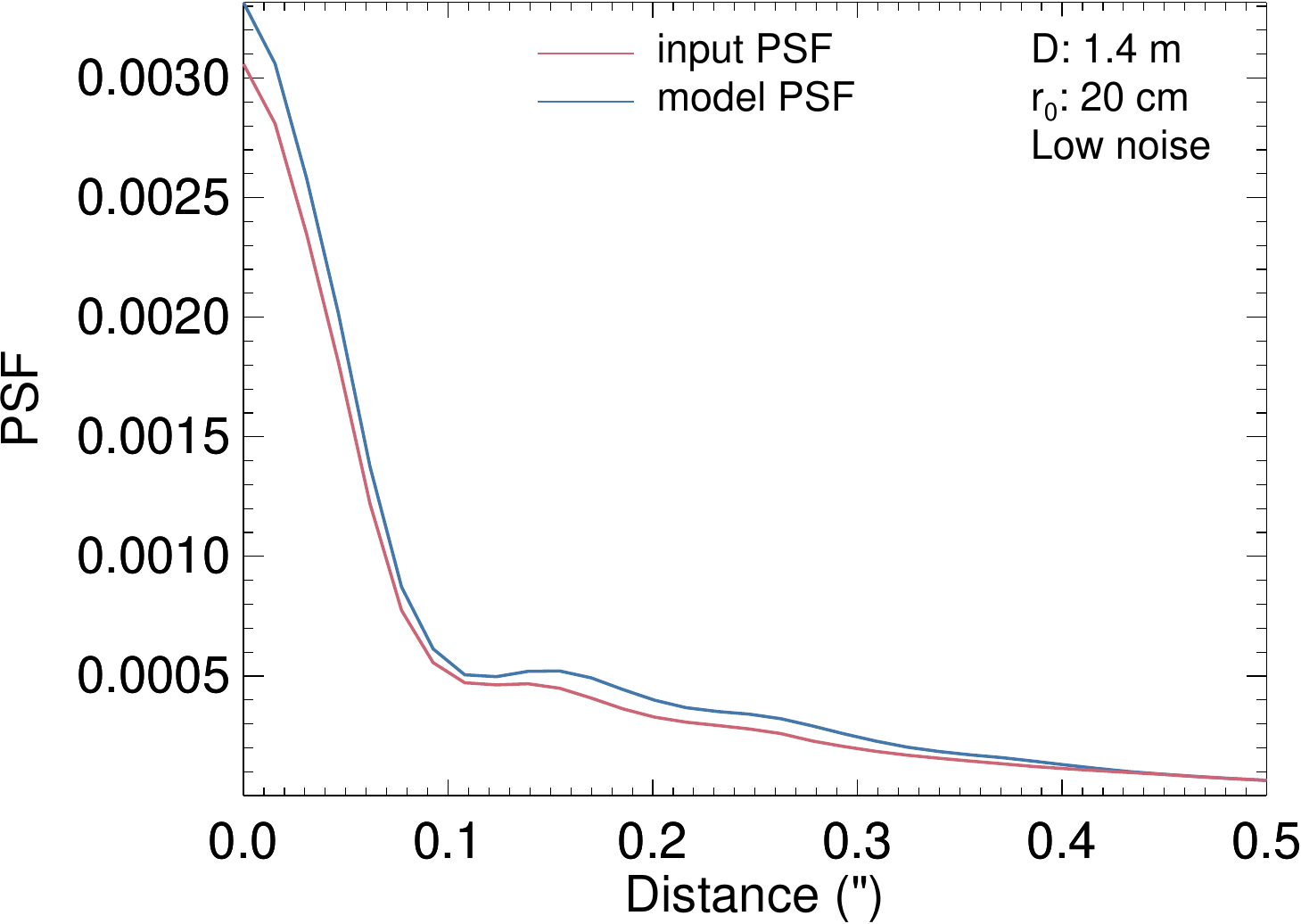}\\[1.5mm]
  \includegraphics[width=0.49\linewidth]{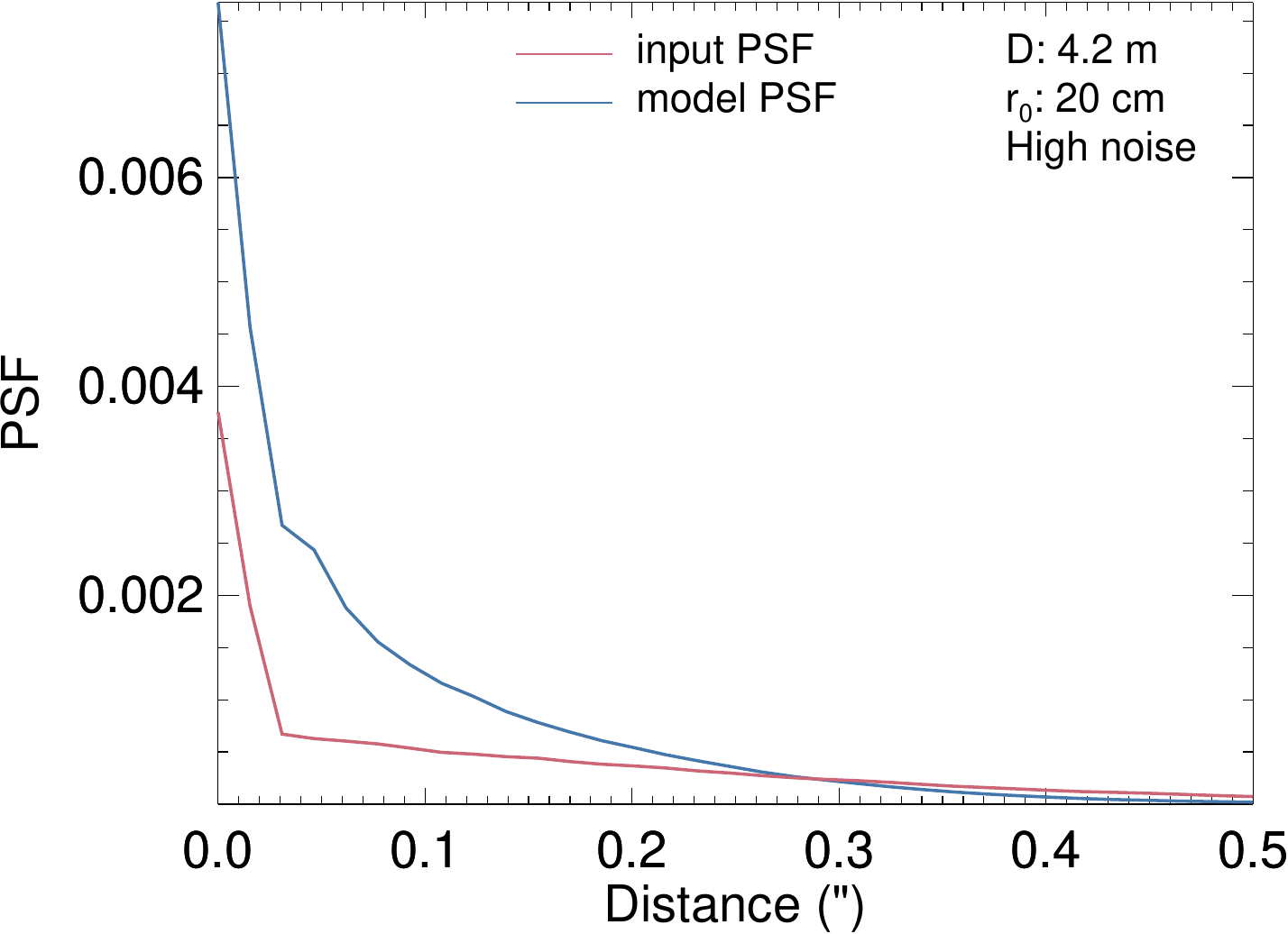}
  \includegraphics[width=0.49\linewidth]{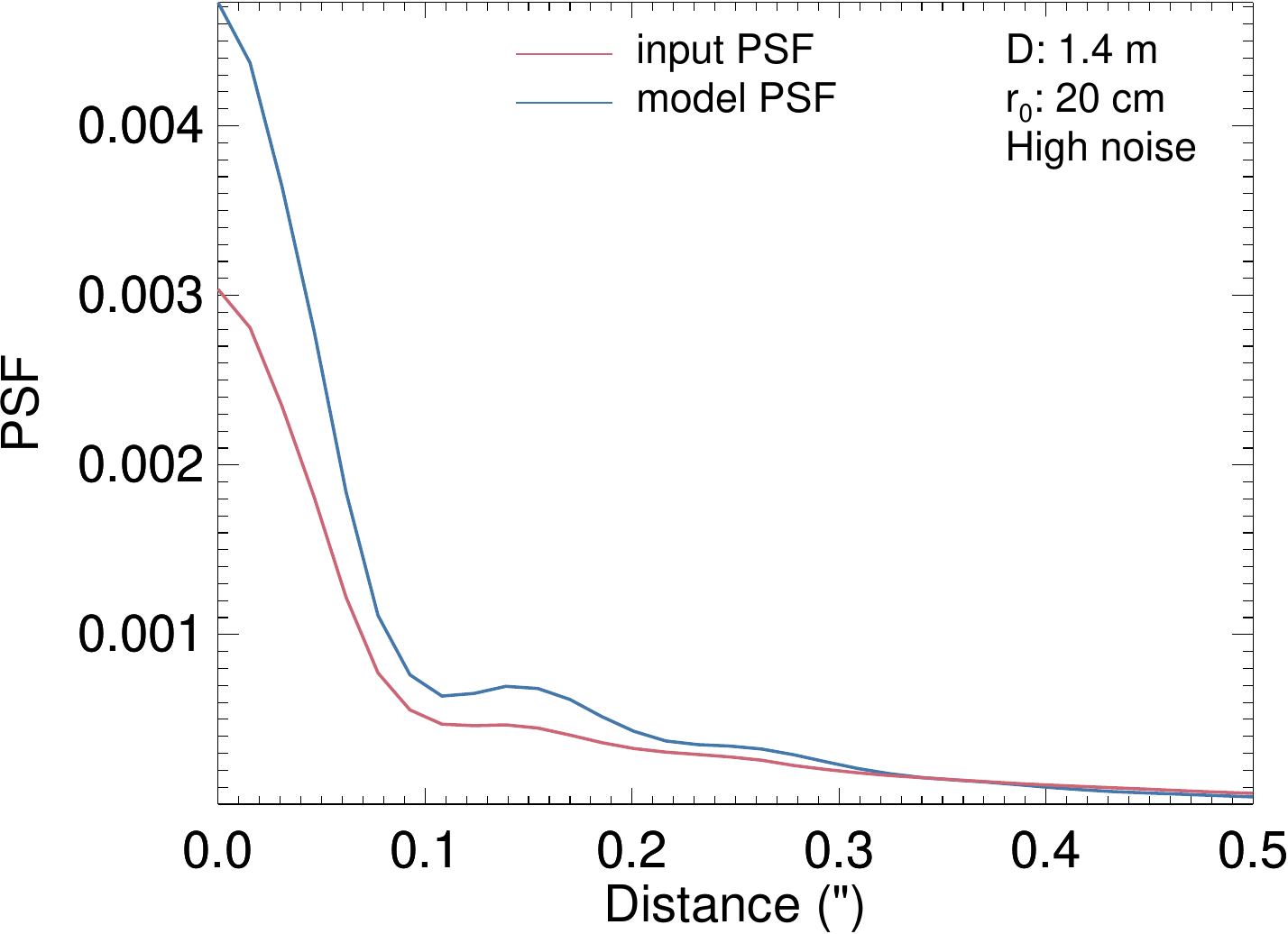}  
 \caption{Comparisons of input and estimated (from MFBD processing) PSFs for a 4.2~m aperture, and 1.4~m aperture. } 
   \label{fig:PSF_comparison}
\end{figure}

\begin{figure}[!t] 
 \centering
\includegraphics[width=0.49\linewidth]{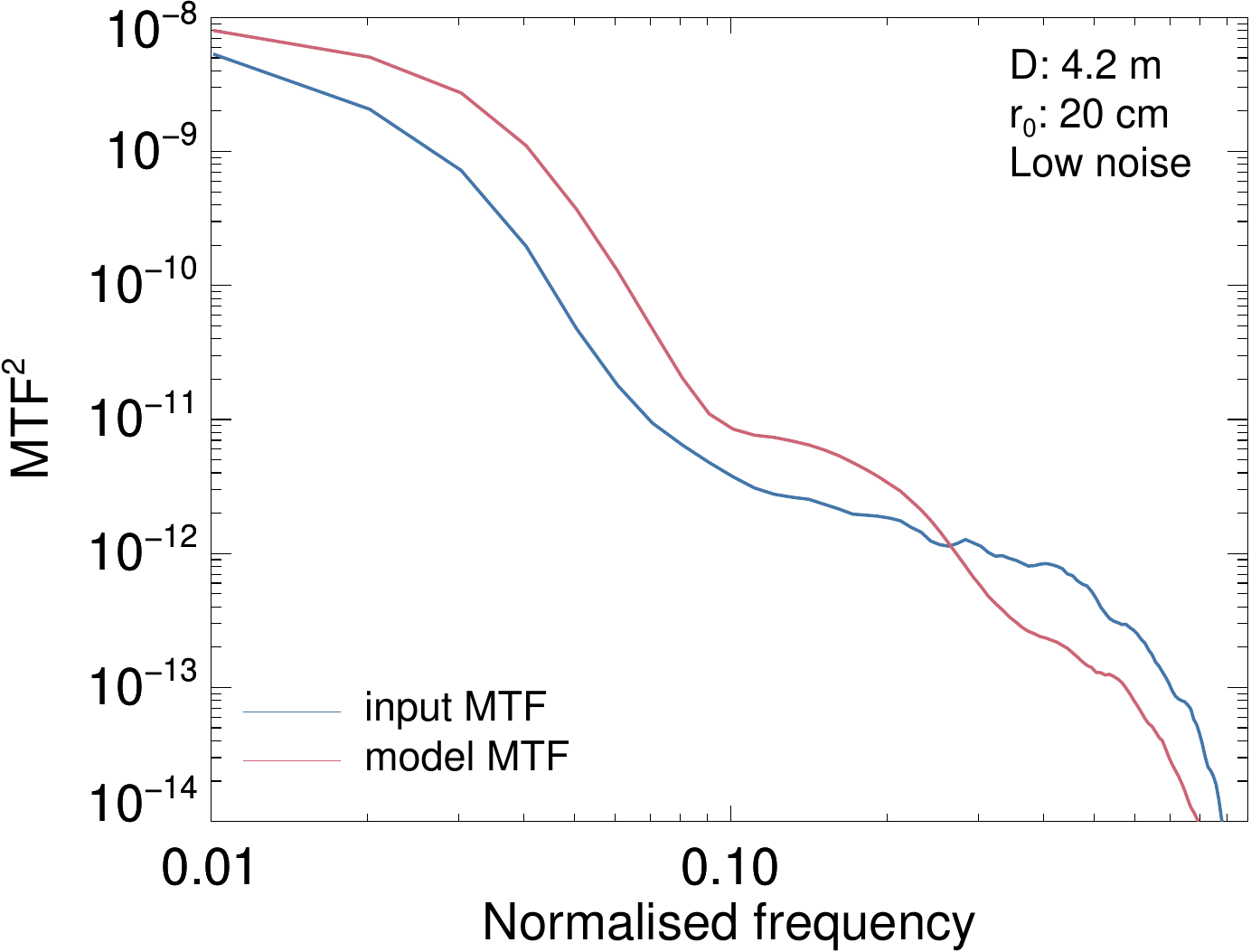}
\includegraphics[width=0.49\linewidth]{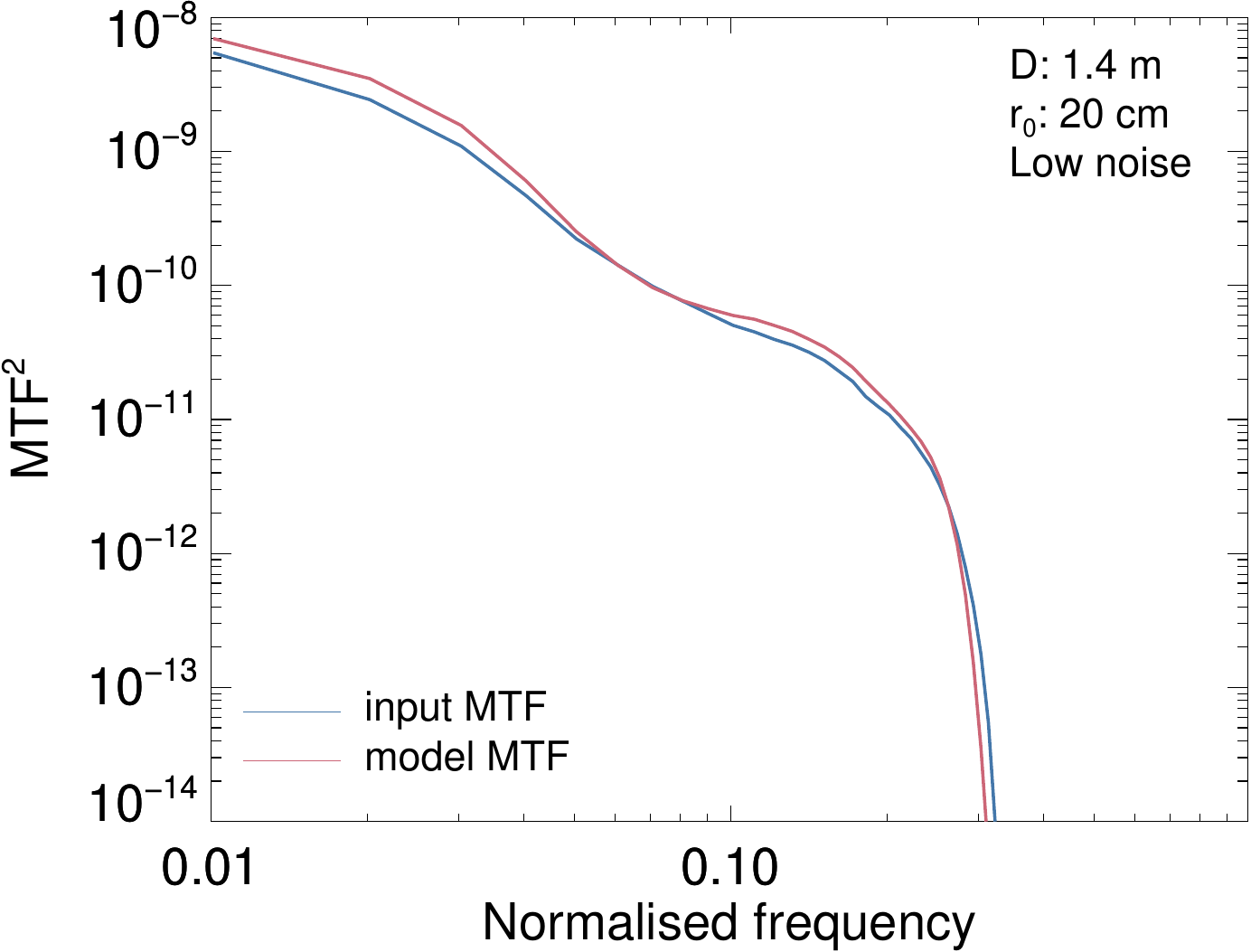}\\[1.5mm]
\includegraphics[width=0.49\linewidth]{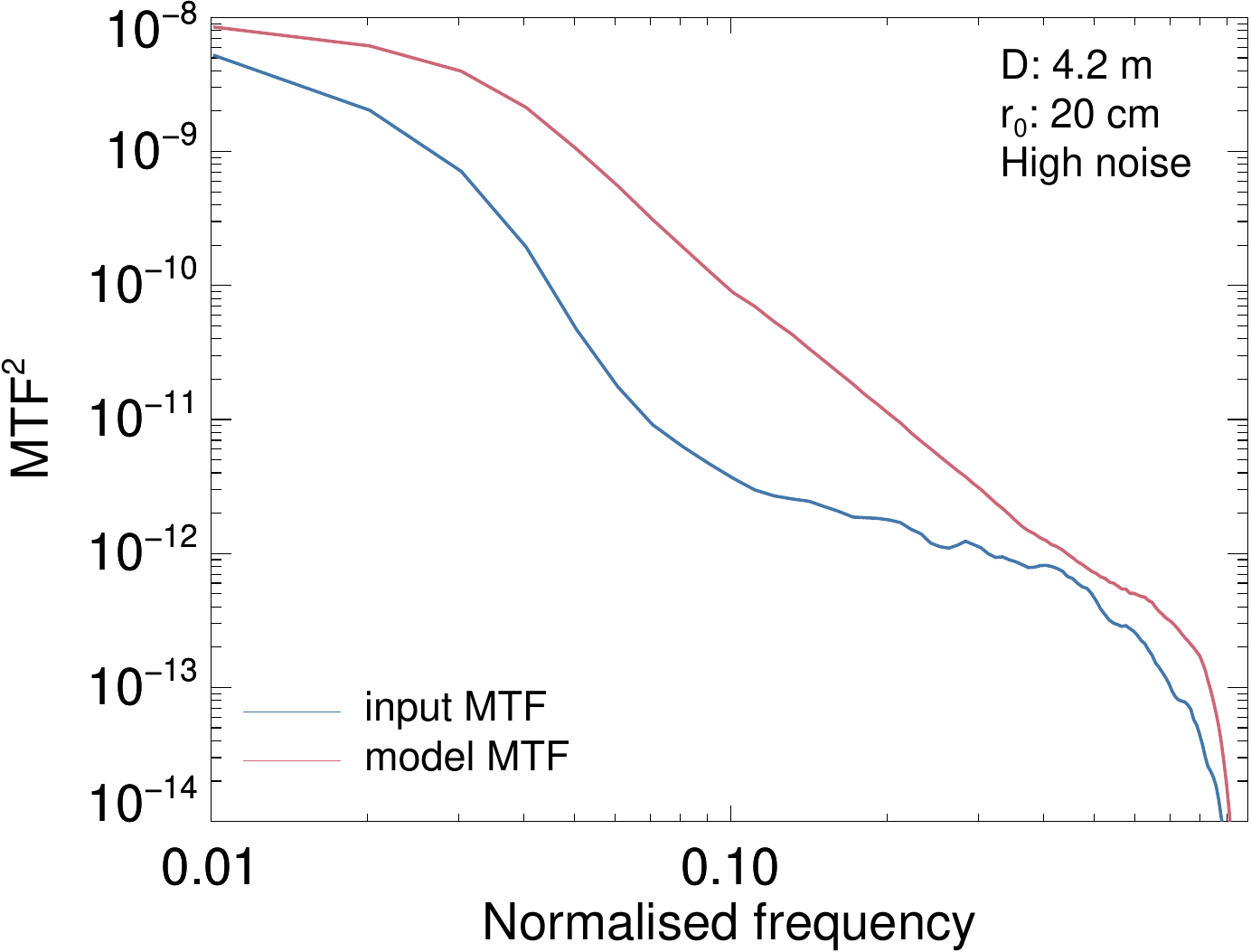}
\includegraphics[width=0.49\linewidth]{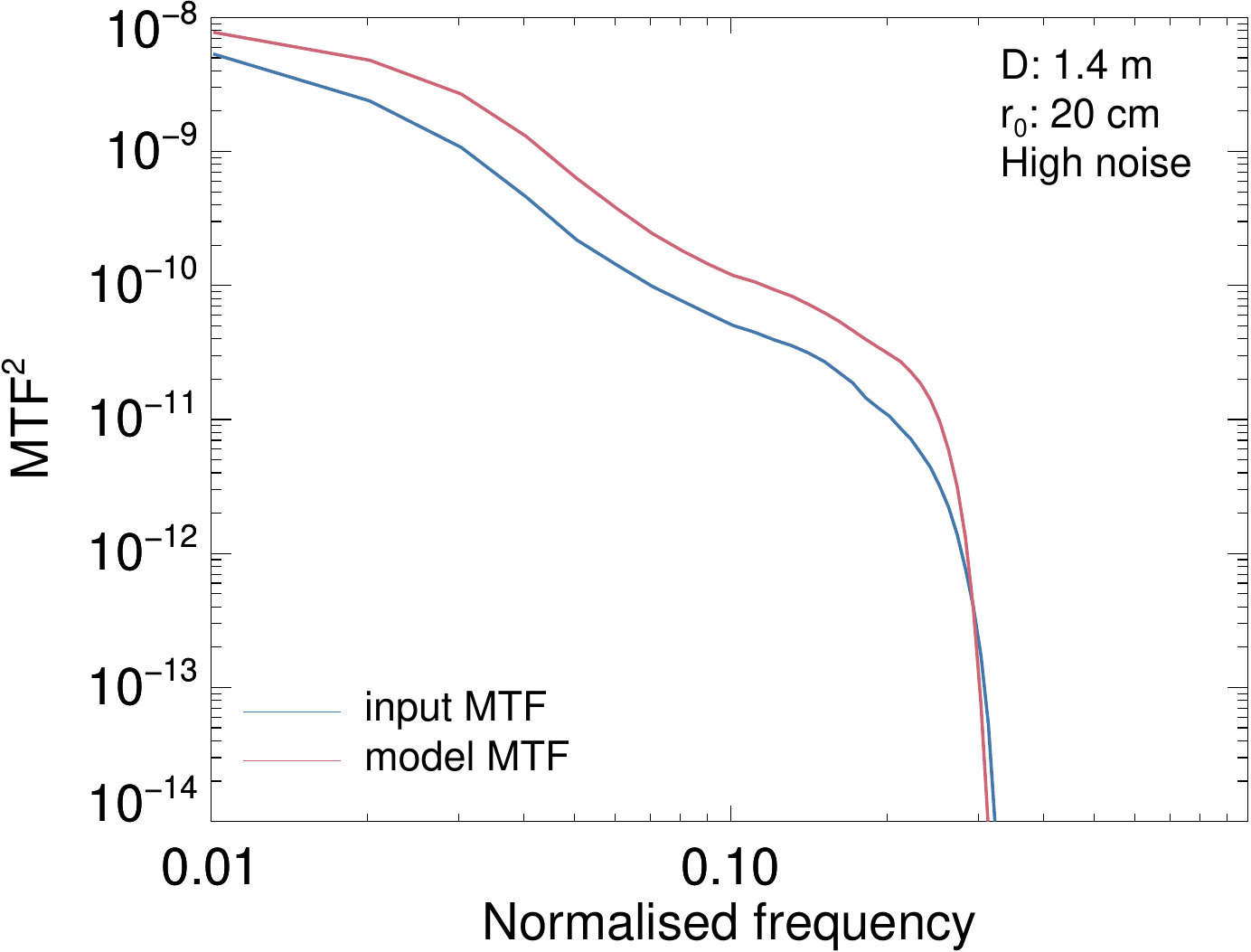}
 \caption{Comparisons of input and estimated (from MFBD processing) squared MTFs of PSFs for a 4.2~m aperture and 1.4~m aperture.} 
   \label{fig:MTF_comparison}
\end{figure}

\subsection{Analysis of PSF data}
\label{sec:psf_analysis}
The MFBD restored images reveal several distinct features that mark important differences between restorations of images from the full EST 4.2~m aperture, and from the much smaller 1.4~m subapertures. In particular, it is of importance to understand the origin of the strongly reduced contrast of the 4.2~m reconstructions, and the differences in image quality. To that end, comparisons of input and output wavefronts are of less relevance -- past investigations \citep{2010A&A...521A..68S} clearly suggest that the modelled MTFs (and thus also PSFs) may well give good representations of the input PSFs, even if the modelled wavefronts are not at all well represented. What matters here is therefore the difference between input and output PSFs -- not the wavefronts.

In Fig. \ref{fig:Strehl_comparison}, we show scatter plots of the Strehl values for the input and output PSFs of the 4.2~m and 1.4~m apertures.  In outstanding seeing ($r_0$=1~m), the correlation between input and output Strehl values is good with the 4.2~m aperture. However, in more realistic seeing conditions, the correlation is poor. In contrast, the correlation is tight between input and output Strehl values for all $r_0$ values with 1.4~m aperture and low noise (\expten{1.0}{-4}). For relatively high (\expten{3.0}{-3}) noise, the plots show two populations of correlation: that of high Strehl is identical to that with low noise, and is apparently insensitive to noise -- this population corresponds to $r_0$ values of 1~m. The other population corresponds to seeing with lower Strehl values and for this population, the output Strehl values are systematically displaced towards higher Strehl values and with a larger spread, when compared to the plot for very low noise. 

The much tighter correlation between input and output Strehl values for the 1.4~m aperture than for the 4.2~m aperture clearly suggests that MFBD processing with 1.4~m aperture is more likely to provide well restored images. But it is also clear, that noise in the images degrades the results both by decreasing the correlation between input and output Strehls. This leads to systematic over-estimates of the Strehls, and thus under-compensated restored images.

Figure \ref{fig:PSF_comparison} shows the average input and output PSFs for the 4.2~m 
 and 1.4 m apertures. The averages were made by centering the individual PSFs on the brightest speckle, and calculating an azimuthal average. These plots demonstrate the much higher robustness of the 1.4~m PSFs compared to those of the 4.2~m aperture, but also that noise degrades the result and introduces a systematic overestimate of the Strehl.
 
Figure \ref{fig:MTF_comparison} shows  the squared MTFs calculated from the PSFs for $r_0=0.20$~m in the presence of low and high noise. These demonstrate that the over-estimated Strehls lead to under-compensated MTFs over a wide range in spatial frequencies. This explains the low contrast of (in particular) the 4.2~m images shown in Figs. \ref{fig:mfbd_50KL} and \ref{fig:mfbd_100-200KL}, as well as the systematically reduced power spectra in these images, as shown in Fig.~\ref{fig:mfbd_power}.

The overall decrease of the estimated wavefront RMS is summarised in terms of wavefront rms for zero noise and \expten{3.0}{-3} noise. For the 1.4 m aperture with $r_0=1$~m, these are (0.44, 0.39) rad, with $r_0=0.20$~m, these are (1.52, 1.29) rad. For the 4.2~m aperture, the corresponding pairs of wavefront rms are (0.91, 0.81) rad and (2.87, 2.01) rad.

\subsection{The importance of stable image quality} \label{sect:stable_image_quality}
A major challenge and priority for EST is to deliver stable image quality over sufficiently long time intervals (many minutes) to enable the recording of spectropolarimetric scans across photospheric and chromospheric spectral lines, and their time evolution. For such scans with narrow-band FPI systems, a noise level of \expten{1.0}{-4} is virtually impossible to reach at diffraction limited resolution, and even \expten{1.0}{-3} is very challenging. For many observing programs, noise levels of  \expten{3.0}{-3} will probably be the target. 

Our simulations demonstrate, that to reach the diffraction limit of EST {\gs over a 1\arcmin{} FOV} without MCAO, $r_0$ must be substantially larger than 0.35~m. This is because the absence of MCAO compensating for the high-altitude seeing suppresses the MTF of the full aperture of a 4-m telescope by more than a factor 10 at high spatial frequencies. The simulations suggest that, under the adopted turbulence model, recovering spatial frequencies approaching 90\% of the nominal 4.2-m diffraction cut-off would require noise levels of order \expten{3}{-4}  and substantially better high-altitude seeing than corresponding to $r_0$=0.35~m.
Without MCAO, the achievable resolution of EST will thus be noise limited, unless the high-altitude seeing is much better than assumed here.

Assuming values of $r_0$ of the high-altitude seeing layer in the range 0.2--0.35~m, our MFBD image restorations quite clearly demonstrate that EST operating in the multi-aperture mode is much more likely to provide stable image quality than with the full 4.2-m aperture. This follows from noting that the image quality demonstrated in Figs. \ref{fig:mfbd_50KL} and \ref{fig:mfbd_100-200KL} shows much smaller variations with the 1.4~m aperture, than with the 4.2~m aperture, when the seeing ($r_0$) and noise vary. The sensitivity to seeing variations is important, because even if the high-altitude seeing likely is relatively stable, the ground-layer seeing will most of the time show rapid and large variations (we remind the reader that MFBD deals with the accumulated ground-layer and high-altitude seeing - it therefore does not matter at what height the variability of the seeing occurs). The sensitivity of image quality to noise may seem less relevant, but actually also is of importance when scanning through strong photospheric and chromospheric spectral lines. 

Our analysis of the PSFs obtained from MFBD inversions in Sect. \ref{sec:psf_analysis} adds further support to the above conclusions. This demonstrates that the PSFs from MFBD inversions are much more accurate with 1.4~m subapertures than with the full 4.2~m aperture of EST. Both the true peak values (Strehls) of the PSFs and their shape are more accurately reproduced by MFBD with the smaller 1.4~m subapertures than with the full 4.2~m EST aperture. 

The above findings are hardly surprising: the much higher variance of the wavefront error with the larger aperture would require modelling with approximately 8 times more KL aberrations than for each of the smaller subapertures to deliver similar residual wavefront errors. This is far beyond the present capability of  MFBD and phase diversity techniques. In theory, {\gs a future method for image reconstruction \citep[such as the wavefront sensor assisted deconvolution method proposed by][] {2011OExpr..19.1975J} could dramatically} improve the quality of the 4.2~m restored images recorded without MCAO. However, the resolution and quality of the images would still be (severely) noise limited. Indeed, MCAO, or much better high-altitude seeing, is needed in order to fully explore the spatial resolution potential of EST with narrowband imaging. Until such an MCAO system is operational, we can fully benefit from the SNR advantage of the large 4.2~m aperture of EST at an excellent spatial resolution, and further boost the SNR advantage of EST by implementing a multi-aperture option.
 
The simulations made here lead to the conclusion that obtaining high-quality spectropolarimetric time sequences over the entire 1 arc min science FOV of EST is much more likely with the EST aperture segmented into 1.4~m sub-apertures than with the full 4.2~m aperture before MCAO is operational, when $r_0$ of the high-altitude seeing layer is in the range 0.2--0.35~m. With (substantially) larger values of $r_0$, EST should operate with its full aperture.

\section{The multi-aperture proposal}\label{sect:multiaperture_implementation}

\begin{figure*}[htbp]
\center
\scalebox{1.00} [1.07] {\includegraphics[angle=0, width=0.75\linewidth,clip]{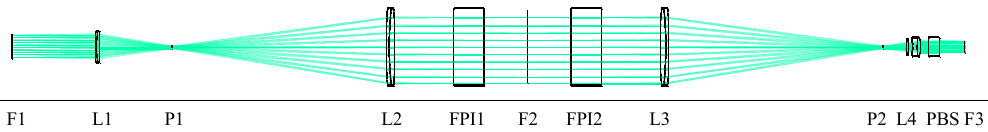}}
 \caption{
Layout of EST-V (overall length 4.5--4.7~m, FPI clear aperture diameter 180~mm). Symbols used: F1-F3 are focal planes, L1-L4 lenses, P1-P2 pupil planes, PBS polarising beam splitter. P2 is the pupil stop with the 2.3 mm prisms, L4 is the camera lens. The vertical scale has been expanded 2$\times$ for clarity.  Adopted from \citet{2026A&A...705A..55S}.}
\label{fig:FPI_layout}
\end{figure*}

\subsection{The implementation}
Our proposal provides an addition to the provisional optical design of three dual FPI systems proposed for EST, referred to as EST-B (380-500~nm), EST-V (500-680~nm) and EST-R (680-1000nm)  \citep{2026A&A...705A..55S, 2025arXiv250521053S}. These designs build on the successful implementations of three FPI systems developed for SST: CHROMIS, CRISP, and CRISP2. Of these, CRISP2 replaced CRISP in July 2025, and CRISP is now installed at the GREGOR telescope on Tenerife. All systems are compact, straight-through telecentric optical systems based on lenses and without folding mirrors. The layout of EST-V is shown in Fig. \ref{fig:FPI_layout}, which also explains the abbreviated symbols used below. 

Of particular interest here are the pupil images at P1 and P2, which for the EST FPI systems has diameters in the range 7--8~mm. For EST-V the pupil diameter is 6.8~mm both on the input and output sides. The lens L2 re-images the pupil plane P1 onto infinity at the location of the etalons (F2), and the lens L3 re-images P1 onto P2, which thus are optical conjugates of each other. At P1, all rays coming out from a particular point in the pupil pass the etalons with the same angle of incidence, but rays from different parts of the pupil pass the etalons with different angles of incidence. Because of the telecentric configuration with the pupil at infinity, this variation of the angle of incidence across the pupil is the same for every point in the FOV. By dividing the pupil into multiple apertures, each subaperture therefore corresponds to a smaller range of angles of incidence on the FPIs than for the full aperture.

Our choice is to only introduce temporary and easily removable minor modifications of the  exit sides of the FPI systems, and only in two ways:
\begin{itemize}
\item We add small (2.3~mm, in the case of EST-V) hexagonal prisms to the second pupil stop (at P2) of the FPI system (the one preceding the camera lens).
\item We replace the camera lens with a lens having three times shorter focal length.
\end{itemize}   
 
The purpose of the prisms is to produce sub-images (the spatial extent of which is defined by the field stop on the input side of the FPI system) that are spatially separated in the focal plane of the camera. The purpose of the lens is to re-image the sub-images with three times reduced spatial sampling, as compared to that of the default FPI setup. Since the pupil stop is optically conjugated to the pupil stop on the input side, the different subapertures correspond to selections of different ray bundles with different angles of incidence impinging on the FPIs. We also add a field stop of adjustable height and width at the focal plane on the input side of the FPI system, if such a field stop does not already exist.

We emphasise that the conversion of the default single aperture FPI system to a multi-aperture system is made in the camera re-imaging system, and this has zero impact on the FPI system itself. Changing between the two modes of operation, will for example not induce any risk of misaligning the system, or invalidating calibrations made -- it is a simple, low-risk operation.

The 7--8~mm pupil images of the EST FPI systems on the output sides means that the hexagonal prisms will have diagonals of about 2.5~mm. Though conventional glass prisms can be manufactured in these sizes and smaller (we are aware of one company that can manufacture prisms with sizes as small as 0.2~mm), this is also in the range of sizes of micro-lenses used in Shack--Hartmann WFS. For example, the microlenses for the SST AO system \citep{2024A&A...685A..32S} were made by Smart Microoptical Solutions (SMOS), a company that can make microlenses with diameters up to 3~mm with wavefront errors as small as 1/10 wave peak to valley. 

As an alternative to the proposed approach, it should also be possible to use a micro-lens array in the pupil plane, with a short enough focal length to re-image the seven subimages as separated from each other. Then this image is magnified several times by a re-imaging system to match the overall dimensions and pixel size of the detector. In the following, the priority is ``only'' to demonstrate the feasibility of our proposal, and we do that by developing one successful optical design for a multi-aperture implementation of the EST-V. We have chosen to do that based on the first concept described above, without claiming that this necessarily is the best approach. 

\subsection{Provisional optical design}\label{optical_design}

The purpose of our design is to demonstrate the feasibility of this approach, and to give an impression of the impact this has on the overall design of an FPI system for EST. Perfecting the design is not needed at this stage. To this end, we simplified the problem by a) restricting our design to that for EST-V,  b) limited the wavelength range to 610--670~nm, and c) limited the design for a pixel size of 6.5~$\mu$m, which is assumed to be the same pixel size as for the default single-aperture system (such that we can use the same cameras). The wavelength range chosen includes the magnetically sensitive 617.3~nm and 630.2~nm lines formed in the photosphere, and H$\alpha$ at 656.3~nm, which are major targets for this FPI system. So, in terms of science potential, this is a highly relevant design. 

Since the multi-aperture system must have a focal length that is about three times shorter than that of the default single-aperture system, a major concern is to ensure that the distance from the last surface of the polarising beam splitter to the focal plane of the camera is long enough for two cameras to be mounted corner to corner, centered on the transmitting and reflecting sides of the polarising beam splitter. This is why we choose a relatively small pixel size, which is more challenging than with a large pixel size.

\begin{figure}
\center
\includegraphics[angle=0, width=0.4\linewidth,clip]{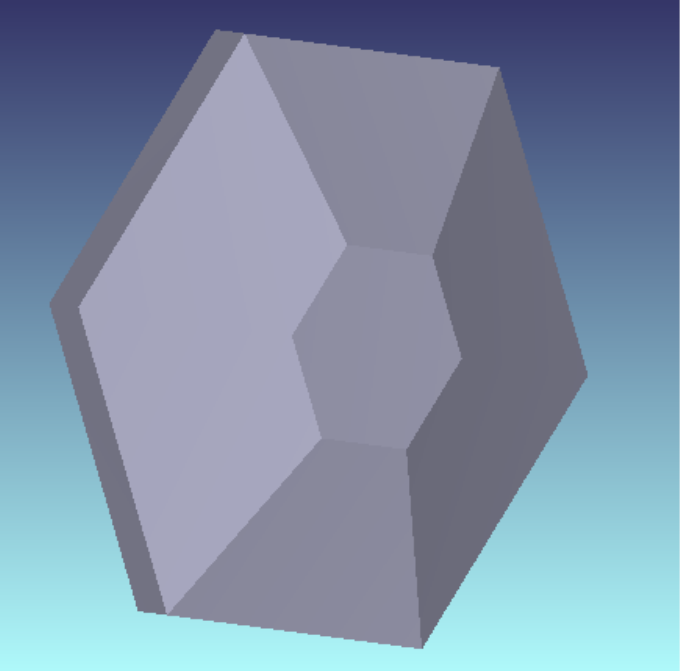}
 \caption{3D visualisation of the 2.3~mm prisms, located at the pupil stop P2. The purpose of the prisms is to deflect and separate the seven possible (six useful) subimages at the focal plane of the detector.}
 \label{fig:prisms}
 \end{figure}
 
 \begin{figure}
\center
\includegraphics[angle=0, width=\linewidth,clip]{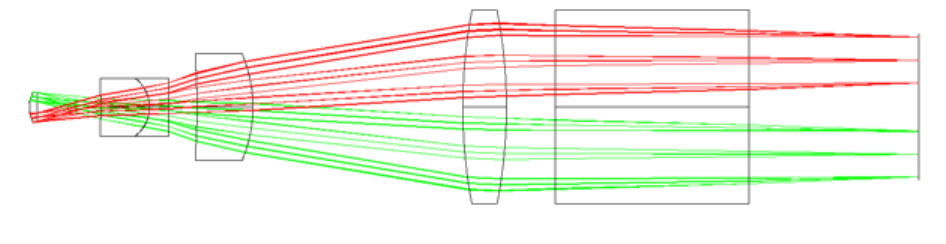}
 \caption{Schematic layout of the multiaperture camera lens (L4), the purpose of which is to reimage the seven possible (six useful) subimages at the focal plane of the detector. The prisms are located at the left side of the image (P2) but are not well shown. The design of the lens is made with a 40$\times$40~mm polarising beam splitter (also shown). The distance from the last surface of the beam splitter to the focal plane as shown is 35~mm, the overall length from the prisms to the focal plane is about 200~mm.}
 \label{fig:camera_lens}
 \end{figure}

 \begin{figure}
\center
\includegraphics[angle=0, width=\linewidth,clip]{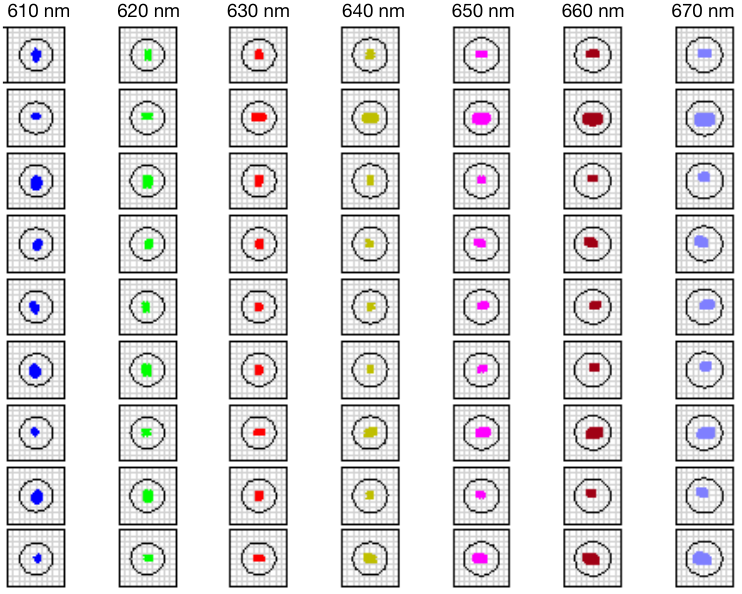}
 \caption{Spot diagrams for the multi-aperture modification of FPI-V for seven wavelengths in the range 610--670~nm, at various distances from the center of the 1\arcmin{} FOV. The lowest Strehl value is 0.95}
 \label{fig:spot_diagrams}
 \end{figure}

The design is illustrated in Figs. \ref{fig:prisms}--\ref{fig:camera_lens}. Figure \ref{fig:camera_lens} shows the layout from the prisms at P2 to the focal plane pf the camera. A 40$\times$40~mm polarising beam splitter is included in the design and is shown in the Figure. The distance from the last surface of the beam splitter to the focal plane is 35~mm, which should be sufficient for mounting two cameras picking up the images from the reflecting and transmitting side of the beam splitter. The image scale is 0\farcs039 per pixel, the FOV diameter 1\arcmin, and the sub-images fit within the same areas as used with the default full-aperture camera lens. The overall length from the pupil plane to the focal plane is 200~mm. In comparison, the distance from the pupil stop to the camera focal plane is 374~mm for the default full-aperture configuration of EST-V. Switching between these modes thus requires the cameras and polarising beam splitter to move about 174~mm.

Figure \ref{fig:spot_diagrams} shows the spot diagrams for the multi-aperture modification of EST-V (which includes prisms at pupil location P2 and the camera lens) for wavelengths in the range 610--670~nm. The lowest Strehl value is 0.95. We conclude that it is perfectly feasible to include a multi-aperture option in the design of EST-V, without sacrificing any aspect of the design or performance of the default full-aperture system. Since both systems use the same cameras, the cost of this added feature of EST-V is limited to the cost of manufacturing the prisms and the camera lens shown in Fig. \ref{fig:camera_lens}. Given the relative simplicity of this design, we see no reason why a similar multi-aperture option should not be feasible for EST-B and EST-R as well. Note, that also the wideband system of each FPI system needs an identical pupil stop with prisms and an identical camera lens.

\subsection{Simulations of performance}
At the location of the FPIs, the re-imaging system emanating from the pupil is telecentric and the angle of incidence is calculated as described by \citet{2011A&A...533A..82L}. Tilting of the HR FPI is implemented as a rotation around the x-axis (see Fig.\ref{fig:multi-aperture_layout}). The angle of incidence $\alpha$ of rays from the pupil is then given by
\begin{equation}
\cos\alpha = \frac{F \cos\theta + y \sin\theta}{(F^2 +x^2 +y^2)^{1/2}},
\end{equation}
where $\theta$ is the tilt angle of the FPI, $F$ the focal ratio of the beam and, as before, $x$ and $y$ are normalised to the pupil diameter. The imaging properties and spectral transmission profile were investigated with essentially the same software as described in \citet{2006A&A...447.1111S}, but modified to allow for pupils that are located off-axis.

The results of the simulations are summarised in Table \ref{table_EST-V}, and Tables \ref{table_EST-B} and \ref{table_EST-R} in Appendix \ref{table_CRISP2}, representing multi-aperture configurations of the three FPI designs developed for EST. At the top of each Table, we give the setup and performance of an ideal system with infinite F-ratio and that of the default single-aperture FPI system. We then summarise the two multi-aperture configurations, with the performance given for each of the six usable sub-apertures plus that of the unusable center sub-aperture (No. 4). The first of the multi-aperture configurations uses a HR etalon that is aligned perpendicular to the incoming optical axis. For both multi-aperture FPI systems, the FWHM is narrower and the peak transmission is higher than for their corresponding default configuration with full apertures. This is hardly surprising, since selecting a small part of the 4.2~m aperture corresponds to narrowing down the variation of angles of incidence on the FPIs within the corresponding sub-apertures. 

The second of the three multi-aperture configurations is obtained by slightly tilting the HR etalon. This reduces the angle of incidence on the etalons for the upper (first two) subapertures and increases the angles of incidence of the bottom two subapertures. The result is that the transmission peaks of the bottom two subapertures are tuned to shorter wavelengths than those at the top, whereas the middle two subapertures are tuned to an intermediate wavelength. Our calculations show, that the amount of possible blue-shift of the transmission peaks without strong degradation of the spectral transmission profile is limited and corresponds to one typical step when tuning through a spectral line. The tilts chosen in the Tables correspond roughly to that situation. This implies that the number of wavelength tunings in this mode can be reduced by about a factor of two, but not significantly more. Whether this is a mode of operation that is attractive needs further exploration, but this option is available at no risk and no cost once the basic multi-aperture configuration is operational. 

There is also a third multi-aperture configuration that is more elaborate and more costly. In this mode, we extend the FOV. This requires the first lens of the FPI system to be replaced by one having longer focal length, with the task of producing a larger pupil image at exactly the same location as for the default configuration. Our simulations lead to the conclusion that the FOV can only be extended by about 30\% linearly (70\% by area) without causing too strong degradation of the performance of the FPI system. Furthermore, such observations with a wider FOV would require the diameter of the pre-filters and polarisation modulator to be increased by 30\%. It seems questionable whether this rather modest increase in the FOV can justify these significant complications. We speculate that a better approach to a wide-field design might be based on an 18 subaperture (0.84~m subaperture diameter) design with an additional lens to shorten the focal length of the EST reimaging system (to avoid the need for large pre-filters). Such a development is beyond the scope of the present paper.

 \begin{table}[tbh]
  \caption{\label{table_EST-V} EST-V FPI system at a wavelength of 617 nm in its default 4.2~m configuration, and in two 1.4~m multi-aperture configurations.} 
  \centering
  \small
 \begin{tabular}{lcC@{\quad}CccCccc}
    \hline
    \noalign{\smallskip}
    \mathstrut
    FPI system  & Ap.  & x_c & y_c & FWHM & Tran &\delta\lambda  \\
                & No. &   &   & (pm) & (\%)  &  \textrm{(pm)}  \\
    \hline
    \hline
    \noalign{\smallskip}
    F-ratio: $\infty$ & - & \phantom-0.00 &  \phantom-0.00 &   5.43 & \llap100.0 & \phantom-0.00 \\
    \hline
    \noalign{\smallskip}
    EST-V default & - &  \phantom-0.00 &  \phantom-0.00 &   6.44 &  84.4 & -1.71 \\
    F-ratio: 147   & &  &   &   &   &  \\
    \hline
    \noalign{\smallskip}
    Multi-aperture & 1 & -0.50          & \phantom-0.87  & 6.01 & 92.9 & -1.71 \\ 
    F-ratio: 441   & 2 & \phantom-0.50  & \phantom-0.87  & 6.01 & 92.9 & -1.71 \\ 
    HR tilt: 0\fdg0   & 3 & -1.00          & \phantom-0.00  & 6.02 & 92.7 & -1.71 \\ 
                   & 4 & \phantom-0.00  & \phantom-0.00  & 5.44 & 99.8 & -0.18 \\ 
                   & 5 & \phantom-1.00  & \phantom-0.00  & 6.02 & 92.7 & -1.71 \\ 
                   & 6 & -0.50          & -0.87          & 6.04 & 92.5 & -1.71 \\ 
                   & 7 & \phantom-0.50  & -0.87          & 6.04 & 92.5 & -1.71 \\ 
    \hline                         
    \noalign{\smallskip}           
    Multi-aperture & 1 & -0.50          & \phantom-0.87  & 5.68 & 97.5 & -0.61 \\ 
    F-ratio: 441   & 2 & \phantom-0.50  & \phantom-0.87  & 5.68 & 97.5 & -0.61 \\ 
    HR tilt: 1\fdg5   & 3 & -1.00          & \phantom-0.00  & 6.30 & 89.5 & -2.56 \\ 
                   & 4 & \phantom-0.00  & \phantom-0.00  & 5.77 & 95.7 & -1.04 \\ 
                   & 5 & \phantom-1.00  & \phantom-0.00  & 6.30 & 89.5 & -2.56 \\ 
                   & 6 & -0.50          & -0.87          & 7.01 & 83.6 & -4.64 \\ 
                   & 7 & \phantom-0.50  & -0.87          & 7.01 & 83.6 & -4.64 \\ 
    \hline
    \noalign{\smallskip}
  \end{tabular}

  \tablefoot{The two multi-aperture configurations use the 4.2~m aperture of EST segmented into seven 1.4 m hexagonal subapertures (of these, aperture no. 4 cannot be used because of the 1.1 m central obscuration), see Fig.\ref {fig:multi-aperture_layout}. For reference, we also give the performance of an ideal FPI system, with an infinite F-ratio. The tilt of the high resolution FPI is given in units of $1/(2F)$ radians, where $F$ is the F-ratio. $x_c, y_c$ are the center coordinates of each subaperture given in units of the subaperture diameter. The wavelength shift given corresponds to that of the peak of the transmission profile.}
  \label{table_EST-V}
\end{table}
  
\begin{figure*}
\center
\includegraphics[width=0.25\linewidth,clip]{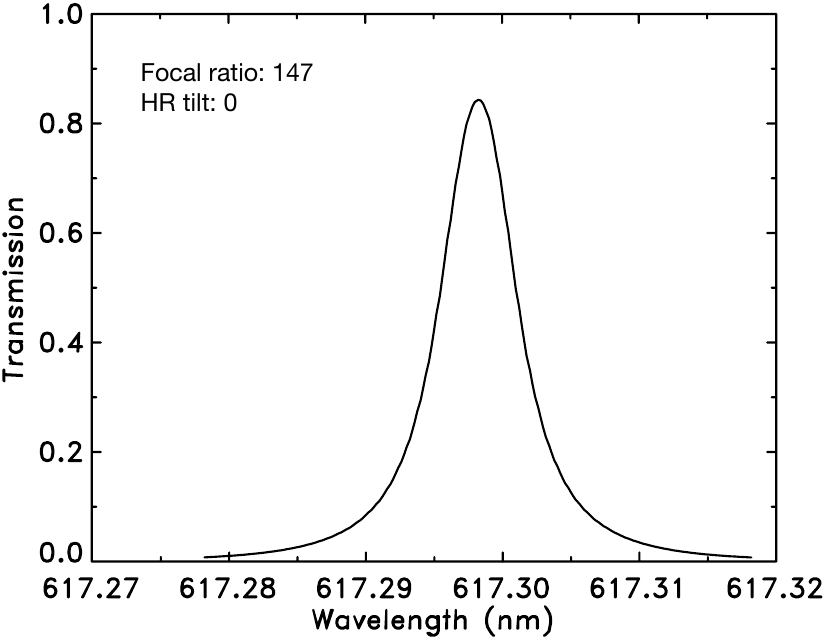}
\includegraphics[width=0.25\linewidth,clip]{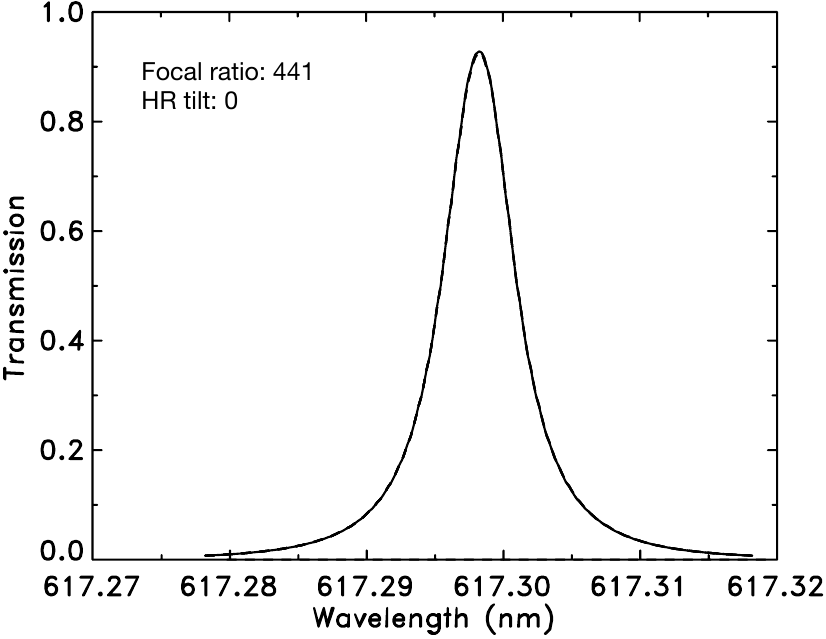}
\includegraphics[width=0.25\linewidth,clip]{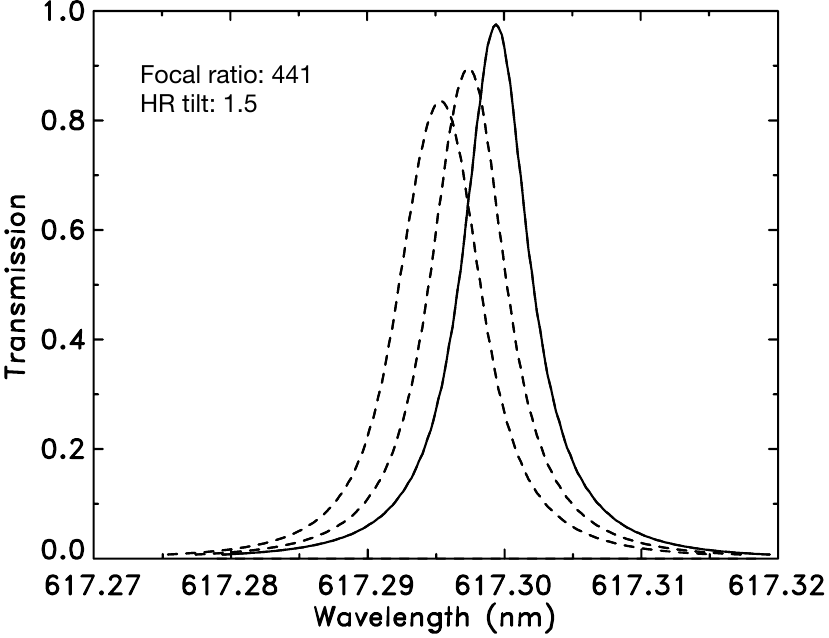}
 \caption{The FPI transmission profiles for EST-V in its default FPI configuration with full aperture, the multi-aperture configuration without HR etalon tilt, and the multi-aperture configuration with HR etalon tilt. Profiles from all six subapertures are plotted for the two multi-aperture configurations. For details, see Table \ref{table_EST-V}.}
 \label{fig:transmission_profiles}
 \end{figure*}

In Fig. \ref{fig:transmission_profiles}, we show the FPI transmission profiles for EST-V left to right, top to bottom, as follows: in its default F/147 configuration, multi-aperture F/441 configuration without tilt, and multi-aperture F/441 configuration with tilt. The corresponding peak transmission and FWHMs are given in Table \ref{table_EST-V}.
 
\section{Discussion}
The image quality obtained with a telescope for which $D/r_0\gg 1$ will be very poor without adaptive optics, even when using short exposures, as is well known from the early work of \citet{1966JOSA...56.1372F}, see also \citet[][Chapter 3]{2004aoa..book.....R}. SST seeing measurements suggest that{\gs , most of the time,} we should not expect the median value of $r_0$ for the high-altitude seeing layers above La Palma to be larger than 0.20--0.35~m (Appendix \ref {SST_AO_measurements}), in particular when the Sun is observed at low elevation. Indeed, most high-quality science data are obtained with the 1 m SST in the morning, and these suggest significant seeing-induced aberrations with an isoplanatic patch of only a few arc seconds. With a much larger telescope, such as EST, the high-altitude seeing will have a much stronger impact.

The 4.2~m EST will operate on La Palma with MCAO to mitigate the effects of high-altitude seeing. Short exposure times will be used for all observations with wideband and narrowband filter systems. However, for operational reasons, EST will during its first years operate with only an SCAO system \citep{managementplan2025,operationsplan2025}, which will compensate mostly near-ground seeing. Simple estimates based on the above values of $r_0$ suggest that such observations{\gs , most of the time,} will be carried out with a very low Strehl over almost the entire 1\arcmin{} science FOV, in particular during the morning. {\gs Diffraction-limited} image quality is only expected within a small part of that FOV, close to the lock-point of the  AO system. 

We therefore propose to divide the 4.2~m aperture into six segments of 1.4~m diameter optically by introducing low-cost and easily exchangeable modifications of the output sides of the three FPI systems designed for EST. These modifications have no impact on the design or use of the actual FPI system, and consist of adding six 2.5~mm prisms and/or microlenses to the second FPI pupil stop, P2, close to the camera lens (Fig. \ref{fig:FPI_layout}). In addition, the camera lens is replaced with a lens that has three times shorter focal length than in the full-aperture configuration. These modifications cannot cause any misalignment or disturb the calibration of the FPI system, and have no impact on the optical design of the FPI system when used in its full-aperture mode. Simulations suggest that this should work very well. A preliminary optical design clearly demonstrates that excellent image quality over the entire FOV is achievable with this approach. Even so, more efforts should be spent on refining this design and exploring various design options.

Simple estimates suggest that optically dividing the 4.2~m aperture into six usable 1.4~m apertures should allow observations to be made with Strehl of about 0.25, which should be sufficient for image reconstruction techniques to recover high image quality over the entire 1\arcmin{} science FOV. In this work, we explore the potential of the above proposal by means of extensive but somewhat simplistic simulations of seeing, PSFs, MTFs and images degraded by seeing, and MFBD restoration of seeing degraded images. These simulations are based on the following assumptions:
\begin{itemize}
\item Perfect AO compensation of ground-layer seeing (but see below)
\item A single high-altitude seeing layer, uncompensated for with MCAO, effectively relocated to the pupil, characterised by values of $r_0$ of 0.2, 0.35 and 1.0~m
\item An SCAO WFS with a FOV that is large enough to filter out the high-altitude seeing\footnote{With a smaller WFS FOV, there will be partial compensation of the high-altitude seeing and higher image quality within that FOV, but at the price of strongly increased wavefront errors in the surrounding FOV.}. The SST WFS uses a FOV of $12\arcsec\times12\arcsec$, which seems sufficient
\item Independent, random phase screens to emulate seeing
\item The use of small  $4\arcsec\times4\arcsec$ sub-fields to simulate the consequences of  anisoplanatism, based on experience from MFBD processing of CRISP data from SST. Note that phase errors from ground-layer and higher-altitude seeing simply are added up, and that the height of the seeing layer(s) is irrelevant for the MFBD processing
\item The use of many (here 3$\times$3) overlapping $4\arcsec\times4\arcsec$ sub-fields for building restored images over a large FOV, based on processing of CRISP data from SST
\item short-exposure imaging with the FPI system
\item MFBD image reconstruction with 50-100 KL modes, based on existing limitations experienced with SST and DKIST images
\item Unobscured EST aperture of 4.2~m diameter, but assuming six useful 1.4~m subapertures because of the EST central obscuration 
\end{itemize}
The above simplifying assumptions allow us to achieve our main goal of the present project, which is to investigate image degradation and restoration of short-exposure images in strong seeing with two vastly different aperture diameters (1.4 and 4.2~m). 

Our simulations provide strong support to the proposal of implementing a multi-aperture option for EST.
Assuming $r_0$ of the high-altitude seeing layer in the range 0.2--0.35~m, the MFBD reconstructed images with the full 4.2~m aperture are of low quality and with high sensitivity to variations in seeing and noise. The origin of this low quality of reconstructed images is discussed, by comparing the input and modelled PSFs. We conclude, that obtaining stable image quality over a large FOV, and over time, with the full aperture of EST could  be extremely difficult without MCAO, at least when the Sun is observed at low elevation. Our simulations suggest that for EST to reach its diffraction limit without MCAO {\gs over a 1\arcmin{} FOV} requires $r_0$ for the high-altitude seeing layer to be substantially larger than 0.35~m, and also requires images with extremely low noise levels.

The recent outstanding three minute long image sequence from DKIST \citep{Kuridze2026}, recorded through a 416~nm  broadband filter, may seem to contradict the above inferences. However, these images have a FOV of only 7\farcs$\times$5\farcs, which is smaller than the FOV of the DKIST AO WFS. The 1\arcmin{}$\times$1\arcmin{} context image shown in Fig. 1b of that paper demonstrates much lower image quality over most of the FOV. This is a direct consequence of the absence of MCAO, in spite of these observations being made near local noon, with the sun at an elevation of  about 74\degr. Moreover, each image is the result of a phase-diversity restoration of 2000 exposed frames, leading to an exceptionally low noise level of around \expten{2.5}{-4}(Sect. \ref{sect:gband_noise}). The remarkable quality of these images, combined with the much lower image quality in the surrounding FOV, implies that the DKIST SCAO system compensated a significant part of the high-altitude seeing at the center of the FOV.  If instead the elevation had been 20\degr, and the noise level an order of magnitude higher, achieving {\gs diffraction-limited} image quality would be far more challenging. 

These observations illustrate how SCAO and post-facto reconstruction can deliver the spatial resolution of a 4-m telescope over a small field of view when excellent ground-layer seeing conditions are present. However, as is the case on La Palma, there are also occasional outstanding days at the site of DKIST, when the differential seeing over a large FOV is significantly weaker than for the above data (F. W{\"o}ger, private communication), and that allows high-quality image restoration over a large FOV \citep[e.g.,][]{2025ApJ...985L..23K}.

Our simulations demonstrate, that observations with the 1.4~m multi-aperture setup is much more likely to deliver consistent, stable and uniform high image quality over the entire 1\arcmin{} science FOV, than with the full 4.2~m EST aperture. The much lower RMS wavefront errors in multi-aperture mode allow better MFBD image reconstruction. Increasing the number of exposed frames by a factor six (from the six useful 1.4~m subapertures) will also have an additional very positive effect on the quality of the restored images.

The importance of  this major difference between expected performance with the full 4.2~m aperture and its 1.4~m sub-apertures can hardly be overstated: achieving many of the EST top-level science goals critically depends on the recording of long-duration (many minutes) high-quality spectropolarimetric data with narrowband FPI systems over a large FOV, and for that stability of the image quality is absolutely crucial. This conclusion is supported by many years of experience with the 1-m SST, which routinely delivers long-duration sequences of spectropolarimetric data, and is located only 60~m from the proposed EST site on La Palma. 

The above discussion focuses on optimising the scientific output of EST during the prevailing seeing conditions, which correspond to morning observations with the Sun at low elevation. However, available seeing data from SST (Appendix \ref{SST_AO_measurements}, Fig. \ref{fig:SST_seeing}) also confirm the existence of days with excellent seeing near noon and median $r_0$ values from high-altitude seeing approaching 0.45~m at 500~nm about 25\% of the time. In such conditions, observations with the full 4.2~m aperture of EST can deliver images with much higher spatial resolution. We also emphasise, that $r_0$ is proportional to $\lambda^{6/5}$, such that $r_0$ varies by a factor 2.6 over the wavelength range 390--860~nm, which represent wavelengths of particular importance for observing chromospheric diagnostics. For example, a value of $r_0$ of 0.45~m at 500~nm corresponds to  0.86~m at 860~nm, but only 0.33~m at 390nm. Obviously, the decision of whether to observe in full-aperture or multi-aperture mode will critically depend on the strength of the high-altitude seeing as well as the operating wavelength of the individual FPI system.

\section{Conclusions}
With the zero impact it has on EST in full-aperture mode, adding the multi-aperture option to the EST FPI systems is a safe decision. This provides EST with the valuable option of either prioritising outstanding image quality over a small FOV, or lower but still excellent image quality over its entire 1\arcmin{} science FOV. By requiring only the addition of micro-optics at the 7--8~mm pupil image and new camera lenses, but not needing new cameras, the cost of adding multi-aperture mode of operation to the FPI system will be a very small fraction (on the order of 1--2\%) of the total cost of an FPI system. 

With properly designed mechanical interfaces, it should be possible to switch from single-aperture to multi-aperture mode and vice versa within an hour, perhaps even 10 minutes, and decisions on what mode to operate in can and should be made individually for each FPI system. Operating one FPI system in single-aperture mode and another in multi-aperture mode is perfectly possible. We do not exclude the possibility that the multi-aperture mode could be the preferred mode in some cases - in particular in the morning, with the Sun at low elevation -  even when MCAO is operational.

We expect the proposed multi-aperture implementation to substantially enhance the amount and quality of EST science data during its first years of operation, in particular when observations are carried out with the Sun at low elevation. In this way, EST can deliver unique high-quality science data over a large FOV, before MCAO is operational. Observing weak chromospheric magnetic fields should be particularly rewarding with this configuration.

We cannot yet make detailed simulations of  SCAO and MCAO performance on EST with respect to high-altitude seeing, and not yet establish optimal strategies for daily decisions on what observing mode to use with each of the EST FPI systems. For that, the height distribution of day-time seeing at the EST site on La Palma needs much better characterisation than is possible with the SST AO data described in Appendix \ref {SST_AO_measurements}. Such characterisation is now in its initial phase.

\begin{acknowledgements}

This material is based upon work supported by the NSF National Center for Atmospheric Research, which is a major facility sponsored by the U.S. National Science Foundation under Cooperative Agreement No. 1852977. We would like to acknowledge high-performance computing support from the Derecho system (doi:10.5065/qx9a-pg09) provided by the NSF National Center for Atmospheric Research (NCAR), sponsored by the National Science Foundation. 
Early on in this project, we used G-band images recorded with the Daniel K. Inouye Solar Telescope (DKIST) of the National Solar Observatory (NSO) and kindly provided by F. Wöger, for simulations of image degradation and MFBD image restoration. 

The Swedish 1-m Solar Telescope is operated on the island of La Palma by the Institute for Solar Physics ofStockholm University in the Spanish Observatorio del Roque de los Muchachos of the Instituto de Astrofísica de Canarias. The Institute for Solar Physics is supported by a grant for research infrastructures of national importance from the Swedish Research Council (registration number 2021-00169).  This work also has received funding from the European Union’s Horizon 2020 research and innovation programme under grant agreement
No 824135.

The European Solar Telescope project is supported by a grant for research infrastructures from the Swedish Research Council (registration number 2023-00169).  

CRISP and CHROMIS were funded by the Wallenberg Foundations, registration numbers 2003.0037 and 2012.1005. 

J. de la Cruz Rodr\'iguez gratefully acknowledges funding by the European Union through the European Research Council (ERC) under the Horizon Europe program (MAGHEAT, grant agreement 101088184). 

G. Scharmer acknowledges valuable comments and suggestions by F. Wöger, M. Collados, the EST AO team, and the anonymous referee, which significantly helped to improve the paper. 
Álvaro Pérez García is thanked for providing details of the design of POP.

\end{acknowledgements}


\begin{appendix}
\section{\label{SST_AO_measurements}High-altitude seeing estimated with the SST AO system}
\begin{figure}[]
\centering
\includegraphics[angle=0, width=0.9\linewidth,clip]{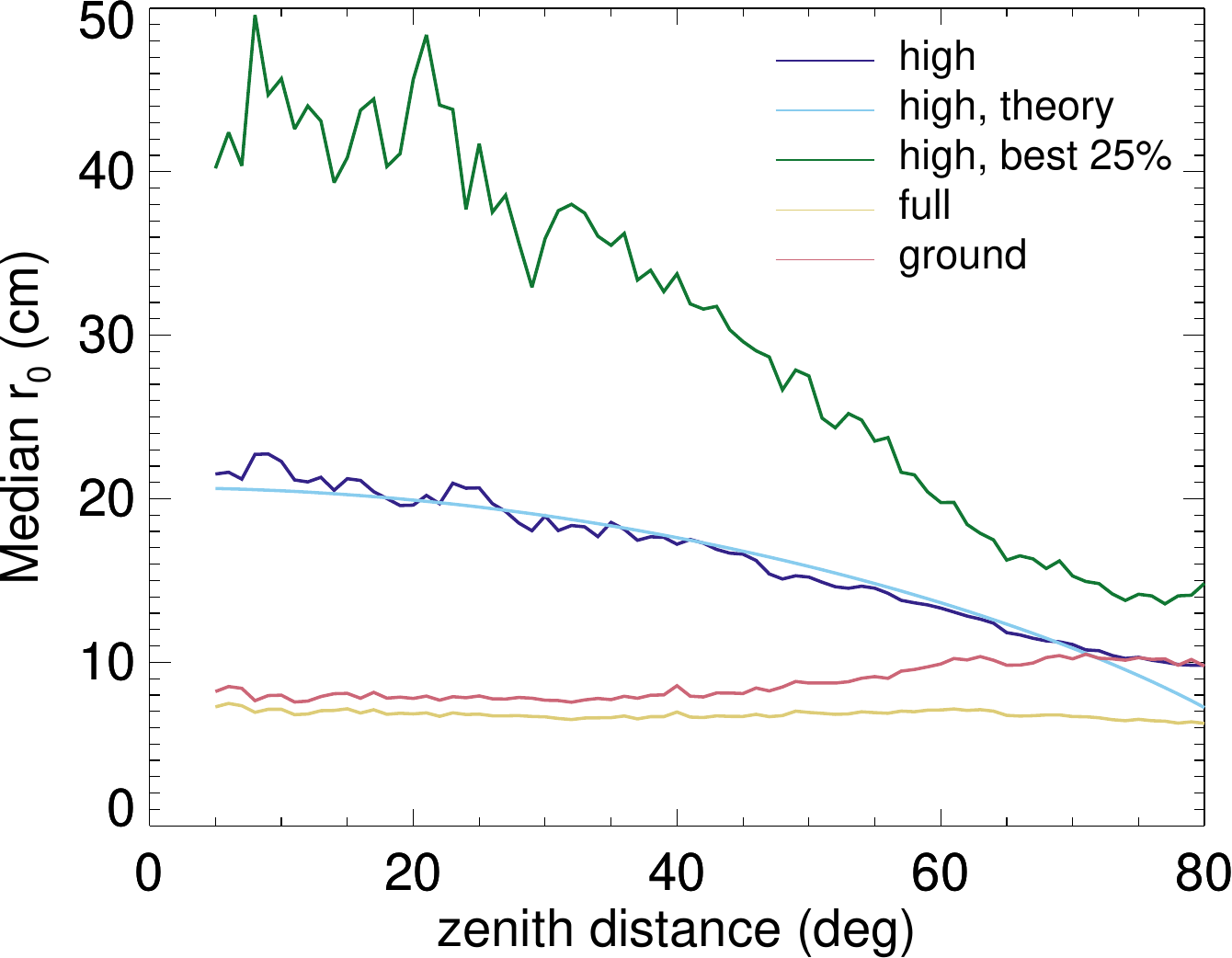}
 \caption{Median values of $r_0$ at a wavelength of 500~nm, obtained by processing all seeing data for 2019 from the SST AO system. Seeing measurements are made by simultaneously measuring differential image motion with 12\arcsec$\times$12\arcsec{} and 4\arcsec$\times$4\arcsec{} sub-fields. The larger sub-field is assumed to deliver measurements of ground-layer seeing and the smaller sub-field a mixture of ground-layer and high-altitude seeing. The high-altitude $r_0$ values are fitted to the expected $[\cos(z)]^{3/5}$ variation, where $z$ is the zenith distance.}
 \label{fig:SST_seeing}
 \end{figure}
 
The SST AO system  \citep{2024A&A...685A..32S} records statistics of seeing ($r_0$) at a wavelength of 500~nm, both with the entire 12\arcsec$\times$12\arcsec{} field of view (FOV) of the AO system  and sub-fields of only 4\arcsec$\times$4\arcsec{}  \citep{2019A&A...626A..55S,2024A&A...685A..32S}. We assume that the larger FOV of the AO wavefront sensor filters out the high-altitude seeing, and thus only measures the ground-layer seeing, whereas the small FOV measures the integrated (total) seeing. These seeing measurements are stored every second when the telescope is open and pointing to the Sun. Since seeing from different layers are uncorrelated, their variances add up. These variances are proportional to $r_0^{-5/3}$, such that $r_{h0}$ for the high-altitude layer can be obtained from the total $r_0$ and that of the low-altitude layer $r_{l0}$ from
\begin{equation}
r_0^{-5/3} = r_{l0}^{-5/3} + r_{h0}^{-5/3} 
\end{equation}
 \citep[here and in the following, we refer to the excellent reviews of ][chapters 2 and 3]{2004aoa..book.....R}. We use this equation to calculate $r_{h0}^{-5/3}$, and select measurements for which $r_{l0}^{-5/3}$
of the ground layer is larger than 0.05~m in order to have more confidence in our estimates of the high-altitude seeing. {\gs Figure \ref{fig:SST_seeing}, which is based on all seeing measurements made with the SST AO system in 2019, shows that at small zenith distances, the full seeing is almost entirely dominated by ground-layer seeing}, such that separating out the high-altitude seeing from the total seeing is sensitive to errors, and therefore challenging. 

Nevertheless, these plots show variations with zenith distances that are reasonable. In particular, the ground-layer seeing peaks when the Sun is at zenith distance in the range 60--80\degr, consistent with SST seeing most often being best in the morning. In contrast, the high-altitude seeing is best when the Sun is at its highest elevation, which is consistent with the expectation that daytime high-altitude seeing is unaffected by solar heating of the ground. Indeed, the variation of $r_{0h}$ is consistent with the expected $[\cos(z)]^{3/5}$ variation {\gs (curve labeled "high, theory")}. The estimates of $r_0$ delivered by the AO system includes real-time estimates of the noise level and corrects for that noise \citep{2019A&A...626A..55S}. We found that assuming a small amount of additional noise of about 0.16--0.19 radians, or 1/40--1/30 wave, slightly improved the fits of the data to the theoretical $[\cos(z)]^{3/5}$ variation with $z$, and this noise correction was adopted.

The upper panel of  Fig. \ref{fig:SST_seeing} implies that the average $r_{0h}$ for the high-altitude layer could be as low as 0.21--0.22~m, when the Sun is at zenith. In the lower panel of the same Figure, we have calculated the median values of  $r_{0h}$ taking into account only the 25\% of the measurements with the highest $r_{0h}$ values. This gives a much more favourable average of close to 0.45~m. Even this more optimistic estimate of $r_0$ is reduced to a mere 0.20~m at a zenith distance of 60\degr, such that taking advantage of the good morning seeing remains an obvious challenge with EST. 

These measurements of $r_0$ do not allow us to accurately establish the actual height, or range of heights, of the high-altitude seeing layer(s). However, this seeing must originate from layers that are high enough that the $r_0$ measurements made with  12\arcsec{}$\times$12\arcsec{} FOV are effectively averaged out but remain visible with a FOV of  4\arcsec{}$\times$4\arcsec{}. Furthermore, the high-altitude seeing layer must be located high enough above La Palma to be unaffected by the heating of the ground by the Sun, since its daily variation shows no such impact. We finally remark that {\gs raw images obtained with SST in variable seeing show uniform image quality variations over the entire FOV. Significant image quality degradation over only parts of the field are not observed, except sometimes when overall seeing conditions are very poor}. Instead, small-scale differential warping and blurring of fine structure is observed when the overall image quality is high. This suggests that the dominant seeing layers are from near the ground and at high altitudes but not at intermediate heights. 

The above limitations of the seeing measurements made with the SST AO system makes it impossible to make accurate estimates of the isoplanatic angle that we should expect with a future 4.2~m aperture EST. At the same time, the isoplanatic angle from a seeing layer well above the ground varies as $[\cos(z)]^{8/5}$, such that we can confidently state that the isoplanatic angle will be a strong limitation for observations made with EST when the Sun is at zenith distances in the range 60--75\degr, which according to vast experience at SST, will frequently offer very good observing conditions.

We emphasise, that Fig. \ref{fig:SST_seeing} shows measured values of $r_0$ at a wavelength of 500~nm, for other wavelengths $r_0$ can be calculated from $r_0(\lambda)=r_0(500~nm) [\lambda/500]^{6/5}$, where $\lambda$ is given in  nm.


\section{\label{table_CRISP2}EST-B and EST-V default 4.2~m and 1.4~m multi-aperture designs}
Table \ref{table_EST-B} shows the design of the narrowband EST-B FPI system and Table \ref{table_EST-R} the EST-R FPI system in their default 4.2~m configurations and two possible 1.4~m multi-aperture configurations. 
\nopagebreak{4}

\begin{table}[tbh]
  \caption{\label{table_EST-B} EST-B FPI system at a wavelength of 393 nm in its default 4.2~m configuration, and in two 1.4~m multi-aperture configurations.} 
  \centering
  \small
  \begin{tabular}{lcC@{\quad}CccCccc}
    \hline
    \noalign{\smallskip}
    \mathstrut
    FPI system & Ap. & x_c & y_c & FWHM & Tran & \delta\lambda \\
               & No. &     &     & (pm)   & (\%)   & \textrm{(pm)}   \\
    \hline
    \hline
    \noalign{\smallskip}
    F-ratio: $\infty$& - &  \phantom-0.00 &  \phantom-0.00 &   6.70 & \llap100.0 &  \phantom-0.00 \\
    \hline
    \noalign{\smallskip}
    EST-V default & - &  \phantom-0.00 &  \phantom-0.00 &   7.79 &  86.4 & -2.01 \\
    F-ratio: 110   & &  &   &   &   &  \\
    \hline
    \noalign{\smallskip}
    Multi-aperture & 1 & -0.50         & \phantom-0.87 & 7.33 & 93.8 & -2.01 \\ 
    F-ratio: 330   & 2 & \phantom-0.50 & \phantom-0.87 & 7.33 & 93.8 & -2.01 \\ 
    HR tilt: 0\fdg0   & 3 & -1.00         & \phantom-0.00 & 7.34 & 93.6 & -2.01 \\ 
                   & 4 & \phantom-0.00 & \phantom-0.00 & 6.70 & 99.8 & -0.24 \\ 
                   & 5 & \phantom-1.00 & \phantom-0.00 & 7.34 & 93.6 & -2.01 \\ 
                   & 6 & -0.50         & -0.87         & 7.35 & 93.4 & -2.01 \\ 
                   & 7 & \phantom-0.50 & -0.87         & 7.35 & 93.4 & -2.01 \\ 
    \hline
    \noalign{\smallskip}
    Multi-aperture & 1 & -0.50         & \phantom-0.87 & 6.87 & 97.9 & -0.70 \\ 
    F-ratio: 330   & 2 & \phantom-0.50 & \phantom-0.87 & 6.87 & 97.9 & -0.70 \\ 
    HR tilt: 1\fdg5   & 3 & -1.00         & \phantom-0.00 & 7.67 & 90.8 & -3.02 \\ 
                   & 4 & \phantom-0.00 & \phantom-0.00 & 7.07 & 96.3 & -1.25 \\ 
                   & 5 & \phantom-1.00 & \phantom-0.00 & 7.67 & 90.8 & -3.02 \\ 
                   & 6 & -0.50         & -0.87         & 8.36 & 85.4 & -5.34 \\ 
                   & 7 & \phantom-0.50 & -0.87         & 8.36 & 85.4 & -5.34 \\ 
    \hline
    \noalign{\smallskip}
  \end{tabular}

\tablefoot{The two multi-aperture configurations uses the 4.2~m aperture of EST segmented into seven 1.4 m hexagonal subapertures (of these, aperture no. 4 cannot be used because of the 1.1 m central obscuration), see Fig.\ref {fig:multi-aperture_layout}. For reference, we also give the performance of an ideal FPI system, with an infinite F-ratio. The tilt of the high resolution FPI is given in units of $1/(2F)$ radians, where $F$ is the F-ratio. $x_c, y_c$ are the center coordinates of each subaperture given in units of the subaperture diameter. The wavelength shift given corresponds to that of the peak of the transmission profile.}
\label{table_EST-B}
\end{table}

\begin{table}[tbh]
  \caption{\label{table_EST-R} EST-R FPI system at a wavelength of 854 nm in its default 4.2~m configuration, and in two 1.4~m multi-aperture configurations.} 
  \centering
  \small
  \begin{tabular}{lcC@{\quad}CccCccc}
    \hline
    \noalign{\smallskip}
    \mathstrut
    FPI system  & Ap.  & x_c & y_c & FWHM & Tran &\delta\lambda  \\
                & No. &  &  & (pm) & (\%)  &  \textrm{(pm)}  \\
    \hline
    \hline
    \noalign{\smallskip}
    F-ratio: $\infty$ & - &  0.00 &  0.00 &   10.42 & \llap100.0 &  \phantom-0.00 \\
    \hline
    \noalign{\smallskip}
    EST-V default & - &  0.00 &  0.00 &   11.54 &  90.1 & -2.38 \\
    F-ratio: 147   & &  &   &   &   &  \\
    \hline
    \noalign{\smallskip}
    Multi-aperture & 1 & -0.50         & \phantom-0.87 & 11.01 & 95.1 & -2.38 \\ 
    F-ratio: 441   & 2 & \phantom-0.50 & \phantom-0.87 & 11.01 & 95.1 & -2.38 \\ 
    HR tilt: 0\fdg0   & 3 & -1.00         & \phantom-0.00 & 11.03 & 95.9 & -2.38 \\ 
                   & 4 & \phantom-0.00 & \phantom-0.00 & 10.43 & 99.9 & -0.31 \\ 
                   & 5 & \phantom-1.00 & \phantom-0.00 & 11.03 & 95.9 & -2.38 \\ 
                   & 6 & -0.50         & -0.87         & 11.04 & 95.8 & -2.38 \\ 
                   & 7 & \phantom-0.50 & -0.87         & 11.04 & 95.8 & -2.38 \\ 
    \hline
    \noalign{\smallskip}
    Multi-aperture & 1 & -0.50         & \phantom-0.87 & 10.66 & 98.7 & -0.79 \\ 
    F-ratio: 441   & 2 & \phantom-0.50 & \phantom-0.87 & 10.66 & 98.7 & -0.79 \\ 
    HR tilt: 1\fdg8   & 3 & -1.00         & \phantom-0.00 & 11.37 & 93.2 & -4.15 \\ 
                   & 4 & \phantom-0.00 & \phantom-0.00 & 10.94 & 96.7 & -2.01 \\ 
                   & 5 & \phantom-1.00 & \phantom-0.00 & 11.37 & 93.2 & -4.15 \\ 
                   & 6 & -0.50         & -0.87         & 12.34 & 88.6 & -7.63 \\ 
                   & 7 & \phantom-0.50 & -0.87         & 12.34 & 88.6 & -7.63 \\ 
    \hline
    \noalign{\smallskip}
  \end{tabular}

\tablefoot{The two multi-aperture configurations uses the 4.2~m aperture of EST segmented into seven 1.4 m hexagonal subapertures (of these, aperture no. 4 cannot be used because of the 1.1 m central obscuration), see Fig.\ref {fig:multi-aperture_layout}. For reference, we also give the performance of an ideal FPI system, with an infinite F-ratio. The tilt of the high resolution FPI is given in units of $1/(2F)$ radians, where $F$ is the F-ratio. $x_c, y_c$ are the center coordinates of each subaperture given in units of the subaperture diameter. The wavelength shift given corresponds to that of the peak of the transmission profile.}
\label{table_EST-R}
\end{table}
\end{appendix}
\end{document}